\documentclass[12pt]{report} 
\usepackage{graphics}
\usepackage{a4} 
\usepackage{epsf}
\usepackage{amsmath}
\usepackage{amssymb}
\usepackage{bbm} 
\renewcommand{\v}[1]{{\mathbf #1}} 
 
\renewcommand{\sp}[2]{#1\! \cdot \!  #2} 
 
\newcommand{\s}[1]{\mbox{sign}(#1)} 
\newcommand{\pdd}[2] {\frac{\partial#1}{\partial#2}}

\renewcommand{\d}[1]{\mbox{d}#1} 
\newcommand{\dd}[2] {\frac{d#1}{d#2}}

\newcommand{\erf}{\mbox{erf}}

\newcommand{\vw }{\v{w}} 
 
\newcommand{\w }[1]{w_{#1}}

\newcommand{\as}{a^{*}}
\newcommand{\bs}{b^{*}}
\newcommand{\vas}{\v{a}^{*}}
\newcommand{\vbs}{\v{b}^{*}}

\newcommand{\vwb}{\bar{\v{w}}}
 \newcommand{\vwt}{\tilde{\v{w}}}
\renewcommand{\c }{\cos  (\theta) } 
\newcommand{\x }{\v{x}} 
\renewcommand{\a }{\alpha} 
\newcommand{\vat}{\tilde{\v{a}}}
\newcommand{\vbt}{\tilde{\v{b}}}
\newcommand{\at}{\tilde{a}}

\newcommand{\lt}{\tilde{\lambda}}
\newcommand{\wb}{\bar{w}}
\newcommand{\wt}{\tilde{w}}

\newcommand{\va}{\v{a}}
\newcommand{\vb}{\v{b}}
\newcommand{\vv}{\v{v}}
\newcommand{\C }{\v{C}}  
\newcommand{\Cm}{\mathcal{C}}
\renewcommand{\r}{\v{r}}

\newcommand{\vx}{\v{x}}
\newcommand{\vy}{\v{y}}

\newcommand{\vxt}{\tilde{\v{x}}}

\newcommand{\hh}{\mbox{$\hat{h}$}}
\newcommand{\htt}{\tilde{h}}
\newcommand{\hb}{\bar{h}}
\renewcommand{\l}{\lambda} 

\newcommand{\bea}{\begin{eqnarray}} 
\newcommand{\eea}{\end{eqnarray}} 
\newcommand{\be}{\begin{equation}} 
\newcommand{\ee}{\end{equation}}

\newcommand{\Wf}{ {\mathbf W} } 
\newcommand{\pif}{\boldsymbol{\pi}}

\begin{document}
\thispagestyle{empty}

\begin{center}
\large

\mbox{}\\[1cm]

\textbf{\Huge
Neural Networks, Game Theory \\[2ex]
and Time Series Generation}\\

\vspace{2cm}

 \epsfxsize= 0.6\textwidth
\centerline{ \epsffile{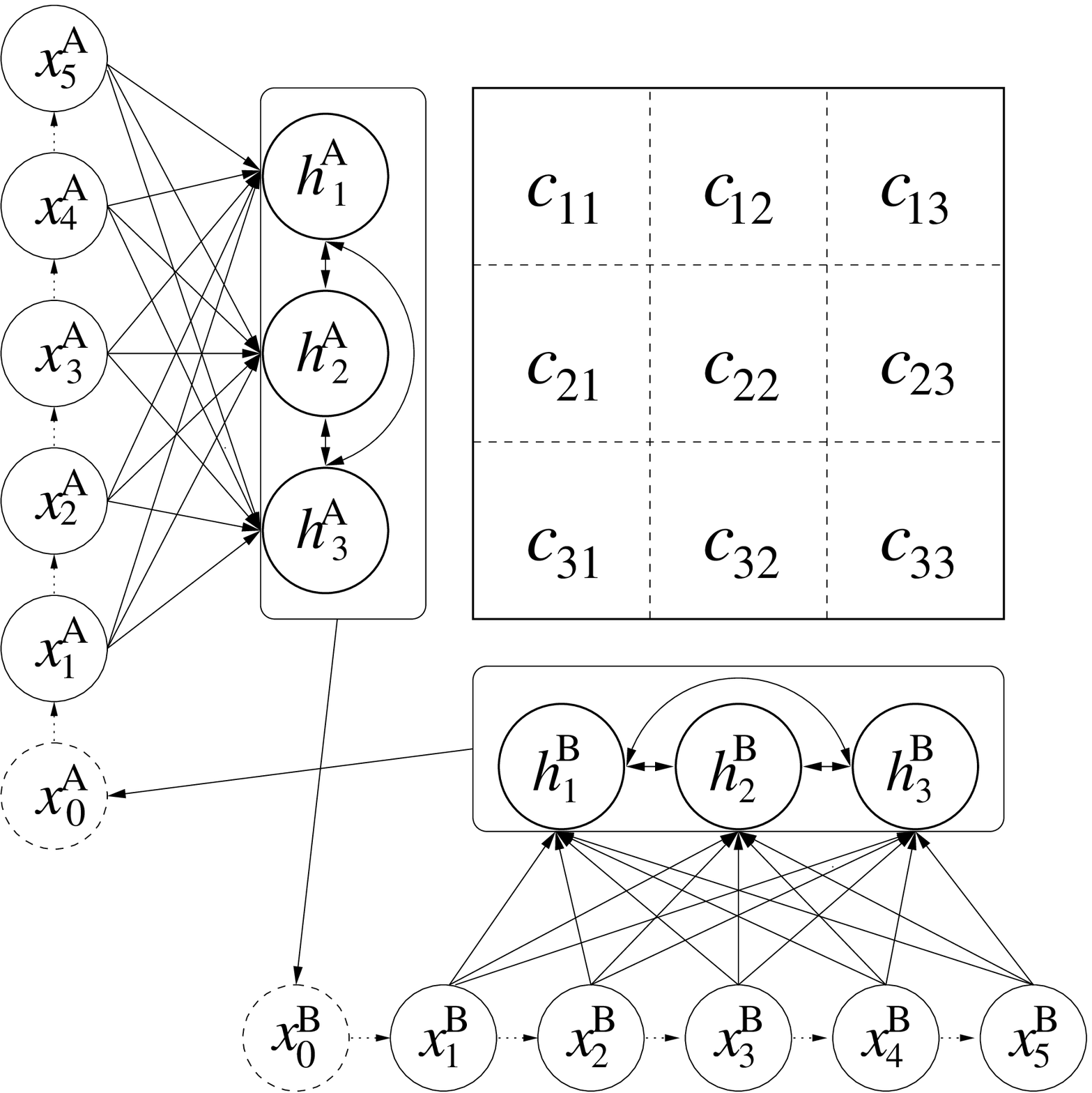}}

\vfill

{\Large Richard Metzler} \\[1ex]

\vfill

\textbf{
Institut f\"{u}r Theoretische Physik und Astrophysik\\
Bayerische Julius-Maximilians-Universit\"{a}t\\
W\"{u}rzburg}\\[1ex]

2002
\end{center}

\newpage 

\mbox{ }

\thispagestyle{empty}

\newpage 

\thispagestyle{empty}

\begin{center}
\large

\mbox{}\\[1cm]

\textbf{\Huge
Neural Networks, Game Theory \\[2ex]
and Time Series Generation}\\

\vspace{2cm}

\vfill

\begin{centering}
Dissertation zur Erlangung des \\
naturwissenschaftlichen Doktorgrades \\
der Bayerischen Julius-Maximilians-Universit\"{a}t\\
W\"{u}rzburg
\end{centering}

\vspace{2cm}

vorgelegt von\\[2ex]

{\Large Richard Metzler} \\[1ex]

aus  W\"{u}rzburg \\

\vfill

Institut f\"{u}r Theoretische Physik und Astrophysik\\
Bayerische Julius-Maximilians-Universit\"{a}t\\
W\"{u}rzburg\\[1ex]

September 2002
 \vspace{1cm}
\end{center}
\newpage
\thispagestyle{empty}

\large
\ \\
\vspace{10cm}

\noindent
Eingereicht am 3.9.2002\\
bei der Fakult\"{a}t f\"{u}r Physik und Astronomie\\
\vspace{0.5cm}
\noindent \\
Beurteilung der Dissertation:\\
\noindent \\
1. Gutachter: Prof. Dr. W. Kinzel\\
2. Gutachter: Prof. Dr. G. Reents\\
\ \\
\ \\
M\"{u}ndliche Pr\"{u}fung:\\
\ \\
1. Pr\"{u}fer: Prof. Dr. W. Kinzel\\
2. Pr\"{u}fer: Prof. Dr. A. Haase\\
\ \\
Tag der m\"{u}ndlichen Pr\"{u}fung: 1.10.2002

\newpage
\thispagestyle{empty}
\normalsize
\section*{Zusammenfassung}
Das kollektive Verhalten von Systemen aus vielen
wechselwirkenden Komponenten ist traditionell ein
Hauptthema der statistischen Physik. In den letzten
Jahren sind verst\"{a}rkt auch \"{o}konomische 
Systeme wie z.B. Aktienm\"{a}rkte untersucht worden,
deren ``Komponenten'' (Menschen und Firmen) durch 
Agenten mit ver\-ein\-fachten Entscheidungregeln 
(beispielsweise durch neuronale Netze) modelliert
wurden. Von besonderem Interesse sind dabei auch die
Zeitreihen, die durch solche Systeme erzeugt werden,
und die F\"{a}higkeit der Agenten, aus den Zeitreihen 
Informationen zu beziehen. 

Kapitel \ref{CHAP-antipredictable} dieser Arbeit widmet sich dem Konzept
der antivorhersagbaren Zeitreihen: zu jedem 
Vorhersagealgorithmus gibt es eine Zeitreihe, f\"{u}r 
die er v\"{o}llig versagt. Die Eigenschaften dieser
Zeitreihen werden f\"{u}r drei spezielle Algorithmen
untersucht und mit denen von gut vorhersagbaren
Zeitreihen verglichen. Aspekte von Interesse sind zum
Beispiel die L\"{a}nge von Zyklen bei diskreten Zeitreihen,
chaotisches Verhalten bei kontinuierlichen Zeitreihen und
die Unterdr\"{u}ckung von Korrelationen, auf die der 
Algorithmus empfindlich ist, bei antivorhersagbaren
Zeitreihen.

Der erste Algorithmus, der untersucht wird, ist das
einfache Perzeptron, bei dem die Dynamik des Gewichtsvektors
eine Reihe von Aussagen \"{u}ber die Autokorrelationen der
erzeugten Zeitreihe zul\"{a}sst. Eine eventuelle 
praktische Anwendung  zur
Erzeugung bin\"{a}rer Zeitreihen mit ma{\ss}geschneiderten 
Autokorrelationsfunktionen wird el\"{a}utert.

Der zweite Algorithmus, ein kontinuierliches Perzeptron,
stellt sich als komplizierte nichtlineare Abbildung heraus, mit
hochdimensionalem Chaos und intermittentem Verhalten. 
Nach einer Mean-Field-Rechnung zur Ermittlung der
statistischen Eigenschaften der erzeugten Zeitreihe
wird das Verhalten in Ab\-h\"{a}ngig\-keit von
Systemgr\"{o}{\ss}e und Verst\"{a}rkungsparameter untersucht
und Lya\-pu\-nov-Exponenten und Attraktordimension ermittelt.

Der dritte Vorhersagealgorithmus benutzt Boolesche
Funktionen. Einige Ei\-genschaften der erzeugten Zeitreihe
lassen sich durch graphentheoretische Ans\"{a}tze 
beweisen, z.B. die L\"{a}nge und Anzahl von Zyklen.
Die Erfolgsaussichten von gleichartigen
Vorhersagealgorithmen mit l\"{a}ngerem oder k\"{u}rzerem
Ged\"{a}chtnis werden untersucht.

Kapitel \ref{CHAP-MG} beschreibt verschiedene Varianten
des Minorit\"{a}tsspiels, bei dem eine gr\"{o}ssere 
Anzahl von Spielern sich unabh\"{a}ngig
voneinander f\"{u}r eine von zwei M\"{o}glichkeiten 
entscheiden soll und diejenigen gewinnen, die in 
der Minderheit sind.
Besonderes Augenmerk gilt den Varianten, die von 
unserer Arbeitsgruppe eingef\"{u}hrt wurden: 
eine korrekte und vollst\"{a}ndigere Betrachtung von 
neuronalen Netzen im Minorit\"{a}tsspiel
wird angestellt; 
die stochastische Strategie von Reents und Metzler 
wird vorgestellt und analytisch gel\"{o}st;
und es wird gezeigt, dass eine Verallgemeinerung
auf mehr als zwei Wahlm\"{o}glichkeiten f\"{u}r alle
g\"{a}ngigen Strategien m\"{o}glich ist und qualitativ
vergleichbare Ergebnisse bringt.

Eine Verbindung zu Kapitel \ref{CHAP-antipredictable}
wird dadurch geschaffen, dass bei vielen Varianten
in bestimmten Parameterbereichen die Zeitreihe 
der Entscheidungen gen\"{a}\-hert werden kann durch 
die antivorhersagbare Zeitreihe eines einzelnen 
Algorithmus, der die Mehrheit der Spieler 
repr\"{a}sentiert. 

Kapitel \ref{CHAP-game} beleuchtet eine andere Verbindung
zwischen neuronalen Netzen und Spieltheorie:
hier versucht ein neuronales Netz, in einem 
Zwei-Spieler-Null\-sum\-men\-spiel durch wiederholtes
Spielen und Anpassen der Gewichte eine gute Spielstrategie
zu entwickeln. Eine Abwandlung von Hebbschem Lernen wird
untersucht, die n\"{a}herungsweise die Nash-Gleichgewichtsstrategie
des Spiels finden kann. Die Eigenschaften dieser Regel
lassen sich f\"{u}r kleine Auszahlungsmatrizen analytisch
angeben, f\"{u}r gro{\ss}e Matrizen mit zuf\"{a}lligen 
Eintr\"{a}gen k\"{o}nnen Absch\"{a}tzungen gemacht werden.
\thispagestyle{empty}

Beim Lernen von Mustern mit 
Vorzugsrichtung (wie zum Beispiel Zeitreihenmustern) muss
dieser Algorithmus jedoch modifiziert werden. Die dann 
naheliegende Variante (die ebenfalls aus Hebbschem Lernen 
abzuleiten ist), stellt sich als Abwandlung eines 
bekannten Lernalgorithmus f\"{u}r Matrixspiele heraus.

 \cleardoublepage

\setcounter{page}{1}
\tableofcontents \cleardoublepage
\chapter{Introduction}
\label{CHAP-intro}
``Neural Networks, Game Theory and Time Series Generation''
-- the title of this dissertation contains three fields
of research that are so broad that bookshelves full of 
literature have been written on each one of them.
Consequently, I will start with a disclaimer: 
this dissertation is not a textbook, and it cannot give exhaustive
accounts of each field. Instead, it is my aim to present 
several projects which connect the three fields. 
Although more elaborate introductions to the ideas
and formalisms needed to understand the projects
will be given in the individual chapters, I will start
with a brief overview over the fields and an outline
of the dissertation.

\section{Neural Networks}
In an attempt to understand the workings of animal
and human nerve systems, simple mathematical 
models of nerve cells (neurons) have been devised, which can be 
combined into artificial neural networks of considerable 
complexity. The simplest building block of these 
networks, the McCulloch-Pitts neuron \cite{McCulloch:Calculus},
is interesting enough when studied in isolation.
This network, which is then called a simple perceptron,
can perform limited, but nontrivial, feats of 
storing, classification, and learning of unknown rules
\cite{Minsky:Perceptrons,Hertz:NeuralComp,Riegler:Dynamics}.
It formalizes two feature of biological neurons: they
receive signals from other cells via synaptic
connections, and they are active if the added input
exceeds some threshold, and inactive if it does not.

For increasing complexity and realism, compared to the
simple perceptron, two directions suggest themselves:
one is to assemble more complex structures out of
simple perceptrons, such as multi-layer
feed-forward networks, associative memory networks
or recurrent networks \cite{Hertz:NeuralComp}. 
This direction tends towards computer science and
machine learning. The other direction is to incorporate
more details from biological mechanisms
\cite{Hodgkin:Quantitative,Tuckwell:Introduction}, such as 
explicit modeling of cell membrane potential and 
firing rates. While these models are hard to approach 
analytically, they can be used to emulate biological
pattern recognition problems realistically 
\cite{Hopfield:WhatI,Hopfield:WhatII}.
In this dissertation, however, the simplest architectures  
that are suitable to a given task -- namely, simple and
continuous perceptrons, and multi-class perceptrons --
will be used. Understanding their interaction with
themselves  or other networks is a sufficient challenge.

\section{Time series generation}
One of the applications that neural networks have been
put to is the prediction of time series 
\cite{Weigend:Time,Freking:Learning}.
The idea here is that many physical systems generate
a stream of observables at regular time intervals
(such as daily temperature measurements) that 
exhibit certain regularities. These properties of
the time series can be used by a suitable algorithm
to predict future values from a finite number of 
past values. Time series that can be predicted 
without error by some algorithm are those that can 
be generated by the same algorithm by feeding the
predicted value back into the history that is used
to make the next prediction. Therefore, it is 
helpful to look at the properties of a time series
generated by an algorithm to learn about its prediction 
capacities.

It has been pointed out in Ref. \cite{Kinzel:Seq.} that
the success of a prediction algorithm depends completely 
on the properties of the time series that it is applied to,
and that there are always time series for which a given
algorithm fails completely. These sequences can again be
generated by that algorithm by inverting its output and
feeding it back to the history. Comparing these
sequences to the ones that are well predictable for
the same algorithm sheds more light on the capabilities
of the algorithm.

\section{Game theory}
Game theory deals with situation in which two or more
players must make decisions, and the payoff which each player
receives depends on the combined decisions of all players.
In such situations, being able to predict what the
others are going to do is extremely useful, whereas 
making decisions that are predictable for others is not.

Two game-theoretical problems will be treated in detail in
this work: in two-player zero-sum games, both players
pick one of their available options independently
at the same time, and the
gain of one player is always the loss of the other.
As the groundbreaking studies of v. Neumann showed 
\cite{Neumann:Theory}, this type of game has a unique
equilibrium set of strategies for both players,
which can be determined by rational calculations
and beyond which no improvements for either player
are possible.
 This allows to quantify the success of a given learning
algorithm  that develops a strategy from repeated 
playing, by comparing the achieved strategy to the
equilibrium strategy.

In the second scenario that will be presented, the
Minority Game \cite{MGhomepage}, a large number of players pick one
of two options, and those in the majority lose.
Although randomly choosing one option would be the
rational strategy, other prescriptions (collectively
labeled ``bounded rationality'') where players try
to predict the majority decision can lead to coordination
among players, and thus improved average gains, compared to
random guessing. On the other hand, attempts to avoid the
predicted majority can lead to over-reactions, which result
in herding behavior and dramatically reduced average gains.
 
\section{An outline of this work}
As mentioned, to understand the properties of prediction
algorithms, it is helpful to study the properties
of sequences that can be predicted well by them 
and especially those are predicted wrongly every time. 
In Chapter \ref{CHAP-antipredictable}, I will do 
this for three prediction algorithms: simple and
continuous perceptrons, and Boolean functions.
Aspects of interest are properties of cycles (for
discrete time series) and attractors (for continuous ones),
and especially the suppression of correlations that the
prediction algorithm is sensitive to in antipredictable
time series.

Chapter \ref{CHAP-MG} will present different aspects of 
the Minority Game,
focusing on the variations of the game that were introduced
by our research group, such as neural networks
(Sec. \ref{SEC-MGNN} -- this is a first link between
neural networks and game theory) and the stochastic strategy 
presented in Sec. \ref{SEC-sto} (where, if a memory
is included, the players can behave like a Boolean 
function with a learning algorithm). 
It will be shown that in Sec. \ref{SEC-PMG} all presented
strategies can be generalized to more that two options.

I will point
out, where applicable, under what circumstances the dynamics
of the Minority Game generate a time series that is 
partly antipredictable for the individual players, and
completely antipredictable for a predictor that represents
the ensemble of players, thus combining game theory and
time series generation. 

Chapter \ref{CHAP-game} presents a different aspect
of connecting game theory and neural networks: 
there, a simple neural network tries to learn a 
two-player zero-sum game by adapting its weights and
learning from experience. If random patterns
are presented to the network, a modified Hebbian 
learning rule can lead to a good strategy
and, under some circumstances, even converge to the
equilibrium strategy. 

If biased patterns are used, the learning rule has to be 
modified. The modified rule turns out to be a stochastic
variation of a well-known learning algorithm for matrix games.

  \cleardoublepage
\chapter{Antipredictable time series}
\label{CHAP-antipredictable}
\section{Prediction of time series}
A time series is a sequence
of values $x^{t_1}, x^{t_2},\dots$ with $t_1<t_2<\dots$\ .
Often, the values are taken at regular time intervals,
and the times $t_1, t_2,\dots$.
 can be denoted by integer numbers $1, 2, \dots$.
In most cases that are of interest to physicists, 
time series are observables of a physical system or 
variables of a mathematical model, and one tries to characterize
the system by studying the statistical properties
of the time series. However, in many cases it is
of practical interest to {\em predict} the time series,
i.e., make good guesses about the future values 
$x^{t+1}$, $x^{t+2},\dots$  from knowledge about a
finite number of past values $x^{t-M+1}, \dots, x^{t}$
\cite{Weigend:Time,Schlittgen}.
For example, considerable effort goes into the
prediction of the weather (temperature, rainfall
etc.), and many amateur and pro stock brokers would
love to be able to predict the future of stock prices.
 
In the absence of information on external factors 
which might influence the time series (like new economic 
data that allow guesses about stock prices, or 
ecological events that influence the climate), a 
prediction algorithm can only take a finite 
stretch of the time series as input data and
produce a guess based on this data. In other words,
a prediction algorithm is a function $g(\v{x},\v{w})$ that maps an
input vector $\vx$ (whose components are the $M$ past 
values of the time series, which may be binary or discrete) 
onto a (continuous or discrete) output.
 \footnote{In game theory, 
``prediction'' can mean predicting the probability 
distribution of possible outputs instead of giving a single
most probable value \cite{Foster:Impossibility}. I will not
use this definition, which is difficult to apply to
systems where only one of the possible realizations of
randomness is observed.}
The function may depend on internal parameters $\vw$, such 
as the weight vectors in a neural network.

\begin{figure}
\epsfxsize= 0.8\textwidth
  \centerline{\epsffile{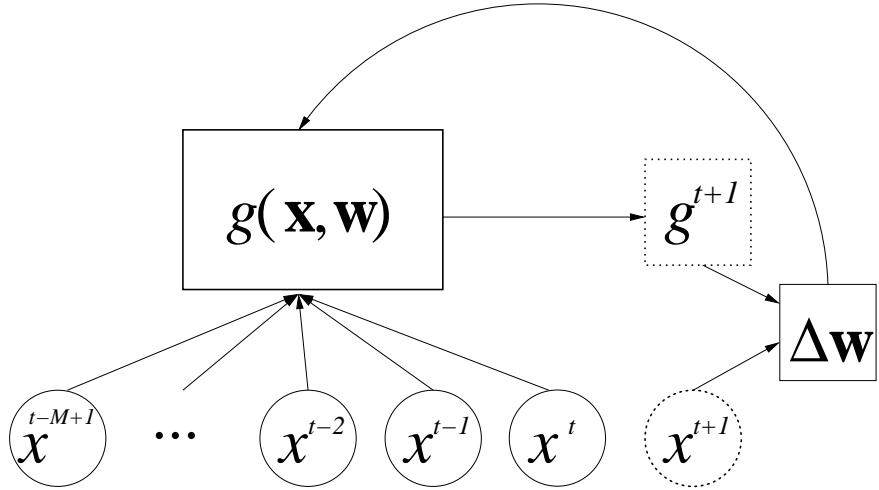}}
\caption{A simple representation of a prediction 
algorithm: external inputs $\vx$ are fed into a function
$g(\vx,\vw)$. Internal parameters $\vw$ are modified
after comparing prediction $g^{t+1}$ and reality $x^{t+1}$.} 
\label{PRED-predict1}
\end{figure}

In the following sections, I will concentrate mainly on
discrete time series $x^t \in \{-1,1\}$, switching to
the binary notation $x^t \in \{0,1\}$ where it is more
convenient. For discrete time series, the concept
of accuracy is easy to define: the error rate is simply
the percentage of predictions that do not agree with
the time series. In some cases, continuous or 
multi-valued discrete values will be considered as well.

Adaptive prediction algorithms are capable of changing 
their internal parameters based on a comparison between
the time series and their prediction of it. One reasonable
demand is that a prediction algorithm does not adapt 
its parameters as long as it is $100\%$ accurate, and
adapts if there are discrepancies between prediction
and reality -- this is for example realized in
gradient-descent-based learning algorithms, where the
norm of the update that is added to the parameters is 
proportional to the error.

Another reasonable assumption is that 
the predictor is deterministic: even though 
the model to be predicted may contain noise,
adding noise to the predictor is unlikely to 
make a prediction better (leaving aside concepts
like stochastic resonance \cite{Gammaitoni:Stochastic}).

The time series that is to be predicted is generated
by some  system 
that can be arbitrarily complex, and can include any amount of 
randomness. The values of the
time series (which represent observables like temperature, 
stock values etc.) reveal some information about the
state of the system. For example, in deterministic 
chaotic systems with a finite number $m$ of degrees of 
freedom, the full information about the state of the
system is contained in a times series of at most $2m$ 
values of a suitable observable \cite{Sauer:Embedology}. 

The prediction algorithm is then a more or less crude 
model of the system that generates
the time series, as sketched in Fig. \ref{PRED-predict1}.
Adjusting the parameters can 
improve the fit between the model and reality, or, 
to speak in the language of neural network theory, 
the overlap between ``student'' and the ``rule'' \cite{Watkin:Statistical} 
or ``teacher'' \cite{Biehl:On-line}.
\begin{figure}
\epsfxsize= 0.65\textwidth
  \centerline{\epsffile{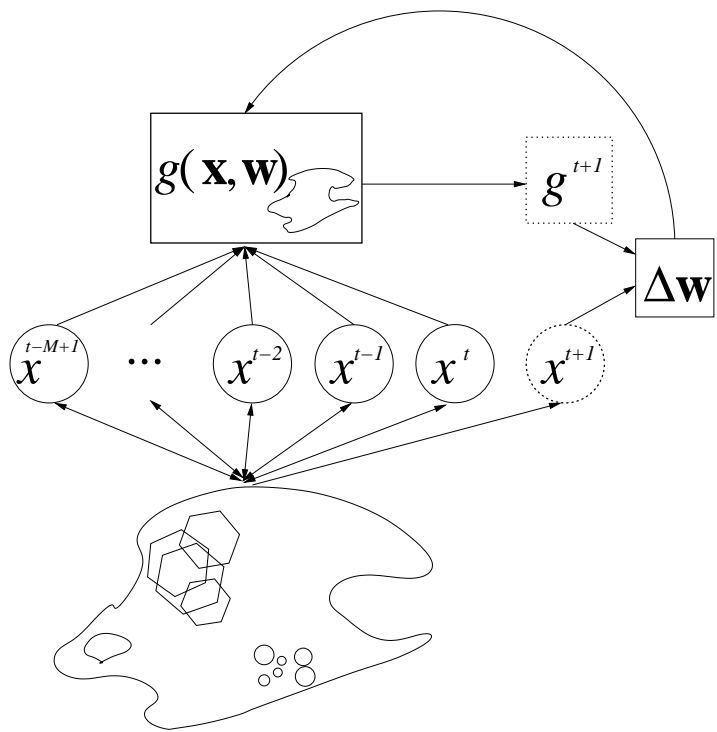}}
\caption{The next step: the system that generates the
time series is included in the consideration. The 
predictor models some (hopefully relevant) properties
of the system, and tries to improve the fit. } 
\label{PRED-predict2}
\end{figure}

One could assume that, if the  prediction algorithm 
has structural similarity to the generator of the time
series, and if the time series is largely deterministic,
the algorithm will predict the
time series with fair accuracy after some training 
time. However, it was pointed out in
Ref. \cite{Kinzel:Seq.} that for any algorithm
that makes a binary prediction, there 
exists a generator of the same degree of complexity that
generates a time series which will make the predictor
fail every single time. This generator can be constructed
easily: it is an exact copy of the predictor, including the 
internal parameters, the adaption algorithm and 
the random number generator, that takes the same input data
and inverts the output.
As the example of the antipersistent binary time series will
show later (see. Sec. \ref{SEC-tabgen}), 
in some cases this scheme  will also 
fool predictors that start with a different set of 
initial parameters, if they ``forget'' their initial 
state quickly enough.

One interesting thing about the antipredictable sequence
generated in this fashion is that it is completely 
deterministic. In the same fashion, deterministic 
sequences can be designed that make the predictor fail
every second, third, etc., prediction \cite{Kinzel:Seq.}.

In the framework outlined above, time series that can be 
predicted without error by an algorithm are those 
where the output of the algorithm is exactly the next
value of the time series. The time series can thus be
constructed (without any external information) 
by feeding the newly generated
bit back into the pattern. Analogously,
antipredictable sequences are those generated by the
algorithm, inverting the bit and feeding it back. In the 
following, prediction and generation are thus very much 
interchangeable, and in order to study the properties of
time series that are predictable or antipredictable,
I will actually study time series {\em generated} by
the algorithm.

\begin{figure}
\epsfxsize= 0.75\textwidth
  \centerline{\epsffile{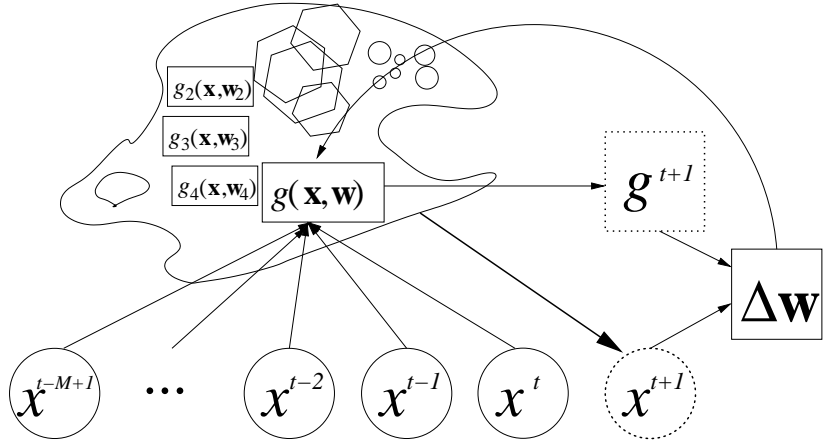}}
\caption{Sometimes, the predicting algorithm is
part of the system that it is trying to predict,
together with other algorithms and a complex 
environment.} 
\label{PRED-predict3}
\end{figure}

The concept of antipredictability may seem highly 
artificial. However, it can turn out to be relevant 
if the predictors are part of the system they are 
trying to predict (see Fig. \ref{PRED-predict3}). 
For example, the concept of the Minority Game that is 
explained in Chapter \ref{CHAP-MG} has a strong element
of self-defeating prophecy.

The following sections will compare the properties of 
predictable and antipredictable sequences for two
simple algorithms: a simple neural network, and a 
Boolean function.  Sections \ref{SEC-AntiHebb} and
\ref{SEC-CSG} are based on from Refs.
\cite{Metzler:CBG} and \cite{Metzler:Diplom}, whereas
Sec. \ref{SEC-Decision} was presented in 
Ref. \cite{Metzler:Antipersistent}. These differences
will shed some light on the considered algorithms
themselves, and the features in the time series
that they are sensitive to, giving a better intuition
about the applicability of an algorithm to a given problem.
 
\section{Simple Perceptrons: the Confused
  Bit Generator}
\label{SEC-AntiHebb}

\subsection{Sequence generation by a static perceptron: the
  Bit Generator}
\label{SEC-BG}
The first example of a prediction algorithm that I am
going to present is the simplest
model of a neural network, the simple perceptron
\cite{Minsky:Perceptrons,Hertz:NeuralComp}. Inspired
by biological neurons \cite{McCulloch:Calculus}, this model has 
received much attention, and in spite of its simplicity,
it shows many interesting features (see
Ref. \cite{Watkin:Statistical} for an overview).

In analogy to a biological neuron that receives impulses
from other neurons through synaptic links and either 
starts firing impulses itself or stays passive depending
on the received inputs, the perceptron performs a
weighted sum over a set of inputs $x_i$, $i=1,\dots,M$.
Each input is multiplied with a synaptic weight $w_i$.
The output is either $+1$ or $-1$, depending on whether
the weighted sum of inputs is larger or smaller than 0.

In a geometrical interpretation, both the weights and the input data
represent $M$-dimensional vectors $\vw$ and $\vx$, and the output 
is the sign of the scalar product between the vectors.

One way to generate a time series using a perceptron (or, for
that matter, any other feed-forward neural network \cite{Priel:Long-term})
is to use an $M$-bit window of the time series as input and
continue the time series with the output on that pattern:
\be
x^{t+1} = \s{\sum_{i=1}^{M} w_i x^{t-i+1}}. 
\ee 
For the simple perceptron with binary output (and, correspondingly, a binary 
time series) and fixed weights, this system, which was named 
Bit Generator (BG), was studied in Refs. 
\cite{Eisenstein:BG,Schroeder:Diss.,Schroeder:Cycles}.
It is a deterministic dynamical system with a finite state space:
there are only $2^M$ possible histories. Once a history appears
the second time, the system is on a cycle. Simulations show that
cycles have a typical length $l<2M$; their average length
$\langle l\rangle$ (averaged over initial conditions) increases
polynomially in $M$ \cite{Schroeder:Diss.}. Typically, 
cycles are dominated by one short sequence that is 
interrupted by ``defects'', such as
$+-+-+-+++-+-+-\dots$\ . The wavelength of the
underlying short sequence usually corresponds to a 
wavelength that is strongly represented in a Fourier
decomposition of the weight vector.

As long as the weight stays fixed, the predictable 
sequence for a given $\vw$ is precisely the anti-predictable
sequence for $-\vw$. Things get more interesting if the 
weight vector becomes adaptive.

\subsection{Hebbian Learning: the Confused Bit Generator}
In order to find regularities in a set of 
data, the weights of a neural network have to adapt to the data. This can 
occur either offline (all patterns and desired outputs
are known and are processed at the same time) or online
(patterns are shown to the network one by one and then 
discarded). Since the patterns and the output to
be learned are generated by the predictor,
offline learning makes little sense. Therefore,
online learning will exclusively be considered from now on. 

Online learning algorithms for perceptrons come in many 
variations, ranging from simple to elaborate 
\cite{Hertz:NeuralComp,Kinouchi:Optimal,Biehl:Adatron,Riegler:D,Riegler:Dynamics}.
However,
most boil down to the same principle: the updated 
weight vector $\v{w}^{t+1}$ is a linear combination of the previous
weight vector $\v{w}^{t}$ and the pattern $\v{x}^t$. This seems 
natural for symmetry reasons, since
these two vectors are the only preferred directions in a
possibly high-dimensional space. 
 \footnote{The interpretation of the pattern as a time
   series, however, breaks the symmetry between input
   dimensions: it may be reasonable to give more weight
   to the more recent past. Secs. \ref{SEC-Bernasconi} and
   \ref{SEC-shaping} provide examples for this.}

The coefficients of the linear combination can depend on the
desired output (in our case, the next step of the time series,
$x^{t+1}$), the so-called hidden field $h^t =
\sp{\v{w}^t}{\v{x}^t}$, a learning rate
$\eta$, and a few other quantities, which should be accessible to 
the neural network.
The two most basic learning algorithms are the Hebbian
rule
\be
\v{w}^{t+1} = \v{w}^{t} + \frac{\eta}{M} \v{x}^t x^{t+1},
\ee
and the Rosenblatt rule,
\be
\v{w}^{t+1} = \v{w}^{t} + \frac{\eta}{M} \v{x}^t x^{t+1} 
\Theta (-x^{t+1} \s{h^t}).
\ee
In the latter case, the correction step is applied only 
if the output of the perceptron and the desired output
disagree. Since this is always the case for a completely 
antipredictable time series, both rules amount to the same
in this case. 
The perceptron which always learns the opposite
of its own output was studied first in Ref. \cite{Kinzel:Seq.},
and examined extensively in Refs. \cite{Metzler:Diplom} and
\cite{Metzler:CBG}, where
it was named Confused Bit Generator (CBG). A very similar
system was studied in Ref. \cite{Samengo:Competing}.
The CBG is defined by the equations
\bea
x^{t+1} &=& - \s{\sp{\x^t}{\vw^t}}; \\ 
\vw^{t+1} &=& \vw^t + (\eta/M) x^{t+1} \x^t. \label{CBG-update}
\eea
This is a deterministic dynamic system with $2M$ variables,
instead of $M$ variables in the BG. Indeed, the dynamics
of the weight vector are very helpful in understanding
the properties of the generated time series.
\subsection{Dynamics of the weight vector}
\label{CBG-Dynamics}
Geometrically speaking, $\vw$ does a random-walk-like
movement on an $M$-di\-men\-sion\-al cubic lattice: in each component
of the weight vector, the learning step has a value 
of $\pm \eta/M$. Although the weight components are real 
numbers, once an initial state $\vw^0$ has been defined,
the components $w_i$ can only take values $w_i^0 \pm n
\eta/M$,
with $n \in \mathbb{Z}$. 

Each learning
step has a negative overlap with the current $\vw$.
The norm of the weight vector fluctuates around 
an equilibrium value that can be estimated by 
 taking the square of Eq. (\ref{CBG-update}) and applying the
usual formalism for online learning
\cite{Riegler:D,Saad:Online}.
Since the patterns are windows of the time series and
are hence generated by a complex dynamical process, the
required averages over scalar products between the
weight and the pattern cannot be done exactly.
However, the generated time series looks sufficiently 
irregular at first glance (no prominent frequency, no
short-term repetitions) that one can try 
replacing $\vx$ with a random vector whose components
are independent random variables of mean 0 and variance 1, 
and see how far that approximation carries:
\be
\langle \sp{\vw^{t+1}}{\vw^{t+1}} - 
  \sp{\vw^t}{\vw^t}\ \rangle = 
-\frac{2 \eta}{M} \langle \sp{\x^t}{\vw^t} \s{\sp{\x^t}{\vw^t}} \rangle
+ \frac{\eta^2}{M^2} \langle \sp{\x^t}{\x^t} \rangle .
\end{equation}
Introducing a time scale $\a$ with $\d{\a} = 1/M$ and averaging
over $\x$, this becomes a deterministic differential equation
for the norm $w = |\vw|$ in the thermodynamic limit
$M\rightarrow \infty$: \footnote{This limit will be tacitly assumed
  every time a differential equation is written down; it
  usually gives good results even for moderate $M$ (on the
  order of $M\sim 20$.)}

\be
\dd{w}{\a} = -\sqrt{\frac{2}{\pi}} \eta + \frac{\eta^2}{2w}.
\label{CBG-dw_da}
\ee
The attractive fixed point of this equation is $w = \sqrt{\pi/8} \eta
\approx 0.6267 \eta$.  The learning 
rate $\eta$ thus only sets a length scale, but does not 
influence the behavior of the system once the weight vector
has reached its fixed-point length. 

However, using the time series generated by the perceptron
as patterns, simulations give a slightly different
value of $w \approx 0.566 \eta$, independent of $M$
(this was already observed in \cite{Kinzel:Seq.}).
Two possible mechanisms for this deviation suggest
themselves: first, the time series windows generated by the CBG
are not isotropically distributed on the $M$-dimensional
hypercybe; some patterns are more likely to appear than
others (see Ref. \cite{Metzler:Diplom}) for details). 
Second, correlations between the outputs and the patterns
might be the cause of the deviations. This can be checked
by using patterns drawn randomly from an anisotropic distribution
that resembles the one generated by the CBG. Since this
leads to values of $w$ compatible with the result for random
vectors, one must conclude that correlations between outputs
and patterns are responsible for the deviation.

Using the approximation of random patterns, the
autocorrelation of the weight vector can be estimated 
as well: 
\be
\langle \sp{\vw^t}{\vw^{t+\tau}} \rangle = 
  w^2 \exp \left ( -\frac{4}{\pi} \frac{\tau}{M} \right ).
\label{CBG-w_autocorr}
\ee
Of course, this does not reflect the fact that the CBG is 
a deterministic system with a bounded number of states,
so cycles are inevitable, and the weight vector must 
return to its original point later (i.e., the
autocorrelation of the weights will return to 
$w^2$ at multiples of the return time). 
The dynamics of the weights can be linked to the 
the autocorrelation 
function $C_j^t$ of the sequence, defined by
\be 
C_j^t = \sum_{i=1}^{t} x^i x^{i-j},
\ee
where $t$ is the number of patterns summed over.
Simply add $t$ update steps according to Eq. (\ref{CBG-update}):
\be
w_j^t  = w_j^0 + \sum_{i=1}^{t} (\eta/M) x^i x^{i-j}
 = w_j^0 + (\eta/M) C_j^t. \label{CBG-w_corr}
\ee
Each value $C_j^t$ for $1\leq j \leq M$ corresponds
to the distance of the weight vector from its starting
point along one axis in the $M$-dimensional weight space,
measured in units of $\eta/M$. This point is important 
and will be exploited in the following paragraphs.
Note that Eq. (\ref{CBG-w_corr}) holds for any perceptron
that learns a time series following the Hebb rule,
regardless whether the time series is predictable,
antipredictable or anything in between. This
means that in the long run, the perceptron is only sensitive
to the first $M$ values of the autocorrelation function of 
the considered time series.  The stronger the
autocorrelation of a sequence is, the
more accurate the prediction will be.

In the CBG, the norm of the weight vector, and therefore
the autocorrelation function of the generated sequence, is
suppressed: as described, the norm of $\v{w}$ stays bounded.
Following Eq. (\ref{CBG-w_corr}), each component of the 
autocorrelation function stays bounded as well, instead of growing with
$\sqrt{\alpha}$ as it would for a random sequence. 

\begin{figure}
\epsfxsize= 0.70\textwidth
  \centerline{\epsffile{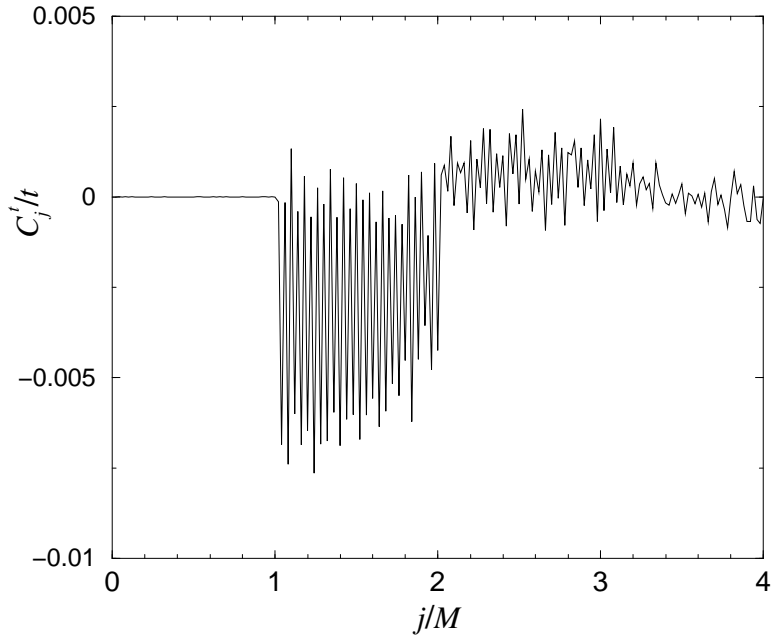}}
\caption{Autocorrelation function of the time series
  generated by the CBG, averaged over $t=2\times10^6$ for
  $M=50$ (taken from Ref. \cite{Metzler:Diplom}). The first 
  $M$ components are suppressed by the dynamics of the
  weights.}                     
\label{CBG-corr}
\end{figure}

\subsection{The {B}ernasconi model}
\label{SEC-Bernasconi}
Bit series with low autocorrelations are of interest in 
mathematics and have applications in signal processing
\cite{Schroeder:Zahlentheorie}.
On one hand, this allows to apply findings on these
sequences to the CGB.  
On the other hand, it is interesting whether the CBG generates sequences
with autocorrelations significantly lower than for random series.
Two measures that indicate low autocorrelations 
are commonly used in the literature:
for periodic sequences of length $l$, an energy function
(which is studied in the so-called Bernasconi  model 
for periodic boundary conditions \cite{Bernasconi:LowAuto}) 
can be defined by
\be 
H_p = \sum_{j=1}^{l-1} (C_j^{l})^2 = 
  \sum_{j=1}^{l-1} \left (\sum_{i=1}^{l}x^i x^{i+j}\right)^2.
\ee
Results on the ground states of this Hamiltonian can be found in 
\cite{Mertens:Bernasconi}. By trial and error, initial conditions for
the CBG can be found which yield cycle s slightly larger
than $2M$, for which all value of $C_j^l$ except one are 0.
However, even for the best sequences we found in 
long simulation runs, 
$ H_p$ was larger than the known ground state energies 
by at least a factor of 2. 

The original model does not use periodic boundary conditions: in a 
sequence of length $p$, only the sum 
over $p-j$ different terms with a lag of 
$j$ can be calculated. The energy for aperiodic
sequences is therefore given by 
\be
H_{ap} = \sum_{j=1}^{p-1} (C_j^{p-j})^2
\ee 
(note the summation limits).
The so-called merit factor $F$ introduced by Golay \cite{Golay:Merit}
is defined by 
\be
F = \frac{p^2}{2 H_{ap}}.
\label{CBG-def_F}
\ee
A merit factor of 1 is expected for a random sequence;
lower autocorrelations yield higher $F$. The theoretical
limit for large $p$ is conjectured to be about $F=12$ 
\cite{Bernasconi:LowAuto,Ein-Dor:Low}, whereas optimization routines 
such as simulated annealing typically find
sequences with $5< F<9$ (see \cite{deGroot:LowAuto} and 
references therein) and exact enumeration for  $p<60$ suggests 
$\lim_{p \rightarrow \infty} F= 9.3$ for the optimal sequence
\cite{Mertens:Exhaust}.

\begin{figure}
\epsfxsize= 0.70\textwidth
  \centerline{\epsffile{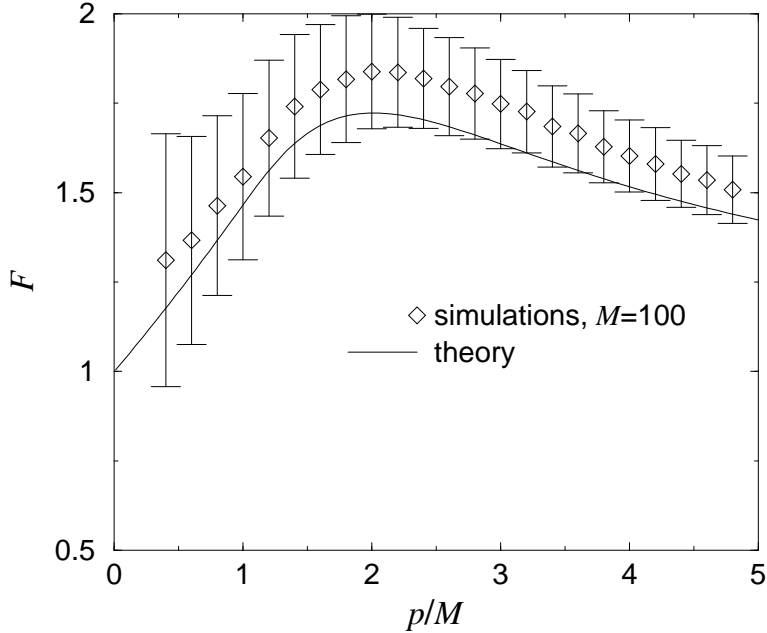}}
\caption{Merit factor $F$ of sequences generated by the CBG,
as a function of scaled 
sequence length $p/M$, compared to Eqs. (\ref{CBG-mf1}) and 
(\ref{CBG-mf2}). 
Error bars denote the
standard deviation of $F$ for $M=100$, 
not the standard error.} 
\label{CBG-meritf}
\end{figure}

To analytically estimate the merit factor of sequences
generated by the CBG, one can solve Eq. (\ref{CBG-w_corr})
for ${C_j^t}^2$ and use the autocorrelation of the weights 
given by Eq. (\ref{CBG-w_autocorr}):
\be
 \langle (C_j^{p-j})^2 \rangle = 
     \frac{M^2}{\eta^2} 
       \langle {w_j^0}^2 + {w_j^t}^2 - 2 w_j^t w_j^0 \rangle
      =    \frac{\pi}{4} M 
       \left (1 - \exp \left (- \frac{4}{\pi}\frac{p-j}{M} \right) \right).   
\label{CBG-csq}
\ee
The energy can be expressed as a sum or approximated by 
an integral in continuous variables $\alpha = p/M$ and 
$\beta = j/M$. Since Eq. (\ref{CBG-csq}) only holds for
$1\leq j\leq M$, ${C_j^{p-j}}^2=p-j$ must be used for $j>M$. 
One gets the expression
\bea
H_{ap} &=& \sum_{j=1}^{p-1} M \frac{\pi}{4} 
  \left (1- \exp \left 
      (-\frac{4}{\pi}\frac{p-j}{M} \right) \right) \nonumber \\
&\approx& \int_0^{\alpha} 
    M^2 \frac{\pi}{4} (1- \exp(-(4/\pi)(\alpha-\beta))) d\beta 
    \nonumber \\
&=& M^2 \frac{\pi}{4}\left [ \alpha - \frac{\pi}{4}
      \left (1- \exp\left (-\frac{4}{\pi} \alpha\right
        )\right )\right]  \label{CBG-mf1}   \\
&&\mbox{ for }j\leq M\mbox{ and }  
 \nonumber     \\
H_{ap} &=& M^2\left[\frac{\pi}{4} 
  \left (  
    1 - \frac{\pi}{4}\left \{ \exp\left (\frac{4}{\pi}(1-\alpha)\right) 
    - \exp(-\frac{4}{\pi}\alpha) \right \} \right ) + 
  \frac{1}{2}(\alpha-1)^2 \right ] \nonumber \\
&& ~~~\mbox{ for } j>M.
\label{CBG-mf2} 
\eea
The corresponding merit factor is compared to
simulations in Fig. \ref{CBG-meritf}: Eqs. (\ref{CBG-mf1})
and (\ref{CBG-mf2}) give qualitatively  correct results, but differ
from the observed values by roughly $10\%$. The
feedback mechanisms of the CBG cause a faster decay of $C_j^t$
than predicted for random patterns. 

A simple extension of the dynamics allows to slightly manipulate this
result: if each weight component gets an individual
(positive) learning
rate, one can write down an update equation for each
component that separates the influence of that component
on the output from all others.
A short calculation (again under the assumption of random 
patterns, which has yielded qualitatively correct
predictions so far) then shows that the mean square value of that
weight is proportional to its learning rate: 
\be
\langle w_i^2 \rangle = 
  \sqrt{\frac{\pi}{8}} \frac{\eta_i}{M} \sqrt{\sum_j w_j^2}
 = \frac{\pi}{8} \frac{\eta_i}{M^2} \left(\sum_{j} \eta_j
 \right).
\label{w-eta}
\ee

Although Eq. (\ref{w-eta}) looks like
a regular on-line learning equation, it does not carry 
quite as far: $ w_i^2 $ is not a
self-averaging quantity \cite{Reents:Self-avg.}, 
i.e., the variance of $w_i^2$
does not go to zero in the thermodynamic limit, and
Eq. (\ref{w-eta}) only makes a statement about the 
long-time average.

What it does state is that a component with a 
higher learning rate has, on the average, a
larger weight and thus a stronger influence on the output. 
This also leads to a faster decay  of the corresponding segment of
the autocorrelation:
\be
\langle w_i^t w_i^{t+\tau} \rangle = 
 \frac{\sum_j \eta_j}{\eta_i} \frac{\pi}{4} 
  \exp \left ( -\frac{\eta_i}{\sum_j \eta_j} \frac{4}{\pi} \tau \right ).
\ee
Returning to the Bernasconi problem, the search for a minimal $H_{ap}$
can be written as an optimization problem in the 
continuous function $\eta(\beta)$, where $\eta_j = \eta(j/M)$.
Solving this problem with a variational approach, one
finds that it is sensible to give the last 41\% of the
weights a learning rate and norm of zero, and increase
the learning rate continuously towards  components
with smaller indices. Unfortunately, even this optimization
does not improve the merit factor beyond $F=1.74$ in theory 
and $\langle F \rangle =1.86$ in simulations. This is
still a lot worse than the results of other optimization
methods \cite{Bernasconi:LowAuto,deGroot:LowAuto}, 
so the CBG is not a competitive generator
of low-autocorrelation sequences. 

The limited use of the CBG in generating sequences with a
high merit factor may be related to phase space arguments:
as seen in Sec. \ref{CBG-cycles}, the CBG can still generate
exponentially many different time series depending on 
initial conditions, whereas there are very few sequences 
with the highest achievable high merit factors (see \cite{Mertens:Density}
for the density of states with cyclic boundary conditions).
The mechanism of the CBG allows for manipulation of the
autocorrelation function only if the constraints on the
desired sequence are not too strong, such as suppressing
all of the elements of $C_j$ on a short time scale.
On the other hand, choosing a given shape for long-time
averages of $C_j$ still allows for many realizations of
the sequence. 

Although it is not very useful in a straight application
to the Bernasconi problem, the Confused Bit Generator 
still has some interesting possibilities:

\subsection{Shaping the autocorrelation function}
\label{SEC-shaping}
In some cases, it may be interesting to generate 
a time series whose power spectrum has a specific 
shape in the long-time limit. 
Using Eqs. (\ref{w-eta}) and (\ref{CBG-w_corr}) in the limit where 
$w^0\approx 0$ and with non-negative learning rates, 
one can obtain the inverse relation 
between the square of the autocorrelation function ${C_j^t}^2$ and the 
corresponding learning rate ${\eta_j}$ 
\be
 \langle {C_j^{t}}^2 \rangle = 
     \frac{\pi}{8} \frac{\sum_{i=1}^{M}{\eta_i}}{\eta_j}. 
\label{C-eta}
\ee
Thus, just about any desired shape of the square
autocorrelation  function is
 achievable by using the appropriate profile for $\eta_j$, which can  
be extracted from Eq.(\ref{C-eta}).

If one is not only interested in the shape, but also the
norm of the autocorrelation function, one can influence 
this by distorting the output with multiplicative noise:
if there is a probability of $p_f$ for flipping the output,
\be
x^{t+1} = \left \{ 
\begin{array}{rl} 
\s{\sp{\x}{\vw}} &\mbox{~~with    probability~~}p_f \\
-\s{\sp{\x}{\vw}} &\mbox{~~with    probability~~}1-p_f
\end{array}
 \right.,
\ee
the learning step has a positive overlap with the weight
with probability $p_f$, leading to a larger norm $w$ 
in the stationary state. For homogeneous learning rates,
Eq. (\ref{CBG-dw_da}) now reads
\be
\dd{w}{\alpha}=  \sqrt{\frac{2}{\pi}} \eta (2p_f-1) + \frac{\eta^2}{2w},
\ee
and the fixed point (which only exists for $p_f<1/2$) 
is correspondingly
\be
w = \sqrt{\frac{\pi}{8}} \frac{\eta}{1-2p_f}.
\ee
However, Eq.(\ref{CBG-w_corr}) still holds -- and 
$\sum C^2_j$ increases accordingly. The same
factor of $1/(1-2p_f)$ also appears in Eq.(\ref{C-eta}) 
if the calculation is repeated for individual 
learning rates. 

If it is desired to enhance certain correlations rather
than suppress them to a lesser or greater degree, it is
possible to give some components a learning rate of 0
and a fixed norm. The sign of that weight component 
determines whether the correlation is positive or 
negative, the norm determines the strength of
the correlations. 

Does all this work in practice? Within certain 
bounds, it does. However, the autocorrelation function
of a sequence cannot be shaped arbitrarily \cite{Kurchan:Elementary},
and imposing strong constraints on some components
of $C$ leads to a strong violation of the assumption
of random patterns that underlies the calculations.
The results are not easy to calculate, so some
trial-and-error is required to get the desired result.

\begin{figure}
\centerline{
\begin{minipage}{0.49 \textwidth}
\epsfxsize= 0.95\textwidth
  \epsffile{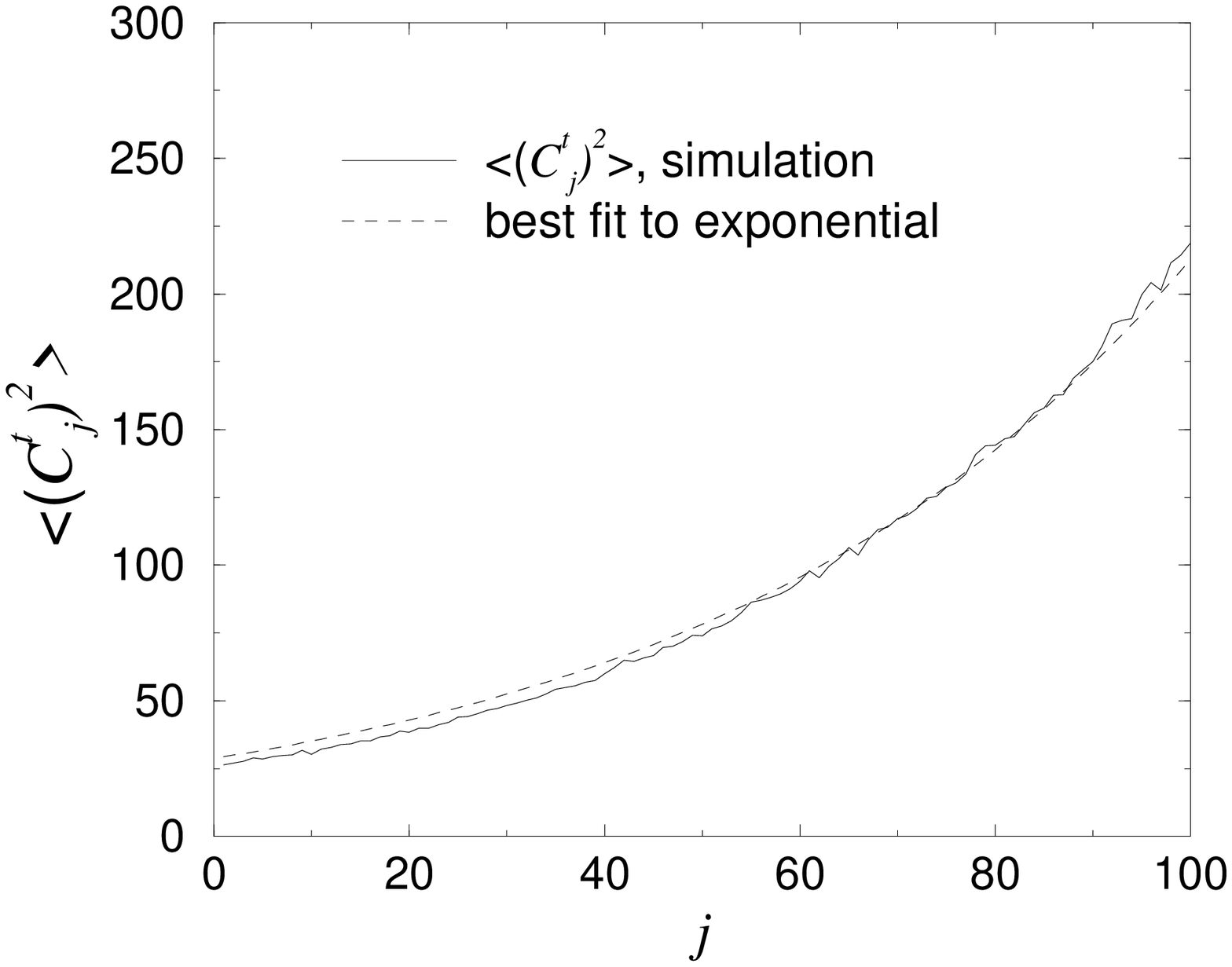}
\centerline{(a)}
\end{minipage}
\begin{minipage}{0.49 \textwidth}
\epsfxsize= 0.95\textwidth
  \epsffile{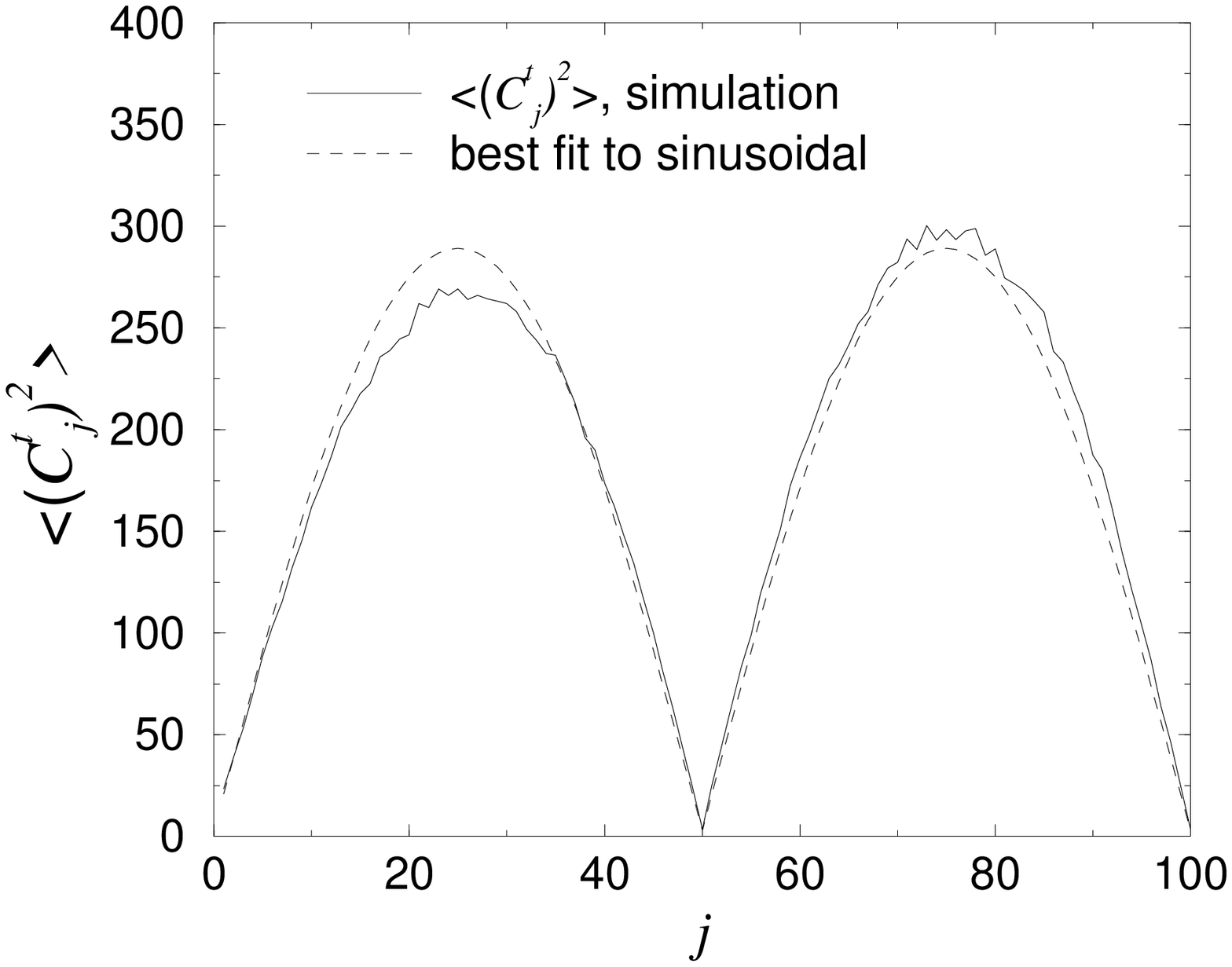}
\centerline{(b)}
\end{minipage}}
  \caption{Mean square autocorrelation function after 
    $t=1000$ steps, generated with a CBG with $M=100$ and
    the learning rate profiles $\eta_j=\exp(-2j/M)$ (plot
    (a)) and $\eta_j = 1/(0.01+|\sin(2\pi j/M)|)$. 
    Dashed lines show the profile predicted from
    Eq. (\ref{C-eta}), with a proportionality constant
    fitted to the data. Simulations averaged over 1000 runs.}
 
  \label{CBG-profiles}
\end{figure}

Furthermore, for each individual set of initial conditions,
the autocorrelation function will show significant
deviations from the calculated profile, and good agreement
is only reached when averaging over sufficiently many
initial conditions. Fig. \ref{CBG-profiles} shows two
examples of autocorrelation functions shaped with 
exponential and sinusoidal profiles, together with the
theoretical profile (the proportionality constant was fitted).

Concludingly, simulations show that with 
some twiddling, the CBG  could be used as an 
alternative mechanism for generating colored 
binary sequences using local rules instead of
nonlocal mechanisms such as Fourier transforms.

\subsection{Distribution of cycles}
\label{CBG-cycles}
Just like the BG with fixed weights, the CBG is 
a deterministic system with a finite number of states,
and thus has to fall into a cycle eventually.
The distribution of cycle lengths $l$ was studied in 
some detail in my diploma thesis \cite{Metzler:Diplom}.
For completeness, I will repeat the 
essential results here.

The connection between the autocorrelation function
and the weights that was exploited in the previous sections
can be used to give a lower bound on cycle lengths of
the CBG: since the weight has to return to its 
original position after $l$ steps, the first $M$ 
components of the autocorrelation function $C_j^l$
have to vanish (see Eq. (\ref{CBG-w_corr})). 
By renaming the indices, one can see that the last
$M$ components $C_{l-j}^l$ must vanish as well. That
would mean that for $l\leq2M$, all components of $C_j^l$ 
would be zero. There is good mathematical evidence
\cite{Mertens:Bernasconi}
that this is not possible for $M>3$.
Furthermore, to get $C_j^l=0$ for any $j$, $l$ must be
divisible by 4.

Extensive simulations indeed show that the lower
bound $l>2M$ holds (with the exception of $M=2$). 
However, the shortest cycle lengths that are observed 
are the smallest multiples of 4 larger than $2M$.
The average cycle length grows exponentially with $M$:
$\langle l\rangle \propto 2.2^M$.

\subsection{Summary of the CBG}
The Confused Bit Generator generates the time series
that is antipredictable for a perceptron, and it is
hardly surprising that the properties of that sequence
differ from one that is easily predictable:
while a perceptron with Hebbian learning is sensitive
to the autocorrelation of a time series, the CBG suppresses
the first $M$ bits of the autocorrelation function -- as
many as it can, with a memory of $M$ time steps.

This suppression can be linked directly to the
dynamics of the weights. These can in turn 
be calculated under the assumption of random patterns,
which allows to estimate quantities like the merit factor of
the generated sequence. Unfortunately, the CBG is not 
a very good source of low-autocorrelation sequences;
however, individual learning rates for different components
allow for a manipulation and customized shaping of the
autocorrelation function.

The suppression of the autocorrelation function 
points to another difference between the CBG and the
BG: the sequence generated by the CBG no longer has
a dominant Fourier component. The Fourier transform
of the weight vector is no longer a very meaningful
quantity either, since it changes constantly as the 
vector moves.

The CBG has a significantly larger number of accessible
states than the BG with equal $M$. However, while the
average cycle length of the BG grows polynomially in $M$, 
and has a typical value of $2M$, $\langle l \rangle$
grows exponentially for the CBG, and has $2M$ as a lower bound.

\section{Continuous Perceptrons: the Confused Sequence Generator}
\label{SEC-CSG}
\subsection{Continuous perceptrons with fixed  and \\ adaptive
  weights}
\label{SEC-SGen}
\begin{figure}
\epsfxsize= 0.70\textwidth
  \centerline{\epsffile{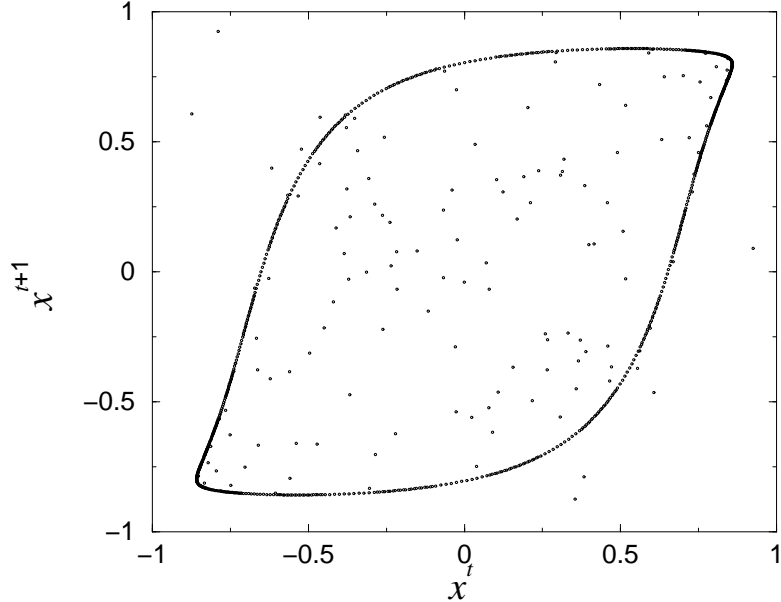}}
\caption{Typical return map of an SGen: after a 
short transient (scattered points) the system enters 
a quasi-periodic limit cycle.}                     
\label{SG-return}
\end{figure}

The perceptron with binary output can readily be
generalized to continuous output by replacing the
sign function with a sigmoid function, such as the error
function. This introduces a new parameter, the amplification
$\beta$:
\be
g(\vx, \vw) = \erf (\beta\,\sp{\vx}{\vw}).
\ee
Just like the simple perceptron, the continuous perceptron
can be used to generate a time series:
\be
x^{t+1} = \erf ( \beta \sum_{i=1}^M w_i x^{t-i+1} ).
\ee
This system (the
Sequence Generator or SGen) was studied in Refs. 
\cite{Priel:Long-term,Kanter:Analytical,Priel:Noisy}.
In the typical case,
depending on the amplification, the system either
converges to the trivial solution $x_{t} \equiv 0$
or, after a short chaotic transient, relaxes into
a stable limit cycle (see. Fig. \ref{SG-return}). 

A careful investigation shows that there are
small areas in parameter space (which is spanned
by $\vw$ and $\beta$) that lead to chaotic
behavior. Since these areas are themselves 
fractals, and the appearance of chaos is therefore
sensitive to small variations in the parameters,
the term {\em fragile chaos} was coined to characterize
the system \cite{Priel:Long-term}. 

It is not quite obvious how to generalize the
Confused Bit Generator to continuous output: 
there is no unique value of the prediction that is ``wrong'',
since any value that is different from the actual value of
$x$ is more or less wrong.
The following simple prescription was chosen 
to keep the dynamics as close as possible to that
of the CBG, and especially to keep the connection between
the weight vector and the autocorrelation function
of the  generated time series. The continuation of
the time series is thus simply taken to be the
negative of the perceptron's output, and the
learning step is proportional to the desired
output (and therefore, to the difference between
the desired and the real output):
\bea
x^{t+1} &=& - \erf(\beta\,\sp{\x^t}{\vw^t}); \label{CSG-upS} \\ 
\vw^{t+1} &=& \vw^t + (\eta/M) x^{t+1} \x^t. \label{CSG-upw}
\eea
This system was introduced in Ref. \cite{Metzler:CBG}, where
it was named Confused Sequence Generator (CSG).

\subsection{Mean-field solution}
\label{SEC-CSG-meanfield}
Similar to the CBG, the weight vector of the CSG 
does a directed pseudo-random walk near the surface of 
a hypersphere of a radius $w$.
Unlike the CBG, the length of the learning
steps, which determines $w$, is not fixed, 
but depends on the magnitude of the output,
which in turn depends on $w$ and the outputs
in previous time steps. To find an approximate 
solution to this self-consistency problem, I will first
ignore correlations between patterns and weights and treat
the patterns as random and independent.

In this approach, the hidden field $h=\sp{\vw}{\vx}$ 
is a Gaussian random variable
of mean 0 and variance $w^2 S^2$, where 
$S^2 = \langle {x^t}^2 \rangle_t$ is the mean square
output of the system. 

The norm $w$ is found by taking the square of (\ref{CSG-upw}):
\be
{w^t}^2 = {w^{t-1}}^2 - 
 \frac{2 \eta}{M} \sp{\vx^t}{\vw^t}
 \erf(\beta\, \sp{\x^t}{\vw^t}) + \frac{\eta^2}{M^2} 
{S^t}^2 \sp{\x^t}{\x^t},
\label {updating}
\ee
and averaging over the input patterns.
The self-overlap $\sp{\x}{\x}$ is on the average $M S^2$,
so the fixed point of $w$ is given by
\be
2 \langle h\; \erf (\beta h) \rangle = \eta S^4.
\ee  
The average on the left hand side can be evaluated
and leads to
\bea
\frac{4}{\sqrt{\pi}}\frac{\beta w^2 S^2}{\sqrt{1+ 2\beta^2 w^2 S^2}} 
&=& \eta S^4, 
\mbox{ or} \\
w^2 &=& \frac{\pi \beta \eta^2 S^6 + 
       \eta S^2 \sqrt{\pi}\sqrt{16 + \pi \beta^2 \eta^2 S^8}}{16 \beta}. 
\label{CSG-w_S}   
\eea

Let us now turn to $S^2$. The probability distribution
of $S$ itself is rather awkward, since it involves 
inverse error functions, and its slope diverges at
$S = \pm 1$. However, $S^2$ can be easily calculated
by using the distribution of $h$:
\bea
\langle S^2 \rangle &=& \int_{-\infty}^{\infty}
  \erf^2(\beta h) (2 \pi w^2 S^2)^{-1/2} \exp \left
    (-\frac{h^2}{2 w^2 S^2}
    \right )\; \d{h}
  \nonumber \\
 &=& \frac{2}{\pi} \arcsin \left 
 ( \frac{2 \beta^2 w^2 S^2}{1 + 2 \beta^2 w^2 S^2} \right ). \label{CSG-S_S} 
\eea  
Plugging $w^2(\eta, \beta,S^2)$ from Eqs. (\ref{CSG-w_S}) 
into (\ref{CSG-S_S}) and solving
numerically, one obtains a self-consistent solution for $S^2$.
A closer look at the equations reveals that if a new quantity
$\gamma = \eta \beta$ is introduced, 
only $\gamma$ enters into the equation for $S^2$,
and $w^2$ is of the form $w^2 = \eta^2 \hat{w}^2(\gamma)$, so only one 
curve must be considered. This is intuitive,
since a higher $\eta$ 
eventually leads to a higher $w$, which
has the same effect on $S^2$ as having a smaller $w$, but 
multiplying $\sp{\vw}{\x}$ with a higher factor $\beta$.  

The map defined by Eqs. (\ref{CSG-upS}) and (\ref{CSG-upw}) always
has the trivial solution $\vx=\v{0}$. Only for a sufficiently high
$\gamma>\gamma_c$ are the outputs high enough to sustain a non-vanishing
solution.
Note that $\vx=\v{0}$ is always an attractive solution for
all $\gamma <\infty$, but its basin of attraction
becomes smaller for larger $\gamma$.
 
\begin{figure}
\epsfxsize= 0.70\textwidth
  \centerline{\epsffile{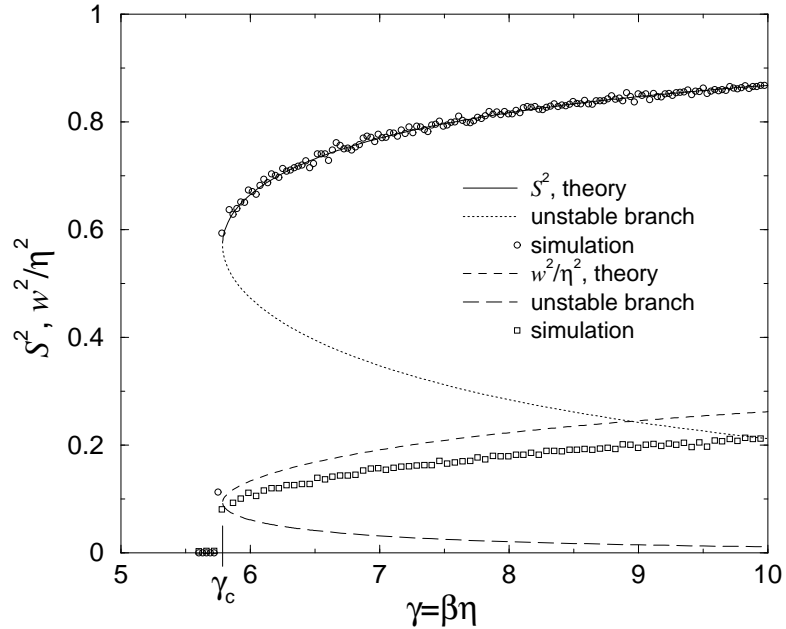}}
\caption{Mean-field solution of the CSG
  (Eqs. (\ref{CSG-w_S}) and (\ref{CSG-S_S})), compared to simulations
  with $M=400$.}                     
\label{CSG-S_beta}
\end{figure}

The numerical solution of Eqs. (\ref{CSG-w_S}) and (\ref{CSG-S_S})
shows that the system undergoes a saddle-node
bifurcation at $\gamma_c \approx 5.785$, which is 
in good agreement with simulations. 
Above $\gamma_c$, two new fixed points exist, only
one of which is stable. While for $S^2(\gamma)$
excellent agreement is found between theory and simulation
(see Fig. \ref{CSG-S_beta}),
$w^2(\gamma)$ shows quantitative differences which are
caused by correlations between $\x$ and $\vw$:
the mean square overlap $\langle (\sp{\x}{\vw})^2 \rangle$
turns out to be $(1.22\pm 0.01) w^2 S^2$ instead of $w^2 S^2$ 
as expected for random patterns. This causes a factor
of roughly 0.82 between the theoretical and observed 
value of $w^2$ , as seen in Fig. \ref{CSG-S_beta}. 
It is the same factor that is is found in the CBG, and
likely caused by the same mechanisms. 

For large $\gamma$, $S^2$ goes to 1 (as it should,
since the system  is identical to the CBG if $\gamma = \infty$), and
the theoretical prediction for 
$w$ goes to $\sqrt{\pi/8}\eta $, just like in the CBG.

\subsection{Autocorrelation function}
The relation Eq. (\ref{CBG-w_corr}) that links
the autocorrelation function to the weights in the CBG
still holds for the CSG. Since the weight
vector is bounded in the CSG as well, the same
argument can be given for the suppression
of the first $M$ values of the autocorrelation
function. Correspondingly, $C_j^p/p$ is almost 
indistinguishable from that of the CBG
shown in Fig. \ref{CBG-corr}.

In principle, this allows to play the same games as
in Sec. \ref{SEC-shaping} to shape the first $M$
components of $C_j^P$. However, the practical 
need for a continuous time series with a rather
odd probability distribution and a custom-tailored
autocorrelation function is probably small. 

\subsection{Cycles and attractors}
\label{CBG-cyc}
The CSG can be seen as a nonlinear mapping that 
maps the vector $\vx^t \oplus \vw^t$ onto $\vx^{t+1} \oplus
\vw^{t+1}$. Again, this is opposed to the
SGen with static weights, where the components of the
sequence window are the only dynamic variables, and
in analogy to the difference between the Bit Generator
and the CBG. The only relevant 
control parameter of this mapping is $\gamma$.

Since both the sequence and the weights now live in
a high-dimensional space of real numbers, the CBG
can display a wide variety of behaviors, depending
on $M$ and $\gamma$:

For $\gamma < \gamma_c$, the zero solution is the
only attractor, and the system will quickly 
reach $\vx^t =\v{0}$ and stop developing.

For $\gamma$ slightly above $\gamma_c$, an
irregular-looking time series with the distribution and variance
that were calculated in Section \ref{SEC-CSG-meanfield}
and displayed in Fig. (\ref{CSG-S_beta}) is generated.
However, the zero solution is still attractive, and
after some time the system will drift close to it
and stay there, i.e. the irregular behavior represents 
a chaotic transient rather than a stable chaotic 
attractor.

The survival time on the transient increases dramatically
with increasing $M$ and $\gamma$.
It is hard to decide from numerical results whether the
average survival time $\langle t_s \rangle$ diverges with a 
power law ($\langle t_s \rangle \propto |\gamma -\gamma_d|^{-a}$),
as one usually finds in scenarios where a chaotic transient
becomes a chaotic attractor \cite{Ott:Chaos}, or whether
$\langle t_s \rangle$ increases exponentially with $\gamma$.
In either case, the system shows chaotic behavior for
sufficiently long times to get stable numerical results --
for example, for $M=20$ and $\gamma=7.0$, the average 
survival time is on the order of $10^6$ steps.
 
If $\gamma$ is larger than some critical value that
depends on $M$, the chaotic transient can eventually
end in a cycle that is related to a 
possible cycle of the discrete CBG. ``Related''  means that 
$x^t$ in the CSG is very close to $\pm 1$ and that clipping
the sequence to the nearest value of $\pm1$ would give
the equivalent attractor of the CBG. More different cycles  
become stable with higher $\gamma$; however, the cycle
lengths are usually of order $2M$ -- short cycles are
apparently more likely to become stable than ones whose 
length is of order $2^M$. A possible explanation for this
is probabilistic: as Sec. \ref{SEC-lyapunov} will show, 
stability is possible only if the transfer function is
almost saturated, which only happens if the pattern and
weight have a high overlap. It is easier to find a short
cycle for which this is fulfilled for every pair of 
pattern and weight, than a long cycle.

\begin{figure}
\epsfxsize= 0.70\textwidth
  \centerline{\epsffile{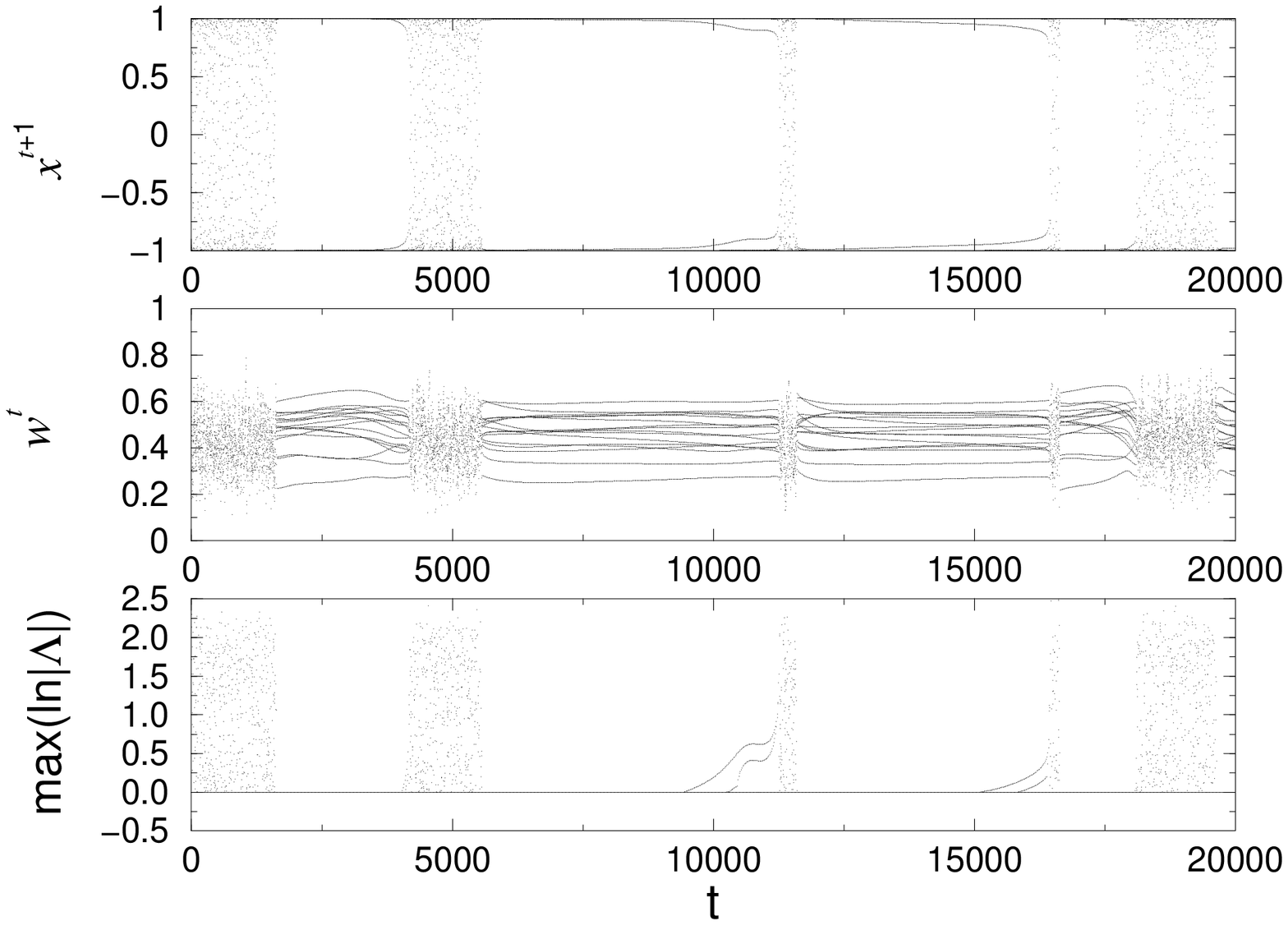}}
\caption{Example of intermittent behavior for $M=6$, $\gamma =8$.
From top to bottom: output $x^{t+1}$, norm of weights $w^t$ and 
largest ``one-step Lyapunov exponent'' $\max(\ln(|\Lambda|)$
(see Section \ref{SEC-lyapunov}).}
 \label{CSG-interm}
\end{figure}

At amplifications $\gamma$ slightly below the 
lowest $\gamma$ for which the first cycle becomes stable
for for a given $M$, 
intermittent behavior is observed: both $\vx^t$ and
$\vw^t$ stay near a cycle for  an extended number of 
steps (typically several thousand steps for $M=6$) before returning
to chaotic behavior for a similar time. An example of this 
is given in Fig. \ref{CSG-interm}.

\subsection{Chaotic dynamics and {L}yapunov exponents} 
\label{SEC-lyapunov}
The term 'chaotic' was used in a loose meaning in Sec. \ref{CBG-cyc}
to describe the irregular time series generated
by the CBG.  However, the system is in fact chaotic 
in the strict sense, i.e., small deviations in the initial 
conditions grow exponentially.

\subsubsection{Propagation of perturbations in one time step}
  
The sensitivity of trajectories of the map given by
Eqs. (\ref{CSG-upS}) and
(\ref{CSG-upw}) to small changes in the initial
conditions  
can be tested by calculating the eigenvalues of the 
Jacobi matrix  
\be
\mathbf{M}^t = \left ( \begin{array}{cc} 
    \pdd{x_i^{t+1}}{x_j^{t}} & \pdd{x_i^{t+1}}{w_j^t} \\
    \pdd{w_i^{t+1}}{x_j^{t}} & \pdd{w_i^{t+1}}{w_j^t}
\end{array} \right ).
\ee
This is to be understood as a $2M \times 2M$ matrix with
indices $i$ and $j$ running from 1 to $M$.
It describes the evolution of infinitesimal 
perturbations $\Delta \vx \oplus \Delta \vw$
during one time step.

The entries of this matrix are of the following form:
\bea
\pdd{x_1^{t+1}}{x_j^t} &=& 
   - \beta w_j^t \frac{2}{\sqrt{\pi}} \exp (-\beta^2 h^2); \nonumber \\
\pdd{x_i^{t+1}}{x_j^t} &=& 
  \delta_{i-1,j}  \mbox{ for } i =2,\ldots,M; \nonumber \\
\pdd{x_1^{t+1}}{w_j^t} &=& 
  - \beta x_j^t \frac{2}{\sqrt{\pi}} \exp(-\beta^2 h^2); \nonumber \\
\pdd{x_i^{t+1}}{w_j^t} &=& 
  0  \mbox{ for } i =2,\ldots,M; \nonumber \\
\pdd{w_i^{t+1}}{x_j^t} &=&
  - \frac{\eta}{M} \erf(\beta h) \delta_{i,j} 
  - \frac{\eta}{M} \beta w_j^t x_i^t \frac{2}{\sqrt{\pi}} 
  \exp (- \beta^2 h^2);  \nonumber \\
\pdd{w_i^{t+1}}{w_j^t} &=& \delta_{i,j} 
  - \frac{\eta}{M} \beta x_j^t x_i^t \exp(-\beta^2 h^2). \label{CSG-jacobi}
\eea
If $|\beta h|$ is large and the transfer function
is saturated, the exponential terms in Eq. (\ref{CSG-jacobi})
are negligible. In that case, the upper left section of $\mathbf{M}$ 
is occupied only on the first lower off-diagonal, and the lower right
section is the $M\times M$ unity matrix. Since the upper right 
section is identically 0, the lower left part does not
enter into the calculation of the eigenvalues either. This
is not completely obvious, but it follows from the
definition of the determinant of a matrix as the sum
over permutations of the matrix elements
\cite{Berendt:Mathematik}: if one of the factors of a
permutation is from the lower left part, at least one 
factor must be from the upper right part, which is 0,
and therefore that permutation gives no contribution to the
determinant.
  
This simplified matrix has $M$ eigenvalues 
$\Lambda =0$ and $M$ eigenvalues 
$\Lambda =1$. The eigenvectors of the latter span the
space of weight vectors, where small changes to $\vw^t$  
are transferred unmodified to $\vw^{t+1}$. The eigenvalues
$\Lambda = 0$ all have the same eigenvector, whose
only non-vanishing component is $x_M$, the component of
the sequence vector that is rotated out at $t+1$.  
This means that the eigenvectors do not span the 
whole space and that thus the eigenvalues are not a
reliable measure of the propagation of a disturbance 
in the system. 

If $|\beta h|$ is small enough for the exponential 
terms to have an appreciable effect, the effect on the
eigenvalues is not easy to calculate. By using values
of $\vx$ and $\vw$ taken from a run of the simulation
and numerically calculating the eigenvalues, one finds that
typically one of the $\Lambda =0$ eigenvalues is changed drastically
and may have an absolute value $|\Lambda| >1$.
This corresponds to a strong susceptibility of the newly generated
sequence component $x_1$ on small changes in $\vw$ or $\vx$.
The other eigenvalues only undergo small corrections,
corresponding to the feedback of the new component
to the weights. 

During the regular phases of intermittent
behavior, the largest eigenvalues of the one-step
matrix are significantly smaller than during the 
chaotic bursts (see Fig. \ref{CSG-interm}) --
corresponding to sequence values that are close to
$x=\pm 1$, and thus a nearly saturated transfer function.

\subsubsection{Long-time behavior}
To find the Lyapunov exponents of the map,
which describe the exponential growth 
or decay of small perturbations of the
initial conditions (see e.g. Ref. \cite{Lam:Nonlin}), 
it is necessary to consider the development of a small
perturbation over a long time, i.e., to calculate the
eigenvalues $\Lambda^T_i$ of $\prod_{t=1}^{T}\mathbf{M}^t$
(of course, the trajectory is determined using the 
full nonlinear map).
The Lyapunov exponents are then defined as 
\be
\lambda_i = \lim_{T\rightarrow \infty} 
(1/T)\ln |\Lambda^T_i| . \label{CSG-lyapdef}
\ee
The straightforward calculation of the product of
Jacobi matrices brings many numerical problems
which can be eliminated by applying a Gram-Schmidt
orthonormalization procedure to the columns
of the product matrix in regular distances,
as described in Ref. \cite{Wolf:Lyapunov}. With this
procedure, it is possible to average over $T>100M$
and get numerically stable results. The largest 
Lyapunov exponent is displayed in Fig. (\ref{CSG-lyapm}).
Typically, there are $M/2$ positive exponents.

\subsubsection{Attractor dimension} 

The Kaplan-Yorke conjecture \cite{Kaplan:Conjecture} states that
there is a connection between the dimension $D$ of a
attractor of a map and the spectrum of Lyapunov exponents,
which are here assumed to be ordered 
($\lambda_1\geq \lambda_2\geq\ldots\geq \lambda_{2M}$):
 \be 
D_{KY} = k + \sum_{i=1}^{k} \lambda_i / |\lambda_{k+1}|, \label{KY-conj}
\ee 
where $k$ is the value for which $\sum_{i=1}^k \lambda_i > 0$
and  $\sum_{i=1}^{k+1} \lambda_i < 0$. Applying this to 
the spectrum of exponents derived from (\ref{CSG-lyapdef})
gives an average attractor dimension between $1.1 M$ and $1.2M$,
slightly depending on $\gamma$.

The attractor dimension is a measure for the effective
number of degrees of freedom of a chaotic system.
It cannot be larger than the number of dynamic variables
of the system (which, in this case, is $2M$). A much 
smaller number is possible, as seen in the case of 
the SGen (Sec. \ref{SEC-SGen}), 
where the quasi-periodic attractor is one-dimensional. 
In that case, all components of the sequence vector
are strongly correlated. The fact that the Kaplan-Yorke
dimension of the CSG is larger than $M$ indicates that 
its pattern components are more or less independent. 

\subsubsection{Lyapunov exponents for large $M$}
An alternative method for measuring the largest 
Lyapunov exponent is to start two trajectories
with infinitesimally different initial conditions,
and propagate both of them using the nonlinear map.   
In regular intervals, measure the distance between
the trajectories, store it, and reset the distance
to the initial value while keeping the direction
of the distance vector. 

Since the largest Lyapunov exponent dominates the
growth of the perturbation, the contribution of all 
other Lyapunov exponents can be neglected for sufficiently
long times. 
The advantage of this method
is that it requires only $\mathcal{O}(M)$ calculations per time
step, rather than $\mathcal{O}(M^2)$ like the previous way,
allowing to go to much higher $M$. 

\begin{figure}
\epsfxsize= 0.70\textwidth
  \centerline{\epsffile{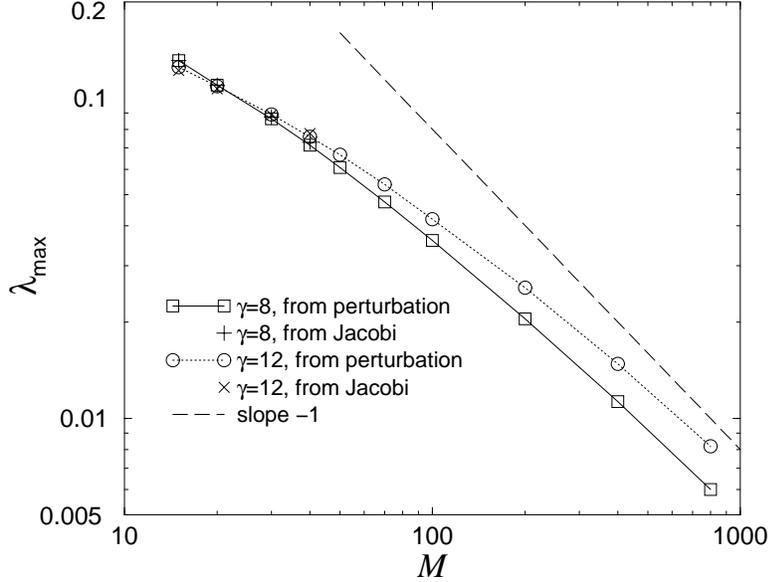}}
\caption{Lyapunov exponents measured from the time development 
of perturbations and from propagating the Jacobi matrix 
(Eq. (\ref{CSG-lyapdef})), for
$\gamma = 8$ and $\gamma =12$.}
 \label{CSG-lyapm}
\end{figure}

The results for $\lambda_1$ are also 
displayed in Fig. \ref{CSG-lyapm}: the values
gained by the two methods agree well within the 
numerical errors. For large $M$, $\lambda_{max}$
decreases with $1/M$, i.e., perturbations 
grow on the $\alpha$-timescale of online learning.

This is slightly astonishing, since the matrix 
Eq. (\ref{CSG-jacobi}) has entries that are proportional
to $1/M$, but also entries that are of order 1. In the 
limit of large $M$, one would expect that it is 
these entries that are responsible for the 
growth of deviations. In that simplifying picture,
the CSG is similar to the SGen with fixed weights,
but is kept in the chaotic transient by the slow movement
of weights. However, even in that picture, it takes
$M$ time steps for a perturbation to spread throughout all
the variables: the matrix entries of order 1 affect the
newly generated sequence component, which is then
rotated through the $M$-component pattern vector step-by-step.
 This would explain the $1/\a$-behavior of
the Lyapunov exponents. 

\subsection{Summary of the CSG}
Just like for the Confused Bit Generator, the step from
fixed to variable weights qualitatively changes
the behavior of the system: while the SGen usually
generates a quasi-periodic sequence, the CSG is a 
rather complicated nonlinear mapping that shows 
many of the trademarks of chaotic systems:
high-dimensional chaos, intermittency, the 
emergence of stable attractors upon variation of 
the control parameter $\gamma$. 
The existence of an  attractive 
absorbing state makes the study of chaotic behavior
difficult for small $M$. For large $M$, chaos is stable
enough for solid numerical measurements of Lyapunov 
exponents and other statistical quantities.

The mean square of the norms and outputs can be 
calculated in a mean-field-like approach, which
correctly predicts the saddle-node bifurcation at
which a fairly stable chaotic regime becomes 
possible. Quantitative differences between the calculation
and simulations have their origin in temporal correlations
between patterns and outputs. 


\section{Decision Tables}
\label{SEC-Decision}
\subsection{Decision tables and {D}e{B}ruijn graphs}
If binary histories are considered, the most general 
prediction algorithm consists of a Boolean function
that has an individual prediction for each possible 
history. The downside of this generality is that no 
generalization is possible: since there is no notion of 
similarity between two histories, one cannot deduce 
correlations along the lines of ``similar histories usually 
lead to similar consequences''. To make this clearer, and
for convenient notation, histories will be denoted by a 
single number $\mu$, written out as a binary string such as
110010, instead of a vector $\v{x} =(1,1,-1,-1,1,-1)$. 
 
\begin{figure}
  \epsfxsize= 0.17 \textwidth
  \centerline{\epsffile{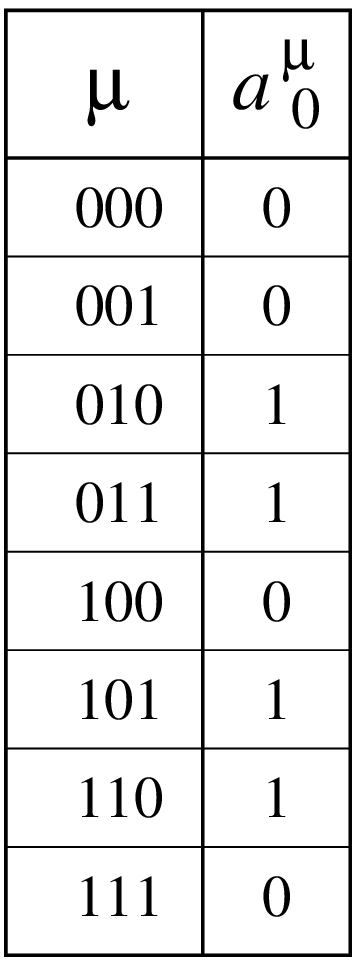}}
  \caption{A Boolean function for $M=3$, 
    represented as a decision table.} 
  \label{AP-Boolean1}
\end{figure}

There are two convenient ways to visualize a Boolean
function: one is a lookup table which contains, in each 
row, a history $\mu$ and a prediction $a^{\mu}$, as in the
example given in Fig. \ref{AP-Boolean1}. 

\begin{figure}
  \epsfxsize= 0.7 \textwidth
  \centerline{\epsffile{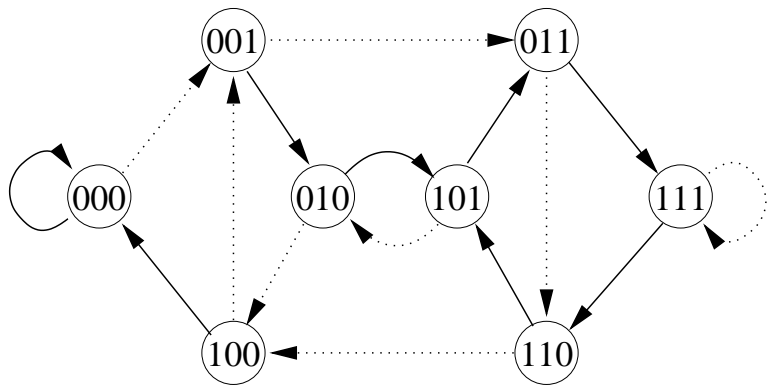}}
  \caption{The Boolean function from Fig. \ref{AP-Boolean1}
    represented as a directed DeBruijn graph of order 3:
    Nodes represent binary strings of length $3$, edges 
    lead to strings that are generated by shifting the
    current string one position and adding either 0 or 1 as the
    new least significant bit. Solid lines denote the active
    exit.} 
  \label{AP-DB}
\end{figure}

The other representation of the Boolean function 
takes a graph-theoretical 
perspective, using directed DeBruijn graphs of order $M$
(see Fig. \ref{AP-DB} for an example of such a graph). 
These graphs have one node for each possible pattern $\mu$. 
Furthermore, each node  
has two edges entering it, coming from the two possible
predecessors, which I denote $^0\mu$ and $^1\mu$. 
For example, if $\mu=1100$, the possible predecessors are
$^0\mu=0110$ and $^1\mu = 1110$.
Each edge also has two outgoing edges, leading to the two
possible successors $\mu^0$ and $\mu^1$ (for $\mu= 1100$, the successors are
$\mu^0 = 1000$ and $\mu^1 = 1001$). The graph is connected, 
since one can reach each node from any other node in a maximum of $M$ 
steps by taking the appropriate exit edges. 

A DeBruijn graph can be modified to 
represent a Boolean function by labeling only the exit 
corresponding to $a^{\mu}$ of each node $\mu$ ``active''.
The sequence can then be easily found by going from 
node to node following the active exits.

\subsection{Static tables}
\label{SEC-static}
Each cyclic sequence in which any $M$-bit pattern appears
at most once can be predicted by a suitably prepared $M$-bit
Boolean function. At most, these sequences have length $2^M$:
namely, if all patterns appear exactly once. These sequences 
are {\em Hamiltonian circuits} or {\em full cycles} 
on the DeBruijn graph of order $M$. These cycles have been
studied extensively; for a  review, see Ref. 
\cite{Fredricksen:Survey}. One of the central (and oldest) results
\cite{Flye:48} is the number of different 
Hamiltonian cycles, which is $2^{2^{M-1}-M}$. 
 
When the system is initialized randomly, there is only a 
slim chance that the cycle indeed has full length:
there are $2^{2^M}$ possible initial states of the 
decision table, of which only $2^{2^{M-1}-M}$ correspond
to a full cycle -- a fraction of $2^{-2^{M-1}-M}$.
Most cycles are significantly shorter. The following 
crude estimate, which ignores much of the structure of the
DeBruijn graph, will nevertheless give an approximation of
average cycle lengths.

Consider a random graph with the following properties:
each of the $N$ nodes has two potential incoming edges and two 
outgoing edges, of which only one is active.
Which node is connected to which is random.
You are starting on one node and following the active exits
until you come back to a node that you have visited 
before (i.e., you enter a cycle). 
Given that you have traveled $\tau$ time steps 
without entering a cycle, the chance of hitting
a cycle in the next step is $p_{h}(\tau)= \tau/2N$ -- there are
$2N$ possible entrances, but each of the 
previously visited nodes is weighted only with one
entrances, because the other one was already ``spent'' --
it is connected to the node you visited one step earlier.

\begin{figure}
  \epsfxsize= 0.7\columnwidth
  \centerline{\epsffile{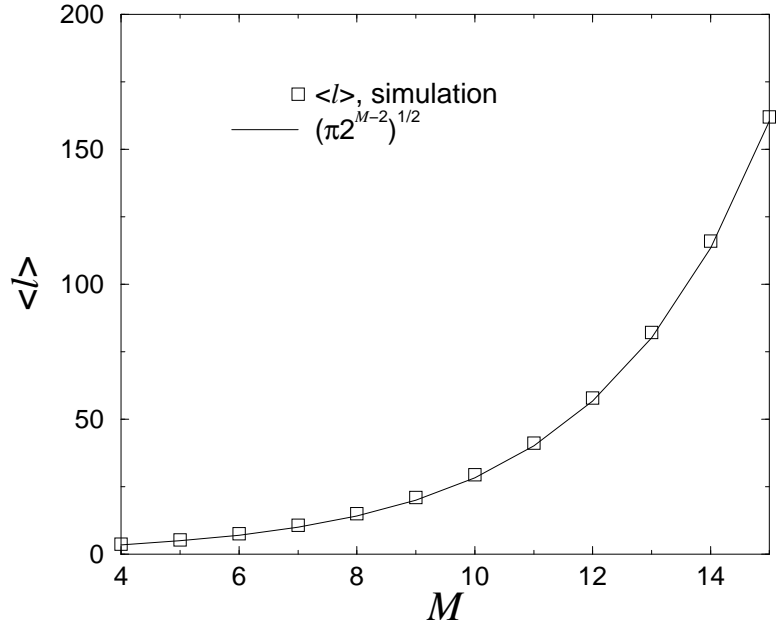}}
  \caption{Average cycle length of a static Boolean table,
    starting from random initial conditions:
    averages over at least $10^4$ runs are compared to 
    Eq. (\ref{ST-l}).} 
  \label{ST-avgl}
\end{figure}

The probability of entering the cycle for the first time
after $t$ steps is 
\be
P_{h}(t) = \prod_{\tau=1}^{t-1} \left(1- p_h(\tau)\right) p_h(t).\label{ST-H} 
\ee
Since at that time there are $t$ nodes that can be 
re-visited (or rather $t-1$, but this is just an
approximation anyway), the cycle that is hit after
$t$ steps has an average length $ l= t/2$.
A further approximation deals with the product in Eq. (\ref{ST-H}): 
\be
\prod_{\tau}^t [1 - \tau/(2N)] \approx [1- t/(4N)]^t \approx
\exp[- t^2/(4N)].
\ee
The average cycle length is then the sum over all 
transient times, weighted with the right probability.
The sum can be approximated by an integral:
\be
\langle l \rangle = \sum_{t=1}^{\infty} P_h(t) t/2 
\approx \int_0^\infty dt \exp\left (-\frac{t^2}{4N}\right
) \frac{t^2}{4N} = \frac{\sqrt{\pi}}{2}\sqrt{N}. \label{ST-l}
\ee
Using $N=2^M$ from the DeBruijn graph, the average 
cycle length scales with $2^{M/2}$. Simulations
give a remarkably good fit to the value calculated in 
Eq. (\ref{ST-l}), as seen in Fig. \ref{ST-avgl}.

\subsection{Dynamic tables: complete antipersistence}
\label{SEC-tabgen}
As pointed out in the introduction, a prediction algorithm 
needs a learning rule in order to adapt to a given problem.
The simplest learning rule for decision tables is to 
set the table entry corresponding to the current history 
to the observed value of the time series.

In order to generate the time series that is completely  
antipredictable for this prediction algorithm, 
the entries in the decision table have to be changed at every 
time step. 
\footnote{The feature that two consecutive appearances
  of a pattern are likely followed by different outputs was
  called ``antipersistence'', e.g. in
  Ref. \cite{Challet:Phase}. Usually this term is applied to 
  continuous time series with a Hurst exponent $<0.5$ 
  \cite{Kantz:Nonlinear}; however, I will use it for
  binary time series, in the sense mentioned above.} 
To be precise, the algorithm is as follows:
At each time step, 
\begin{itemize}
\item a new bit is generated by taking the 
table entry corresponding to the current history $\mu_t$:  
$s_{t+1} = a^{\mu_t}_t$;
\item the history is updated: $\mu_{t+1} = (2\mu_t +
  s_{t+1}) \bmod 2^M$; i.e., all bits are shifted 
  one position to the
  left (multiplication with 2), the newly generated bit is
  added, and the oldest (most significant) bit is dropped (division
  modulo $2^M$);
\item the table entry $a^{\mu_t}_t$ that was used for making the 
decision is changed, such that the sequence will be
continued with the opposite decision when the pattern
$\mu_t$ occurs the next time: 
$a^{\mu_t}_{t+1}= 1-a^{\mu_t}_t$.
All other entries remain unchanged.
\end{itemize}
It should be noted this definition of the dynamics is
different from the definition as it was used for the CBG:
the output of the generator is used for continuing the
sequence, rather than the inverse of the output. However, the same
time series could be generated by taking always the inverse 
of the output, but using the inverse of the decision
table for initial conditions. I chose the
convention presented above to avoid the paradox 
that one would travel through the
graph following the {\em inactive} exits. 

\begin{figure}
  \epsfxsize= 0.7\columnwidth
  \centerline{\epsffile{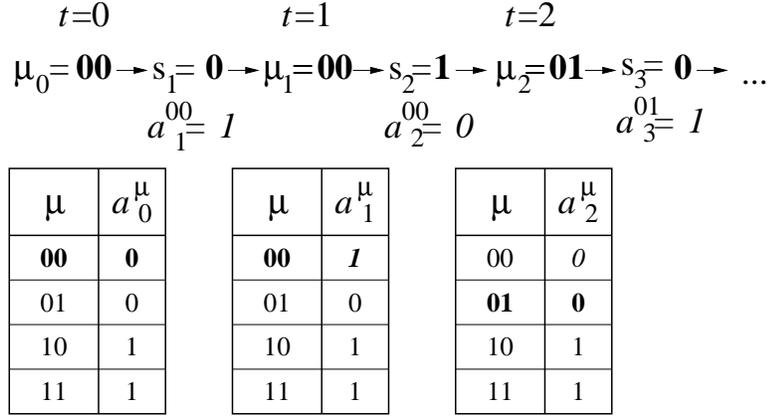}}
  \caption{An example of a decision table with $M=2$ and two
steps of the dynamics. Boldface numbers indicate the current
history and the table entry used for continuing the
sequence;
italic numbers denote the last table entry that was changed.
The sequence generated in this example is $010\dots$.} 
  \label{AP-Tab2}
\end{figure}

As in the case of the confused perceptron, the internal parameters
of the prediction algorithm (in this case, the table entries) become
dynamical variables. As an example, three steps of the combined
dynamics of sequence and table for $M=2$ are shown in Fig. \ref{AP-Tab2}.

\begin{figure}
  \epsfxsize= 0.7\columnwidth
  \centerline{\epsffile{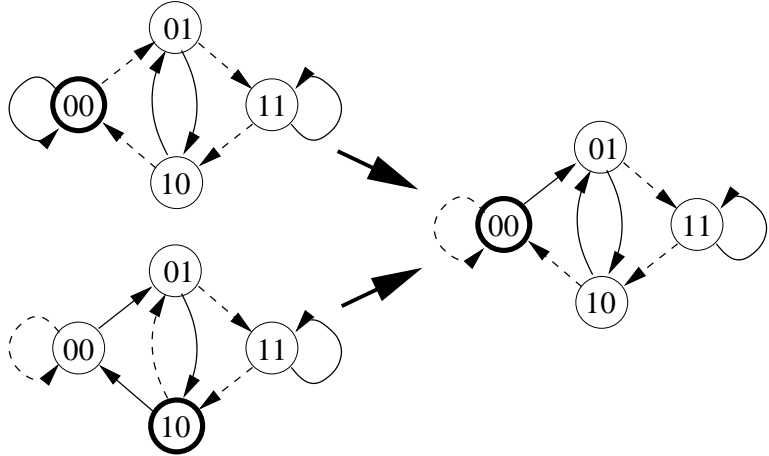}}
  \caption{Example for a step of the dynamics on a graph
    of order 2. Active exits are denoted by solid lines,
    inactive ones by dashed lines. The bold circle indicates
    the node currently visited. Both the upper left
    configuration (which happens to be part of a cycle, and
    corresponds to the example in Fig. \ref{AP-Tab2}) and
    the lower one (which is part of a transient) 
    lead to the configuration on the right, which shows that
    the dynamics is irreversible (see Sec. \ref{AP-propcyc}).} 
  \label{AP-DB2}
\end{figure}

In terms of DeBruijn graphs, history moves from one node to the
next along the currently active exit. After the move, the 
exit of the node that was just left is switched,
as shown in Fig. \ref{AP-DB2}.

\subsection{Properties of cycles}
\label{AP-propcyc}
The introduced dynamics is deterministic, and the
combined system of pattern and table has a finite number  
$\Omega = 2^M \cdot 2^{(2^M)}$ of different
states, so the dynamics necessarily leads into a cycle
eventually. The dynamics is irreversible: if a 
currently visited node has two inactive entrances, it is
impossible to tell which path the system took to get to its
current state (for an example, see Fig. \ref{AP-DB2}).
 This means that not every state can be part
of a cycle, so we will have to consider the necessary
conditions for being in a cycle. I will show, step by step, 
that all cycles are of length $2\cdot 2^M$ and touch all nodes
exactly twice.

Some of the proofs that now follow are redundant; on the
other hand, they help to understand the properties of the
system, and some of them are applicable to generalizations
of the problem, whereas others are not.

Let us assume that at time $0$, the system is already moving
on a cycle of length $l$. We count the number of times that
a history $\mu$
has occurred between time $0$ and time $t$ by a visit number 
$v^{\mu}_t$. Since the definition of a cycle is that after
$l$ steps the system must be in the same state again, it is 
necessary that {\em $v^{\mu}_l$  is even for all $\mu$},
since the table entries $a^{\mu}$ return to their original
state only after even numbers of visits.

Also, {\em all possible nodes are part of the cycle}. Let us prove
this by assuming the opposite, namely that there are some 
nodes that are not touched by the cycle. Since the graph is 
connected, there must be unused connections between the 
part of the graph involved in the cycle and the part that is
left out. But as we have seen in the paragraph before, the
visit number of each of the nodes that are actually part of
the cycle must be at least
2 (larger than 0, and even), so each of its two exits is 
used, including the one leading to the part of the graph 
supposedly not included in the cycle. This is a
contradiction, so all nodes are involved.

An even stronger statement is possible: the total number of visits to the
predecessors $^0\mu$ and $^1\mu$ of $\mu$ must be equal to 
twice the number of visits to $\mu$, since exactly half of the
visits they get are followed by $\mu$, while the other 
half is followed by the so-called conjugate state
$\bar{\mu}$ of $\mu$. (For example, 001 is the conjugate
state of 000.) Thus, we have
\be
v^{\mu}_l = (v^{^0\mu}_l + v^{^1\mu}_l)/2 \mbox{~~ for all }\mu.
\label{AP-visnum1}
\ee
This can be written as a linear equation for an eigenvector
with eigenvalue 1 of a matrix $\mathbf{M}$, with entries 
$a_{\nu\mu} = 1/2$ if $\mu$ is a possible successor of
$\nu$ and $a_{\nu\mu}=0$ otherwise. 
For example, for $M=2$, the set of Eqs. (\ref{AP-visnum1}) looks as
follows:
\bea
\left [ \left ( \begin{array}{cccc}
    \frac{1}{2}~~ & 0~~ & \frac{1}{2}~~ & 0 \\
    \frac{1}{2}~~ & 0~~ & \frac{1}{2}~~ & 0\\
    0~~ & \frac{1}{2}~~ &  0~~ & \frac{1}{2} \\
     0~~ & \frac{1}{2}~~ &  0~~ & \frac{1}{2}
    \end{array}
\right) - \mathbbm{1}_{4} \right ] 
\left (\begin{array}{c}
v^{00}_l\\
v^{01}_l\\
v^{10}_l\\
v^{11}_l
\end{array} \right ) = \left ( \begin{array}{c} 
0\\
0\\
0\\
0\end{array}\right). 
\eea
Since the sum of columns in matrix $\mathbf{M}$ is always 1, and the
individual entries are $\geq 0$, and it describes
transitions on a connected graph, we can apply results from
the theory of stochastic matrices to state that it has one 
unique eigenvector with eigenvalue 1 \cite{Feller},
and we easily guess that $v_l^{\mu} = \mbox{ const}$
fulfills Eq. (\ref{AP-visnum1}). That means that in a cycle,
{\em all states are visited with the same frequency}. 

The next step is to show that in a cycle, each node is 
exactly visited twice, i.e., all cycles of length
$4\cdot 2^M$, $6\cdot 2^M$, and so on, are in fact two,
three or more repetitions of a $2\cdot 2^M$-cycle. Again,
assume the system is moving on a cycle. If this cycle were truly
longer than $2\cdot 2^M$, there must, at the point $t=2\cdot 2^M$, 
be nodes that have been visited three or more times 
while others have not been visited for the second time --
the visit numbers must add up to $2\cdot2^M$, and if all
visit numbers were equal to 2, the cycle would be complete.
More specifically, there must be an earlier time when 
all visit numbers $v^{\nu}_t$ are either 0, 1, or 2, 
and one node is about to be visited for the third time.
It suffices to show that this cannot happen to prove that 
the cycle cannot be longer than $2\cdot 2^M$.

The third visit to a node $\mu$ with $v^{\mu}=2$ cannot come
from a predecessor (let us say, $^0\mu$) with a visit number
of $v^{^0\mu}=0$, for the obvious reason that this
predecessor has not been visited yet. It also cannot come
from a predecessor with $v^{^0\mu} =1$: if $v^{\mu}=2$, 
either it must have been visited before from $^0\mu$
(which it cannot -- the predecessor has only had one
visit so far), or it must have been reached twice from
$^1\mu$ -- this is impossible as well, since it means that
$v^{^1\mu}\geq 3$. For similar reasons, we can exclude a
visit from a node with $v^{^0\mu}=2$: either $v^{^1\mu}\geq
3$ as before, or the first visit to $^0\mu$ led to $\mu$ -- then the 
second cannot. This means that all nodes must receive two
visits -- thus finishing a cycle -- before one of them 
can be visited for the third time. The question arises why
this line of reasoning does not hold true during the
transient. The key lies in the observation that the first
node to receive three visits is the node where the system 
was started -- it did not get its first visit {\em from}
anywhere on the graph, so the arguments are not applicable.

Using all previous conclusions, the cycles turn out to be
of the kind mentioned before: since every combination of $M$-bit 
pattern and following bit, i.e., every $M+1$-bit pattern,
occurs exactly once, the cycles are Hamiltonian paths, or
{\em full cycles}, 
on the $M+1$-graph. Again using results from Ref. \cite{Flye:48},
we know that there are  $2^{2^M - (M+1)}$ distinct Hamiltonian paths.
 
All possible full cycles on the $M+1$-graph can be generated by the
antipersistent walk on the $M$-graph: write down the
desired sequence starting at some arbitrary point, look for the first
occurrence of each $M$-bit pattern $\mu$, and set the
corresponding table entry to the bit that follows it. 
Starting the antipersistent walk at the first pattern of 
the  desired sequence, the antipersistent walk will
reproduce it. 

\begin{figure}
  \epsfxsize= 0.7\columnwidth
\centerline{  \epsffile{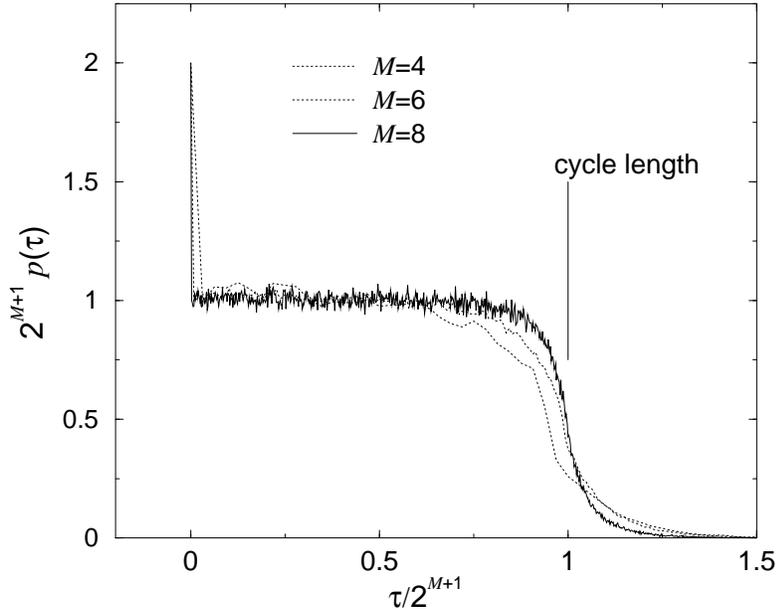}}
  \caption{Distribution of transient lengths $\tau$, 
rescaled by the cycle length $2^{M+1}$.} 
  \label{AP-trans}
\end{figure}

Since all cycles are of length $2\cdot 2^M$,
a total of $2\cdot 2^M \times 2^{2^M - (M+1)}= 2^{(2^M)}$ 
states is part of a cycle.
As mentioned before, the total number of possible states is 
$\Omega = 2^M \cdot 2^{(2^M)}$, which means that a fraction 
of $2^{(2^M)}/\Omega = 2^{-M}$ of possible states is part 
of a cycle. It it thus interesting to check how long it 
takes for the system to reach a cycle, i.e. study the
distribution of transient lengths $\tau$.
This distribution is not easily accessible to
analytical approaches, but easy to measure in computer 
simulations, either by complete enumeration for small
systems or by Monte Carlo simulations for larger ones. 
The following picture emerges, as shown in
Fig. \ref{AP-trans}:

The probability
for transient length $\tau=0$ is just the probability of 
hitting a cycle right away, and thus the fraction of state
space filled with cycles. As mentioned, this is equal to 
$2^{-M}$. 

The probability distribution is more or less flat 
for $1\leq \tau \leq 2^{M+1}$, the cycle length. From
normalization constraints, it follows that $p(\tau)\approx
2^{-(M+1)}$ in that range.

Near $\tau=2^{M+1}$, there is an exponential drop
reminiscent of a phase transition, which gets steeper with
increasing $M$. Even for small $M$, no transients longer
than $2^{M+2}$ have been observed.

\subsubsection{Antipersistence on different timescales}
\label{AP-time}
Obviously, a sequence that is antipredictable for one
algorithm can be perfectly predictable for another.
Is it necessary for that other algorithm to be completely
different in its structure, or is it maybe enough to
adjust some parameters? To answer this question for this
particular algorithm, it suffices to let an observer
predict the antipersistent time series using a decision table,
and  vary the memory length
of the observer. For simplicity's sake, we consider the
long-time limit, in which both the generator and the
observer move on a cycle.

If the observer looks at the same time window as the
generator ($M_{obs}=M$), it is obvious that the success
rate will be $0$ -- since each pattern is continued with 
alternating bits on each visit. For an observer with
a slightly  larger window, the picture changes: as mentioned
above, the antipersistent cycle corresponds to a
Hamiltonian cycle on the $M+1$-graph, which is completely
persistent and predictable with $100 \%$ accuracy. For even
larger $M_{obs}$, the antipersistent cycle looks like a
closed path which includes only a fraction of
$2^{M-(M_{obs}+1)}$ of nodes on the $M_{obs}$-graph.
Prediction is again $100\%$ reliable, and the observer does
not even need all of his storage capacity to handle the
cases that occur.

\begin{figure}
\centering 
  \epsfxsize= 0.55\columnwidth
 \centerline{  \epsffile{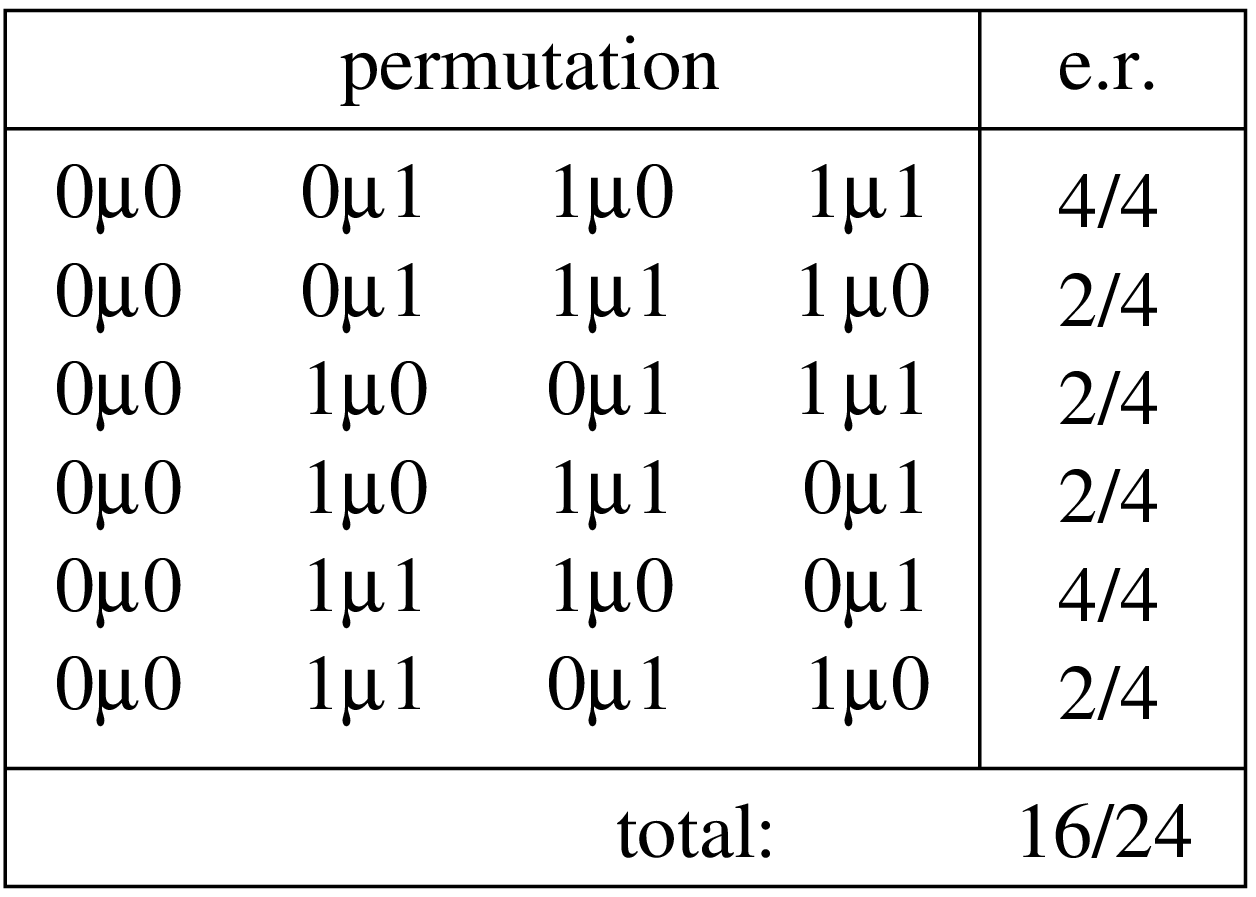}}
        \caption{Possible permutations of predecessors and 
successors of a pattern $\nu$ on a cycle. The error rate,
given in the last column, is the rate of flips between 
0 and 1 in the sequence of successors.}
        \label{AP-tab1}
\end{figure}

If the observer has a shorter time window than the
generator, more than one of the generator's patterns will 
affect the same table entry for the observer. For example,
an $M-1$-bit pattern $\nu$ corresponds to either of the
$M$-bit strings $0\nu$ or $1\nu$, both of which occur twice
in the $M$-cycle, each time followed by a different
successor. The success rate of the predictor depends on the
sequence in which these combinations occur; if each
permutation of $0\nu 0$, $0\nu 1$, $1\nu 0$ and $1\nu 1$
has the same probability, the success rate for all patterns
is the average over the different permutations. Fig.
\ref{AP-tab1} shows that this average is $1/3$ for 
$M_{obs}=M-1$.

For $M_{obs}=M-2$, all permutations of 
eight combinations of predecessors and successors have to be
taken into account -- a task best left to computer algebra 
programs, which yield $\langle s(M,M-1)\rangle =  3/7$, in excellent 
agreement with simulations (see Fig. \ref{AP-predict}). 
Larger differences in the time
window are beyond even the scope of computer programs; 
however, it can be argued that for larger $M-M_{obs}$,
the visits to the $M_{obs}$-nodes become more and more
random, and $s(M,M_{obs})$ will tend to $1/2$.

To answer the question raised above, it is thus not always
necessary to go to structurally completely different algorithms or 
updating methods to turn a negative overlap into a positive
one. However, as the next section shows, increasing the 
amount of information that is processed does not always 
improve prediction accuracy.

\subsection{Stochastic antipersistence}
\label{AP-stoch}
All of the observations from the last sections relied on the fact that the
generator is on a cycle with well-known properties. It is 
thus interesting to ask how stable these results are if the
sequence is not completely antipersistent. The simplest 
generalization is to introduce a probability $p$ for 
changing the table entry/ exit when visiting a node:
$p_{AP}=1$ reproduces the completely antipersistent walk; 
$p_{AP}=0$ is equivalent to using a constant (quenched)
decision table, and $p=1/2$ generates a completely random
time series. 

\begin{figure}[t]
  \epsfxsize= 0.7\columnwidth
 \centerline{ \epsffile{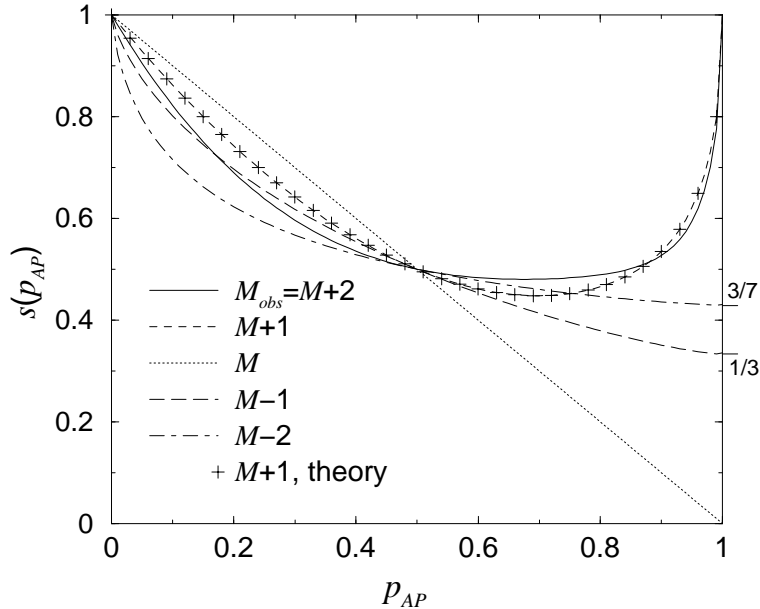}}
  \caption{Success rate of an observer keeping a table of
    the recent occurrence of $M_{obs}$-bit strings and the
    respective following bit, for $M=16$. $p_{AP}$ is the
    probability of flipping the exit in the generating
    graph. The symbols labeled ``theory'' were calculated 
    using Eq. (\ref{AP-s_p}) with approximate probabilities
    $\pi_l(p_{AP},M)$ taken from the simulation itself.} 
  \label{AP-predict}
\end{figure}

A first intuitive guess would be that even a small deviation
from deterministic dynamics completely destroys all predictability:
after all, on a path of length $2^{M+1}$, there are
on the average $(1-p_{AP})2^{M+1}$ occasions where the the
sequence is continued persistently, thus leaving the cycle.
Indeed, a single ``error'' is usually enough to move the 
system from one cycle to another; however, much of the
local structure remains untouched.
It turns out that the functions $s(M, M_{obs},p_{AP})$ of
prediction rates converge for large $M$ (meaning roughly
$M>12$) to a set of curves that depend only on $p_{AP}$ and $M-
M_{obs}$, which is displayed in Fig. \ref{AP-predict}.

The limit values for $p_{AP}=1$ have been explained
above, and they are approached continuously for
$p_{AP}\rightarrow 1$. For $p_{AP}=0.5$, the curves intersect at $s=0.5$ -- no
prediction beyond guessing is possible. For small $p_{AP}$, 
all curves converge to 1: the system is dominated by 
short loops in which only a small fraction of the possible
states participate, and those are predicted with high accuracy.

Interestingly, between $p_{AP}=0.5$ and roughly $p_{AP}=0.85$, 
all shown curves are below 0.5, meaning that even observers
with longer memory predict the sequence with less than 
$50\%$ accuracy.
I will give an analytical argument why this is the case
for $M_{obs}=M+1$. An $M+1$-bit pattern $\nu$ is a
combination of an $M$-bit pattern $\mu$ and one of its 
predecessors, let us say $^0\mu$, whereas the companion
state $\hat{\nu}$ is a combination of $\mu$ and the 
other predecessor $^1\mu$. A visit to either $\nu$ or
$\hat{\nu}$ switches the exit of $\mu$ with probability $p_{AP}$.
Consider two subsequent visits to $\nu$, with some number 
$l$ of visits to $\hat{\nu}$ between them. The probability
$s(p_{AP}, M, M+1)$ of continuing with the same bit after these
two visits is a sum of two probabilities:
either the exit of $\mu$ was switched upon leaving $\nu$
the first time and then switched an odd number of times
during the $l$ visits to $\hat{\nu}$, or it was not
switched the first time and switched an even number of
times in between. Given $p_{AP}$ and the probability 
$\pi_l(p_{AP},M)$ of having $l$ intermediate visits to $\hat{\nu}$, one then
obtains by basic combinatorics
\be
s(p_{AP},M, M+1) = \sum_{l=0}^{\infty} \frac{1}{2} \pi_l(p_{AP},M) 
[1+ (1-2p_{AP})^{l+1}].
\label{AP-s_p}
\ee  
Unfortunately, $\pi_l(p_{AP},M)$ does not seem to be 
analytically accessible for general $p_{AP}$. It can be measured
in simulations, and the accuracy of Eq. (\ref{AP-s_p})
can be verified (see Fig. \ref{AP-predict}); also, for $p=1/2$, 
since the system does a completely random walk on the
graph, one gets the simple distribution $\pi_l(1/2,M) =
2^{-(l+1)}$. Assuming that this distribution does not 
change discontinuously near $p+{AP}=1/2$, Eq. (\ref{AP-s_p})
yields the  approximation $s(1/2+\delta p, M, M+1) \approx
1/(2+2\delta p)
\approx (1/2)(1-\delta p)$. This is obviously $<1/2$ for
$\delta p>0$, i.e., $p_{AP}>1/2$.

Numerical evidence suggests that near the point of 
random guessing, the approximation $s(1/2+\delta p, M, M_{obs}) \approx
1/2 +  2^{-|M-M_{obs}|}\delta p$ holds, i.e., the prediction
success is below $1/2$ for all small $\delta p >0$, but the
deviation from guessing decreases as the difference in time scales  
increases. As an educated guess, one can extrapolate  that as $M_{obs}-M
\rightarrow \infty$,  the success rate will be close to
$1/2$ for all values of $p_{AP}$ except very close to $p_{AP}=0$ and 
$p_{AP}=1$, where it will approach $1$.

 \ \\
The error rate $1-s(p_{AP},M,M_{obs})$ is a measure of the
antipersistence of the time series on the scale $M_{obs}$ --
for $M_{obs}=M$, $1-s$ is completely equivalent to the
antipersistence parameter $p$ of the underlying dynamics.
However, even if $s(p_{AP}, M, M+1)$ could be calculated, 
it would not be possible to recursively apply this function
to find the antipersistence on the scales $M+2, M+3$ etc.
In other words, 
\bea
1-s(p_{AP}, M, M+1) &\neq& \nonumber \\
 1-s( \{1-s(p_{AP}, M, M+1)\},M+1, M+2).
\eea
It is thus not sufficient to give a single parameter
$p_{AP}$, or $1-s$, for some $M$ in order to characterize 
the behavior of a time series completely and to 
calculate its predictability on other scales of 
observation. The scale on which the dynamics work is 
important as well.  

\subsection{Summary of antipersistent time series}
Every prediction algorithm is sensitive to specific
features in the time series. For the perceptron, this
feature was autocorrelations. For decision tables, 
it is fairly obvious: it is the probability that
a given pattern is followed by a certain bit. 
Apparently, antipredictable sequences suppress the 
sensitive feature of their corresponding prediction 
algorithm: in a completely antipersistent time series 
generated by an $M$-bit decision table, the
probability of continuing an $M$-bit pattern with $1$
is $50\%$. 

This means that all quenched preferences for patterns vanish 
if antipersistence is introduced: all patterns appear with
equal probability. In the deterministic antipersistent
time series, it is even possible to prove that all 
patterns appear exactly twice during one cycle.
This is a recurring theme of antipredictable time
series: cycles are longer than for predictable algorithms,
because the state space of the generating algorithm becomes
larger.

I have also shown that observers that keep track of the most recent occurrence
of $M_{obs}$-bit strings can predict the completely 
antipersistent cycle with $100\%$ accuracy if $M_{obs}>M$,
and with less than $50\%$ success rate if $M_{obs}\leq M$.
If the stochasticity is introduced by means of a probability
$p_{AP}$ of  flipping the exit edges, the success rate even of
observers with $M_{obs}>M$ can drop below $50\%$,
which shows that larger memory does
not necessarily give better results.  The rate
of antipersistence on one scale is not sufficient to
calculate the rate for other scales, which again shows
that a time series is antipredictable only for a 
specific algorithm with specific parameters, including
memory length.


  \cleardoublepage
\chapter{The Minority Game and its variations}
\label{CHAP-MG}
\section{Introduction}
In recent years, physicists have applied methods from
statistical physics and time series analysis to 
problems in sociology, biology, and economics
\cite{Weidlich:Sociodynamics,Bar-Yam:Dynamics,Farmer:Physicists}.
One of the important techniques
was the invention and analytical treatment of toy 
models that still capture essential properties of 
vastly more complex real-world problems. The Minority 
Game \cite{MGhomepage}, which has inspired more than 100
publications so far,
has become one of the most influential models.

The inspiration for the Minority Game (MG) comes from
the El-Farol bar problem brought up by W. B. Arthur
\cite{Arthur:ElFarol}: a popular bar has a limited capacity 
for patrons. If fewer people attend, they have a good time. 
If the capacity is exceeded, the bar is crowded, and
the potential patrons who decided to stay at home that night
made the better choice. Arthur's hypothesis was that 
people have a number of possible prediction algorithms to 
decide whether to go or stay home, and that they pick
the algorithm they use according to their individual success
with that algorithm. As a result, attendance fluctuates
around the saturation value.

The underlying idea of the El-Farol scenario is 
competition for limited resources, and can be applied to
a large number of different fields. One often-cited 
example is stock markets, where it pays off to sell if 
everyone else wants to buy, and to buy if there are 
many bids to sell. Other possible scenarios include alternative
roads between two locations, the specialization of 
animals on food sources or habitats, 
a university student's choice of 
field with a view to job prospects later on, and many more.

One obvious feature of these scenarios is that no 
``best strategy'' can exist: if it did, it would be the
same for all players (since there are no a priori 
differences between them), every player would make the
same choice, and all would lose. 
The game thus relies on coordination between the players: 
the aim is to find a niche that few others occupy, and
the means cannot be deduction and careful planning, but only
inductive thinking -- learning from experience and 
trial and error. 

In Arthur's paper \cite{Arthur:ElFarol}, each potential
patron has a repertoire of fundamentally different prediction 
algorithms, which use the time series of previous
attendances to predict how crowded the bar will be. 
He monitors how well each algorithm predicts the
actual history, and chooses the most successful one.

This idea was simplified and formalized in the 
original MG by Challet and Zhang (see Sec. \ref{SEC-SMG}),
where the threshold for overcrowding was set to half
the number of players, and the different prediction algorithms
were modeled by randomly chosen decision tables (Boolean
functions). However, other rules of 
behavior are conceivable, like fine-tuning the parameters
of one prediction algorithm such as a neural network (see
Sec. \ref{SEC-MGNN}) or changing individual entries in a 
decision table if they do not seem to bring success 
(Sec. \ref{SEC-sto}). 
Another interesting generalization is to allow for more than
two different options, as detailed in Sec. \ref{SEC-PMG}. 
Again, all sorts of strategies are conceivable in this 
situation.
The following sections give an
overview over the different strategies, with plenty of
detail on my contributions to the field. Where possible,
I will try to point out where the behavior of the
players leads to antipredictability in the generated
time series.
  
\section{Rules of the {M}inority {G}ame}
\label{SEC-generalMG}
In all variations of the Minority Game that will be 
described, the following rules hold:
\begin{itemize}
\item there is an (odd) number $N$ of players $i=1,\dots,N$.
\item at each time step $t$ each player makes a decision 
$\sigma_i^t \in \{+1,-1\}$ (buy/ sell, go/ stay at home, take 
the highway/ take the country road etc.). 
\item the minority is determined: $S^t= -\s{\sum_i
    \sigma_i^t}$; the players in the minority 
   win ($\sigma_i^t =S^t$), the others lose.
\item global efficiency can be measured by
\be
\sigma^2 = \left \langle \left(\sum_i \sigma_i^t
  \right)^2\right \rangle_t.
\ee
Random guessing would result in $\sigma^2=N$.
Smaller values indicate good coordination, whereas
larger values (often of order  $N^2$ )
indicate herding behavior: finite fractions of the
player population correlate their decisions and
effectively act as a single agent \cite{Hart:Crowd}.
\item players cannot make contracts or otherwise
communicate with each other. The only exchange of
information is through the global minority.
\item consequently, players can use a window
of the global history for making their decision.
In many cases, however, very similar or identical
results are achieved if the actual history is 
replaced by an artificial (random) history \cite{Cavagna:Memory}.
\end{itemize} 

\section{The standard {M}inority {G}ame}
\label{SEC-SMG}
In a series of publications by Challet and Zhang 
\cite{Challet:Emerg.,Challet:Analytical} and 
Challet, Marsili and Zecchina \cite{Challet:Theory},
the following model, which became known as the 
``standard Minority Game'', was introduced and 
studied numerically and analytically:

Each player is equipped  with two decision tables $A_+$ and $A_-$,
each of which prescribes an action $a_+^{\mu}$ or
$a_-^{\mu}$ for each state of the world $\mu$. The state
is often taken to be the string of the last $M$ global
minority decisions; e.g., if $M$ is set to $3$, and in the
last three steps the global decision was $-1, -1, 1$,
the
current state, or ``history'', is $\mu= (-1,-1,1)$. Each 
decision table thus has $2^M$ entries, which are determined
randomly at the beginning of the game. Of course, each
player can only use one decision table at any given time,
so he keeps scores for each table and follows the table
with the highest score.
  
The sign of the sum of all players' choices 
$A^t = \sum_i \sigma_i^t$ then determines who wins.
Those in the minority ($s_i^t = - \s{A^t}$) gain a 
point, the others lose. Players also update the
scores of their decision tables: Tables that would have
predicted the minority sign ($a_{\pm}^{\mu} = - \s{A^t}$)
gain a point, regardless of whether this table was used or
not. Tables that gave ``bad advice'' lose a point on their
score. Then the game is repeated, and players make 
choices based on the updated state of the world and
possibly use a different decision table than before.

As mentioned above, success or failure of coordination can be measured 
with the standard deviation of $A$, usually referred to as
$\sigma^2 = \langle A^2 \rangle$ in the literature.
Random guessing of all players leads to $\sigma^2=N$;
a value $\sigma^2/N<1$ therefore indicates successful
coordination, whereas $\sigma^2/N>1$ is a sign of 
``herd behavior'': agents adapt the same strategies
that others are using as well, leading to larger 
fluctuations.

Closer analysis shows that the relevant parameter
in the original MG is the ratio $\alpha = 2^M/N$ of 
entries in the decision table to the number of players. 
Is is also possible to replace the time series with 
a random state of the world, with $p$ possible states.
The control parameter is then $\alpha =p/N$.
For 
$\alpha>\alpha_c \approx 0.33740$ (i.e., few players,
compared to the number of fundamentally different
strategy tables available), there is fairly good
cooperation ($\sigma^2/N<1$). However, the frozen 
disorder in the decision tables causes some preference
in the outputs generated by the system:
each pattern $\mu$ leads to outputs $+1$ or $-1$ 
with probabilities $P^{\mu}(\pm 1) \neq 1/2$. 
The average deviation of this probability from $1/2$
can be quantified by defining
\be
H= \frac{1}{p} \sum_{\mu}{p} (P^{\mu}(+)-1/2)^2.
\ee

\begin{figure} 
  \epsfxsize= 0.7\textwidth
  \centerline{\epsffile{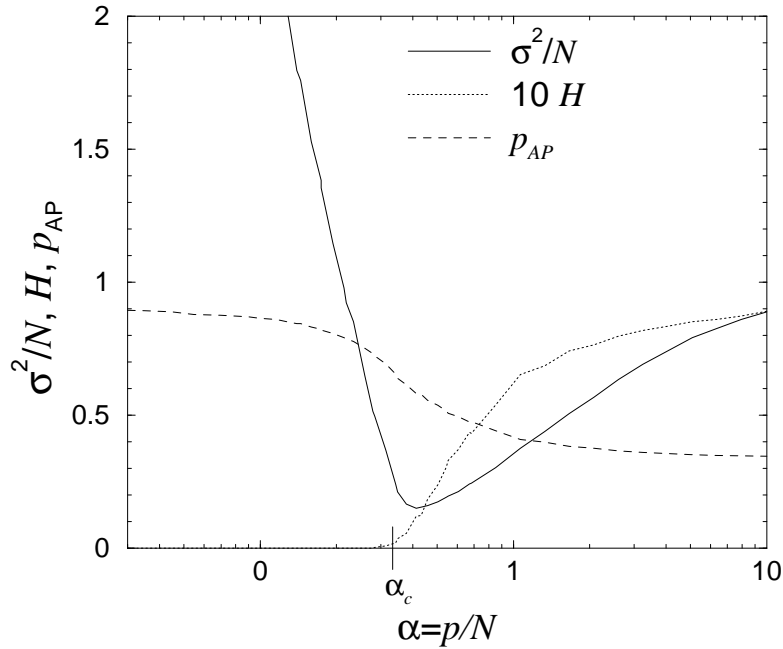}}
  \caption{Standard minority game: results of
    simulations for $\sigma^2/N$, $H$ and the
    antipersistence parameter $p_{AP}$ as a function of 
    $\alpha =p/N$. $H$ is multiplied
    by 1- for better visibility. Simulations use $N=201$
    and random patterns.} 
  \label{CZMG-ordpar}
\end{figure}

At $\alpha=\alpha_c$, the information that can be
extracted from the time series in this fashion vanishes
in a second-order phase transition. Instead, dynamics
become ergodic (i.e., each pattern $\mu$ is visited with
the same frequency) and increasingly antipersistent (i.e., if a 
pattern $\mu$ results in $\s{A}=+1$, the probability
$p_{AP}$ of getting $-1$ on the next occurrence of $\mu$
is larger than $0.5$).
The interpretation is that players overreact to occurrences
of a pattern, switching to a strategy that prescribes
the opposite reaction for the next time that this pattern
occurs. This can lead to dramatic crowding, with 
$\sigma^2 \propto N^2$ as $\alpha \rightarrow 0$.
These results are illustrated in Fig. \ref{CZMG-ordpar}.

The standard MG can be solved analytically by 
treating the choice between the two decision tables
as a spin variable and applying the replica method to
find the ground state of the system 
\cite{Challet:Analytical,Challet:Theory}. This analysis 
reveals that the quenched preference for specific
outputs is necessary to achieve coordination. 
If players have two completely opposite decision 
tables instead of two random (and hence, approximately 
orthogonal)
tables, there is no bias to be exploited, and 
$\sigma^2/N$ goes to 1 for all values of $\eta$.


Newer studies \cite{Garrahan:Continuous,Heimel:Generating} 
have shown that the dynamics in the
crowded phase depend strongly on initial conditions:
if the initial difference between the scores of the
two decision tables are drawn randomly from a 
sufficiently wide distribution, only few players 
switch from one table to the other in response to one
pattern, and coordination becomes better.

\section{Neural Networks in the MG}
\label{SEC-MGNN}
An alternative strategy for the Minority Game 
was first presented in Ref. \cite{Metzler:Diplom}
(although with an erroneous calculation) and published 
in Refs. \cite{Kinzel:Dynamics} and \cite{Metzler:Interact}:
each player is equipped with a simple neural network 
(a perceptron, to be precise) that is fed with the vector 
of past minority decisions. 

In a sense, the ensemble of perceptrons is an inverted 
committee machine: the regular hard committee machine
is a two-layer neural network where the output is 
given by the majority of outputs of the perceptrons 
that make up the first layer 
\cite{Watkin:Statistical,Schwarze:Generalization,Engel:Storage},
whereas in our case the minority decides who wins.
Another difference is that a large number of training
algorithms for the committee machine are not compatible
with the rules of the MG: since the ``members'' of the
committee machine can communicate and cooperate, it is for example 
possible to select the perceptron with the smallest 
hidden field and modify only it -- the so-called
least-action algorithm. The players in the
Minority Game cannot compare notes, and they are
assumed to be greedy -- foregoing an opportunity to 
join the winning side in favor of a co-player is
not possible in the model.

Therefore, only  learning algorithms that include
quantities locally available to the neural networks are
allowed. The simplest (and the only one for which an
analytical solution exists so far) is Hebbian learning of
all networks. 

The calculation I will now present is a correct, more
complete version of that in my Diploma thesis
\cite{Metzler:Diplom}. It is included for the sake of 
completeness and because it facilitates understanding of
the slightly more involved calculation for 
multi-choice neural networks in Sec. \ref{SEC-PMG_NN}.

\subsection{Ensembles of vectors}
As mentioned, each player in this variation of the MG
has a perceptron with an individual weight vector $\vw_i$
to make his decision:
\be
\sigma_i= \s{\sp{\x}{\vw_i}}.
\ee
When considering a population of players, it is helpful 
to split the weight vectors into a center-of-mass vector 
\be
\C = \frac{1}{N} \sum_i^N \vw_i
\ee
and relative vectors
\be
\r_i = \vw_i - \C. 
\ee
These relative vectors are automatically anti-correlated:
if the initial conditions are  chosen such that 
$|\r| =r =1$, one gets from the condition $\sum \r_i= 0$
\be
\langle \sp{\r_i}{\r_j} \rangle_{i\neq j} = 
\frac{1}{N-1} \sum_{i\neq j} \sp{\r_i}{\r_j} = 
- \frac{1}{N-1}
\mbox{~~~ for~~} i\neq j. \label{MG-NN_anticorr}
\ee
A different norm of $r$ can, of course, be included 
in the calculation by replacing $|\C|=C$ by $C/r$ in the
appropriate places.
If the initial vectors are chosen at random such that
$\langle r \rangle =1$, it is a fair approximation that each
pair of vectors individually is anti-correlated as in 
Eq. (\ref{MG-NN_anticorr}) if the dimensionality of
the space they live in is large enough: the variance
$\langle ( \sp{\r_i}{\r_j} + 1/(N-1))^2 \rangle$ is
of order $1/M$. 

The average scalar product between two different weight vectors
is now
\be
\langle R \rangle_{i,j} = \frac{1}{N(N-1)}
\sum_{i, j\neq i}\sp{\vw_i}{\vw_j} =  C^2-1/(N-1),
\ee
and their average norm is 
\be
\langle w \rangle_i =  \frac{1}{N} \sum_{i} \sqrt{\vw_i^2}
\approx C^2 +1.
\ee
The correlation between the output of two 
perceptrons on the same random input pattern $\x$ is
\be
\langle \s{\sp{\x}{\vw_1}}\s{\sp{\x}{\vw_2}} \rangle_{\x} = 
1- \frac{2}{\pi} \arccos\left(\frac{\sp{\vw_1}{\vw_2}}{w_1 w_2}\right).
\ee
With these relations, the global efficiency
$\sigma^2 = (\sum_i \sigma_i)^2$ can be calculated:
\bea 
\frac{\sigma^2}{N}  &=& \frac{1}{N} \left \langle \sum_{i=1}^{N} 1 +
\sum_{i,j\neq i}^{N} \s{\sp{\x}{\vw_i}} \s{\sp{\x}{\vw_j}}
\right \rangle_{\x} \nonumber \\
&=& 1 + (N-1) \left (1 - \frac{2}{\pi} 
\arccos \left (\frac{C^2 - 1/(N-1)}{C^2+1}
\right ) \right ). \label{MIN-sm2}
\eea
If $C$ is set to 0 and $N$ is large, a linear expansion of 
the arccos term in Eq. (\ref{MIN-sm2}) gives $\sigma^2_{opt}/N 
\approx 1- 2/\pi \doteq 0.363$. The small anticorrelations
(of order $1/K$) between the vectors suffice to change the 
prefactor in the standard deviation.

If $C$ is much larger than $r$, there is a strong correlation
between the perceptrons. Most perceptrons will agree with
the classification by the center of mass $\s{\sp{\x}{\C}}$.
As $C \rightarrow \infty$, $\sigma^2/N$ saturates at $N$.

What these considerations have shown is that, assuming
symmetry between the perceptrons, the system can be 
described by the number $N$ of players and one
order parameter, $C$ (or, strictly speaking, $C/r$).
The next section describes what happens to that order
parameter during a simple learning process. 

\subsection{Hebbian Learning}
\label{SEC-MGHebb}
The most remarkable property of neural networks is 
the ability to adjust their parameters to 
unknown rules using efficient learning algorithms, 
and this is what the players in this model do:
At each round of the game, each perceptron is trying 
to learn the decision of 
the minority according to the Hebbian learning rule. 
$S$ denotes the minority decision, and the superscript $+$ denotes
the updated quantity after the learning step:
\be 
\vw_{i}^+ = \vw_{i} - \frac{\eta}{M} \x\,
\s{\sum_{j=1}^{N}  \s{\sp{\x}{\vw{j}}}}
  =  \vw_{i} + \frac{\eta}{M} \x\, S.
\ee
As the same correction is added to each weight vector, their mutual
distances remain unchanged. Only the center of mass is
shifted, and a simple equation for its movement can be found:
\bea
\C^+ &=& \sum_{i=1}^{N}\frac{ \vw{i}}{N} + 
       \frac{\eta}{M} \x\, S. \\
{C^2}^+ &=& C^2 +\frac{2 \eta}{M} \sp{\x}{\C} S + \frac{\eta^2}{M}.
\eea
To average over $\sp{\x}{\C} S$ in the thermodynamic limit,
it is helpful to split the hidden fields $h_i =
\sp{\x}{\vw_i}$ into contributions from the center-of-mass,
 $h^C = \sp{\x}{\C}$, and relative fields $h^r_i =
 \sp{\x}{\r_i}$.
For a given $h^C$, one can then average over $\x$:
\be
\sp{\x}{\C} S = - |h^C| \s{\sum_{i=1}^{N} \s{h^C}\s{h^r_i+h^C}}.
\label{MIN-xCS}
\ee
The quantity $\s{h^C} \s{h^r_i +h^C}$ is a random variable with mean 
$\erf(|h^C|/\sqrt{2})$ and variance $1- \erf(|h^C|/\sqrt{2})^2$.
In a linear approximation for small $|h^C|$, I can replace this by
mean $\sqrt{2/\pi}|h^C|$ and variance 1. It becomes clear 
in Eq. (\ref{MIN-indep}) why this linear approximation is 
admissible: in the range where it is no longer valid, the 
effect of $h^C$ has already saturated.

\begin{figure} 
  \epsfxsize= 0.65\textwidth
  \centerline{\epsffile{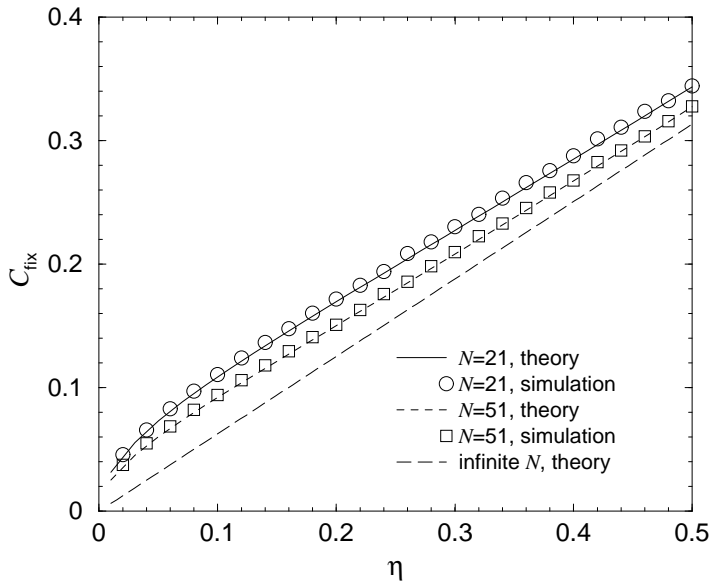}}
  \caption{Fixed point of $C$ vs. $\eta$: simulations with $M=100$
    agree well with Eq. (\ref{MIN-Cfix}). The limit for $N \rightarrow
    \infty$ is $C = \sqrt{2 \pi} \eta/4$. } 
  \label{MIN-C_eta}
\end{figure}

\begin{figure} 
\epsfxsize= 0.65\textwidth
\centerline{  \epsffile{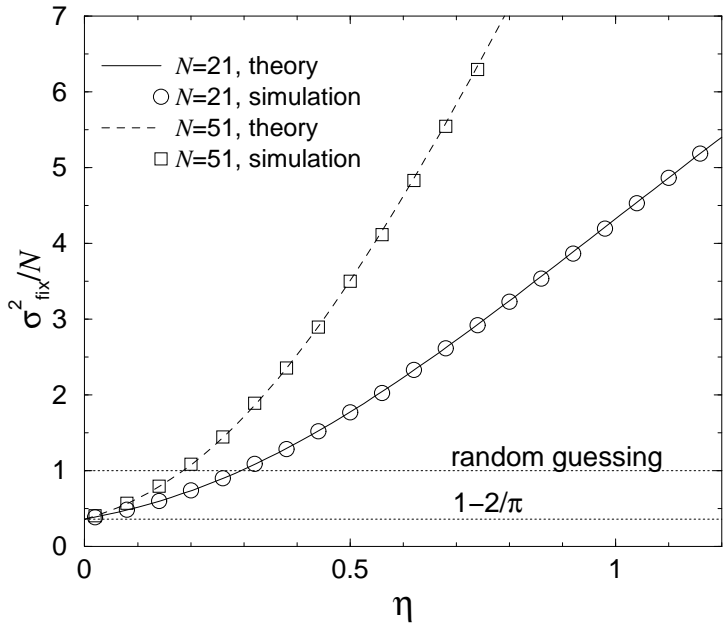}}
  \caption{Fixed point of $\sigma^2/K$ vs. $\eta$: the
    combination of Eqs.(\ref{MIN-sm2}) and (\ref{MIN-Cfix})
    shows that sufficiently small learning rates lead to
    $\sigma^2/K <1$.} 
  \label{MIN-sm_eta}
\end{figure}

For sufficiently large $N$, one can use the central limit theorem 
\cite{VanKampen}
to show that $\sum_{i=1}^{N} \s{h^C}\s{h^r_i+h^C}$
becomes a Gaussian random variable with mean $\sqrt{2/\pi} N |h^C|$.
Since the terms of the sum in (\ref{MIN-xCS}) 
are anticorrelated rather than independent,
the variance turns out to be $(1-2/\pi) N$ rather than $N$, 
analogously to Eq. (\ref{MIN-sm2}). 
   \footnote{This was the
  source of confusion in Ref. \cite{Metzler:Diplom}: the
  anticorrelation was not taken into account properly. In a 
  guessed fit to simulations, the factor $\pi-2$ that
  appears in the calculation was then replaced by 1.}
This yields
\be
\left \langle \s{\sum_{i=1}^{N} \s{h^C} \, \s{h^r_i + h^C}}
 \right  \rangle_{h^r_i} = 
\erf (\sqrt{N/(\pi-2)}|h^C| ). \label{MIN-indep} 
\ee
Since $h^C$ is a Gaussian variable with mean $0$ and variance $C^2$,
the average over $h^C S$ can now be evaluated. The 
following differential equation for the norm of the center 
of mass results:
\be 
\dd{C^2}{\a} = -\frac{4 \eta}{\sqrt{2 \pi}} 
    \sqrt{\frac{2 N/(\pi -2)}{1 + 2 N(\pi -2)C^2}} C^2 +\eta^2.
\ee
The fixed point of $C$, which can be plugged into Eq.  
(\ref{MIN-sm2}) to get $\sigma^2/N$ as a function of $\eta$
and $N$, is 
\be
C_{fix} = \frac{\sqrt{\pi}}{4} \eta \sqrt{1 +\sqrt{1+ 
  \frac{16 (\pi-2)}{\pi N \eta^2}}} \label{MIN-Cfix}
\ee
(see Figs. \ref{MIN-C_eta} and \ref{MIN-sm_eta}).

For small $C$, $h^C$ has only a small influence on the
decision of the majority. 
If the majority disagrees with $\s{h^C}$, 
the learning step has a positive overlap 
with $\C$. This leads to $C \propto \sqrt{\eta}$ as $\eta \rightarrow 0$.

If $C$ is large, 
the majority of perceptrons will usually make
the same decision as $\C$, which then behaves like the single confused
perceptron: $C \rightarrow \sqrt{2 \pi} \eta/4$ if 
$K \eta^2 \rightarrow \infty$ -- compare to Eq. (\ref{CBG-dw_da}).

For small $C$, the majority may not coincide with $\s{\sp{\x}{\C}}$.
In that case, the learning step has a positive overlap 
with $\C$, leading to $C \propto \sqrt{\eta}$ as $\eta \rightarrow 0$.

\begin{figure} 
  \epsfxsize= 0.70\textwidth
  \centerline{\epsffile{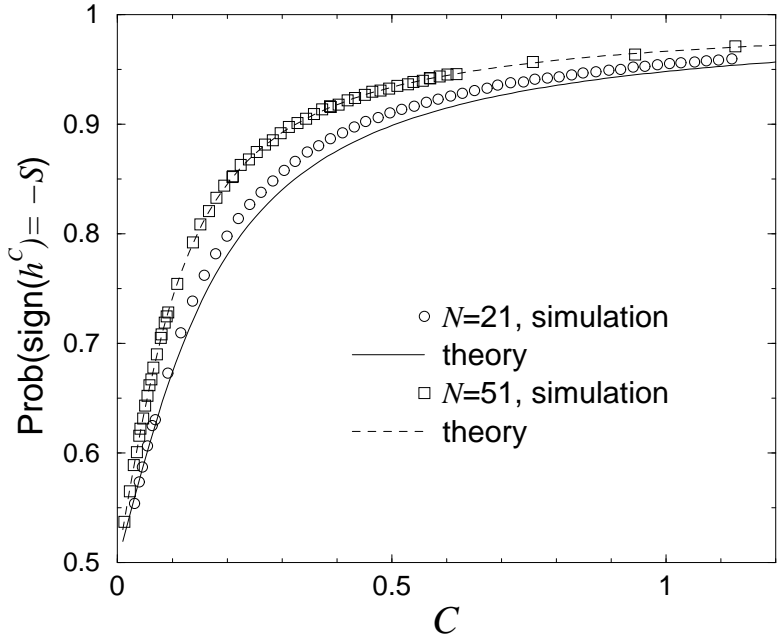}}
  \caption{Probability that the center of mass determines
    the decision of the majority: Comparison between 
    Eq. (\ref{MIN-ap_C}) and simulations with
    $M=100$. Agreement is excellent for $N=51$, but not so
    good for $N=21$.}
  \label{MIN-ap_C-fig}
\end{figure}

The last points are important, since they relate the Minority
Game to antipredictability. They are therefore worth a 
second look. The probability that the majority agrees with 
the output of the center of mass can be calculated easily.
The most important step has already been done, in
Eq. (\ref{MIN-indep}). This can be integrated over the
distribution of $h^C$ to get
\be
\mbox{Prob}(\s{\sum_i \sigma_i} = \s{\sp{\x}{\C}})= 1 -
\frac{1}{\pi} \arctan\left( \sqrt{\frac{\pi-2}{2NC^2}}\right).
\label{MIN-ap_C}
\ee
This result is compared to simulations in
Fig. \ref{MIN-ap_C-fig}. It shows the crossover of the
time series from essentially random behavior to 
a high probability of agreeing with the center of mass.
In that limit, the ensemble of players could almost be
replaced by a single effective player (represented by 
a confused perceptron) for the purpose
to time series generation. 

Concludingly, neural networks that learn the output of 
the minority with Hebbian learning can coordinate well in the 
Minority Game, reducing the measure of global loss
$\sigma^2/N$ by a factor of up to  $0.363$. However, if
learning rates are not small enough, learning causes
a positive overlap between the agents, leading to herding
behavior and large global losses. The transition between
anti-correlated and correlated behavior is a smooth
crossover, not a phase transition.

\section{Johnson's evolutionary MG}
\label{SEC-EvolMG}
N. F. Johnson and coworkers introduced another simple set of
rules in Ref. \cite{Johnson:Segregation}: each agent has 
access to a history table that records the outcome
that followed each possible history pattern $\mu$ upon the 
last occurrence of $\mu$. Also, each agent $i$ has an
individual probability $p_i$. In each round, each agent
looks at the entry in the history table corresponding to 
the current pattern $\mu^t$, and chooses this entry
with probability $p_i$; otherwise, he
chooses the opposite of the last ``winning option''. 

\begin{figure}  
\epsfxsize= 0.7\textwidth
  \centerline{\epsffile{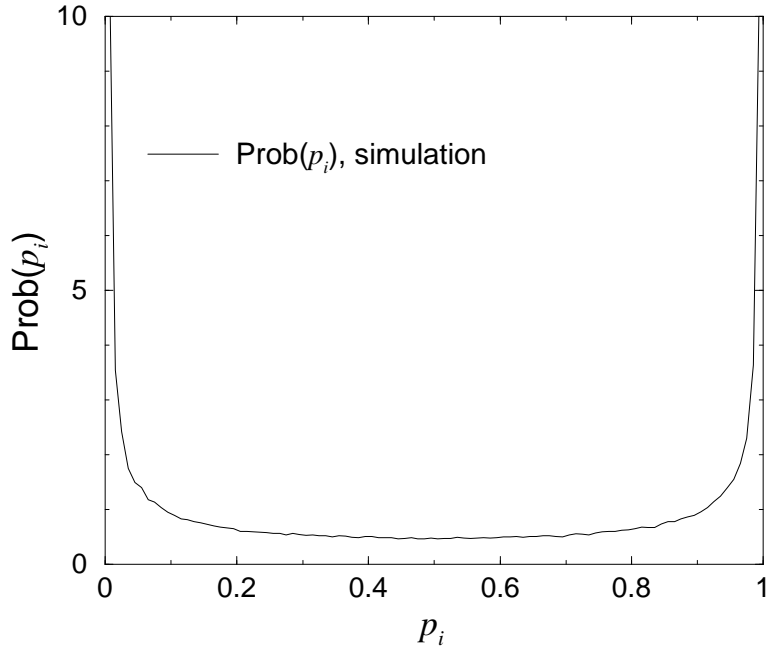}}
  \caption{Stationary distribution of switch probabilities
    $p$ in Johnson's evolutionary MG. Simulations used
    $N=1001$ and a threshold $d=-10$; however, results
    hardly depend on these parameters if $|d|$ is large 
    enough.} 
  \label{EMG-Q2}
\end{figure}

The learning algorithm in Johnson's scenario is 
evolutionary: agents who are in the minority gain one
point, whereas those in the majority lose one. If an agent's
score falls below a certain threshold $d<0$, the agent's 
strategy is modified, i.e., $p_i$ is changed. The
distribution of $p$s converges to a stationary state
where extreme probabilities $p\approx 0$ and $p\approx1$ 
are much more likely than intermediate probabilities,
as seen in Fig. \ref{EMG-Q2}.

As Burgos and Ceva pointed out in
Ref. \cite{Burgos:Selforg.}, the history table is essentially
unnecessary in this variation of the game: all players 
use their individual $p_i$ for each of the possible
histories,  and there is complete symmetry between $+1$ and 
$-1$. If one is only interested in the distribution of
$p$s and not in the time series generated by the game,
one can therefore examine an equivalent game in which
each player chooses $+1$ with probability $p_i$ and
$-1$ with $1-p_i$. One analytical approach was given in 
Ref. \cite{Lo:Theory}; another theory that models pairs of
players whose decisions cancel each other with a given 
probability was proposed in Ref. \cite{Horn:Diplom}.
Temporal oscillations were taken into account in a 
newer study, Ref. \cite{Nakar:Semianalytical}.

\section{Reents' and Metzler's stochastic MG}
\label{SEC-sto}
Similar in spirit is the strategy suggested by
Reents, Metzler and Kinzel in Ref. \cite{Reents:Stochastic}:
players like to stick to the option they chose in 
the previous round. If a player wins, he has no 
motivation to change anything, and will choose the
same action again. If a player $i$ loses, he will
still choose the same action with a probability $1-p_i$ --
after all, times could get better again. However,
there is a probability $p_i$ for the player to decide
that things have to change -- he chooses the opposite
action.

Since each player only remembers his current state,
the probability of finding the 
system at a given state at $t+1$ only depends on the state of the
players at $t$ and the parameters $p_i$ and $N$. The problem
is thus a one-step Markov process, and the standard 
treatment is to calculate the probabilities of finding the
game in each possible state.

If each player has the same probability $p$, the 
game is fairly easy to describe mathematically: 
it is unnecessary to keep track of the whole set
of $\{\sigma_i\}_{i=1}^N$; instead, one can consider
\be
K(t) = \frac{1}{2}\sum_i \sigma_i(t)\,.
\label{STO-defK}
\ee
The possible values $k$ that $K(t)$ can take are
half-integer and run from $-N/2$ to $N/2$ in steps of $1$.
This choice of variables may seem awkward, but it is the
most convenient. The other possibility,
$A= \sum_i \sigma_i$, results in a large number of factors
of $1/2$ in the calculation. Furthermore, one player
switching sides results in a change in $K$ of one unit,
which is easy to remember.

The probabilities to be calculated  are
\begin{equation}
\pi_k(t) = {\sf prob}\left(K(t) = k\right),
\end{equation}
and the dynamics is defined by the transition probabilities
\begin{equation}
W_{k\ell}={\sf prob}(K(t+1)=k \ | \ K(t)=\ell).
\end{equation} 
To shorten notation we
consider the probabilities $\pi_k(t)$ as components of the state vector 
$\pif(t) = (\pi_{-N/2}(t),\ldots,$ $\pi_{N/2}(t))^T$. 
The number of players in the majority at time $t$ is $N/2
+|K(t)|$. Since the individual players perform independent Bernoulli
trials, the transition probability $W_{k\ell} = W(\ell \to k)$
from a state with $K(t)=\ell$ to $K(t+1)=k$ is given by the binomial 
distribution: 
\begin{eqnarray}  \label{mat-binom}
W_{k\ell} &=& 
  \binom{\frac{N}{2} + \ell }{\ell - k} p^{\ell - k} (1-p)^{\frac{N}{2}+k} 
  \mbox{\ \  for\ } \ell > 0\,, \nonumber \\
W_{k\ell} &=& 
  \binom{\frac{N}{2}-\ell}{k-\ell} p^{k-\ell} (1-p)^{\frac{N}{2}-k}  
\mbox{\ \  for\ } \ell < 0\,. 
\end{eqnarray}
This stochastic process may be considered a random walk in one
dimension, where steps of arbitrary size with probability (\ref{mat-binom})
are allowed only in the direction of the origin $K=0$.

Given the initial state $\pif(0)$, the state $\pif(t)$
is updated at each time step by multiplying it by the transition matrix
$\Wf$:
\begin{equation} \label{dynamics}
\pif(t+1)= \Wf \pif(t)\,. 
\end{equation}
The mathematical
theory dealing with this kind of problems is that of Markov chains with 
stationary transition probabilities \cite{Feller}. Since 
$(\Wf^2)_{k\ell} > 0$, the chain is irreducible as well as
ergodic \cite{Gantmacher}, which implies that regardless of the initial 
distribution the state $\pif(t)$ converges for $t \to \infty$ 
to a unique stationary state $\pif(\infty) \equiv 
\pif^s$. In view of Eq.\,(\ref{dynamics}),\ 
$\pif^s$ corresponds to an eigenvector 
of $\mathbf{W}$ with eigenvalue $1\,$:
\begin{equation} \label{eigenvector}
\Wf\,\pif^s = \pif^s \ \ \ \mbox{and}
\ \ \ \sum_k\pi^s_k = 1.
\end{equation}
This eigenvector, which characterizes the equilibrium state 
of the system, can, in principle, be calculated analytically
for any $N$. However, the results are generally rational
functions in $p$ whose degree is roughly $N$ -- they become rather awkward 
for larger $N$, and the computational effort to find them 
is intolerable. Analytical solutions in a more agreeable
form can be found in the two limiting cases of 
small $p$ ($p = 2x/N$, $N\rightarrow \infty$ and large
$p$ ($p= \mbox{const}$, $p\cdot N \gg 1$). 

In intermediate regimes, while analytical results are hard
to come by, numerical solutions are easy to
obtain. The problem can be simplified by exploiting the
symmetry $W_{-k,-\ell}=W_{k\ell}$, which implies the symmetry 
$\pi^s_{-k} = \pi^s_k$ of the stationary state. By rewriting the
eigenvalue problem for the independent components of
$\pif^s$ and exploiting the knowledge that the largest 
eigenvalue is 1, the stationary state is the solution of a
linear equation with $N/2$ variables.
It can be calculated
numerically up to $N\approx 1200$ in reasonable time 
with standard linear algebra packages.     

\subsection{Solution for small $p$}
\label{SEC-STO-smallp}
If $p$ is so small that only $\mathcal{O}(1)$ players
switch sides, it is fairly intuitive that the stationary
state is localized within $\mathcal{O}(1)$ of the 
origin; i.e., only states with $|K| =\mathcal{O}(1)$ have
an appreciable probability. This suspicion 
is confirmed by the analytical 
calculation in the limit $p = 2x/N$, $N\rightarrow \infty$.
The following approximation is good as long as $x\ll N$.
 
In this case the matrix elements $W_{k\ell}$ can be approximated 
by Poisson probabilities \cite{Feller}:
\begin{eqnarray} \label{poisson}
W_{k\ell} &\rightarrow& W^P_{k\ell} = e^{-x} \frac{x^{\ell-k}}{(\ell-k)!} 
   \mbox{\ \ for } \ell > 0, \nonumber \\
W_{k\ell} &\rightarrow& W^P_{k\ell} = e^{-x} \frac{x^{k-\ell}}{(k-\ell)!} 
   \mbox{\ \ for } \ell < 0,
\end{eqnarray} 
where, again, $1/m!$ for negative $m$ has to be interpreted as zero.
In the limit $N\rightarrow \infty$ the solution of the
problem is an infinite 
component vector $\pif^s$ satisfying the eigenvalue 
equation together with the proper normalization:
\begin{equation} \label{pi-stat}
\Wf^P\,\pif^s = \pif^s \ \ \ \mbox{and}
\ \ \ \sum_k\pi^s_k = 1.
\end{equation}

This problem was solved by G. Reents, who noticed that the 
moments of the stationary distribution follow simple rules:
\begin{eqnarray} \label{moments}
\left\langle |k|-\frac{1}{2}\right\rangle &=& 
\frac{x}{2}\,, \nonumber \\  
\left\langle\left(|k|-\frac{1}{2}\right)
\left(|k|-\frac{3}{2}\right)\right\rangle &=& \frac{x^2}{3}\,, \\
\left\langle\left(|k|-\frac{1}{2}\right)
\left(|k|-\frac{3}{2}\right)
\left(|k|-\frac{5}{2}\right)\right\rangle &=& \frac{x^3}{4}\,,\nonumber \\
& \mbox{etc.} &\nonumber
\end{eqnarray}   
These in turn determine the characteristic function of $\pi^s_k$, 
and a Fourier transform finally leads to 
\begin{equation} \label{pi-stat-sol} 
\pi^s_k = \frac{1}{2\,(|k|-\frac{1}{2})!}\sum_{j=0}^\infty
\frac{(-1)^j\,x^{j+|k|-\frac{1}{2}}}{j!\,(j+|k|+\frac{1}{2})}\,.
\end{equation}
Ch. Horn proved in a rather lengthy calculation
that Eq. (\ref{pi-stat-sol}) indeed satisfies the
eigenvalue equation (\ref{pi-stat}) \cite{Horn:Diplom}.

\begin{figure}[h]  
\epsfxsize= 0.7\textwidth
  \centerline{\epsffile{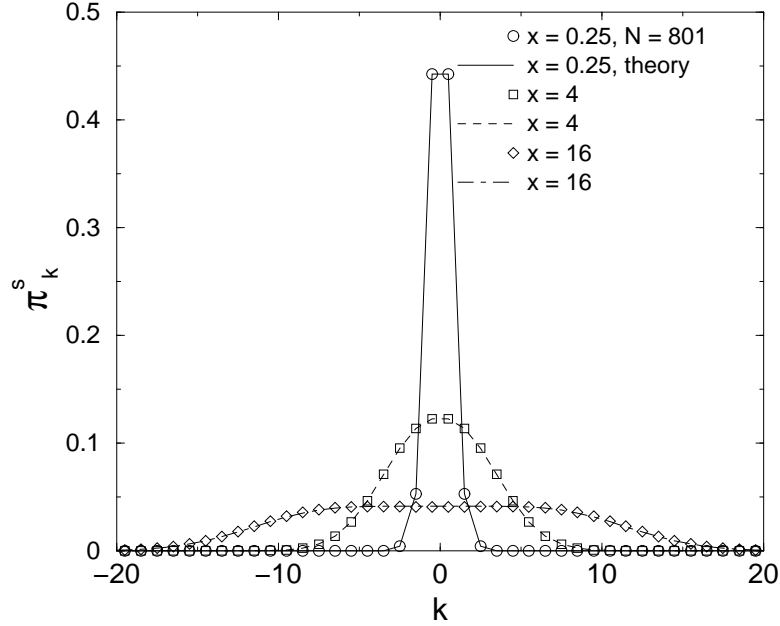}}
  \caption{Stationary solution $\pi_k^s$ for $p = x/(N/2)$. The 
    numerical solution for $N=801$ (symbols) is in very good agreement with
the analytical solution for $N\rightarrow \infty$.} 
  \label{STO-gammfig}
\end{figure}

 $\pi^s_k$ can also be expressed by the incomplete gamma
function:
\begin{equation}
\pi^s_k = \frac{\gamma (|k|+\frac{1}{2},x)}{2\,x\,(|k|-\frac{1}{2})!}\,.
\end{equation}
A comparison with numerically determined eigenvectors of the 
matrix (\ref{mat-binom}) for $N=801$ gives excellent agreement, as seen 
in Fig.\,\ref{STO-gammfig}. 
The distribution is roughly flat for
small $|k|$, has a turning point near $|k| = x$ and falls off 
exponentially with $k$ for larger values of $|k|$. 
From Eq. (\ref{pi-stat-sol}), the variance $\sigma^2 = 
\left\langle (2\,k)^2\right\rangle$  can be
calculated:
\begin{equation}
\sigma^2 = 1 + 4\,x + \frac{4}{3}\,x^2. \label{STO-smx}
\end{equation}
For small $x$, this approaches the optimal 
value $\sigma^2=1$ that occurs if the majority is always as narrow as 
possible, but even for larger $x$, $\sigma^2$ does not increase with $N$.
\subsection{Solution for large $p$}
\label{SEC-STO-largep}
\begin{figure}[h]  
\epsfxsize= 0.7\textwidth
  \centerline{\epsffile{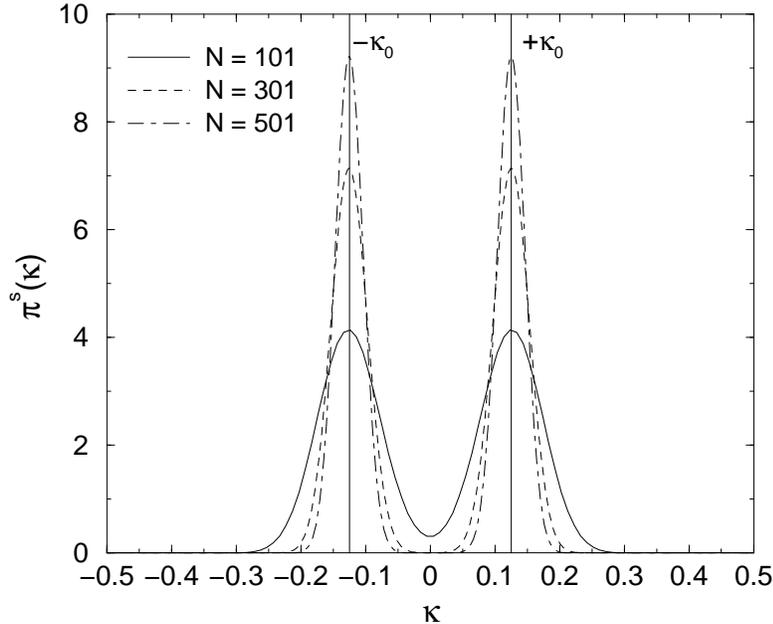}}
  \caption{Stationary solution $\pi^s(\kappa)$ for $p = 0.4$. 
With increasing $N$, the width of the peaks becomes
narrower.} 
  \label{STO-p04}
\end{figure}

As mentioned before, an (approximate) analytical
solution can also be found if $p$ is of order 1
and  $p\,N \gg 1$. To handle this regime, 
I introduce a rescaled (continuous) coordinate 
$\kappa = k/N = \sum_i \sigma_i /(2N)$, the range of which is 
$-1/2\leq \kappa \leq 1/2$.
Multiplied by $N$, the stationary
state $\pi^s_k$ for large $N$ turns into a probability density
function $\pi^s(\kappa)$, and the matrix $W_{k\ell}$ becomes an 
integral kernel $W(\kappa,\lambda)$; 
consequently, the eigenvalue equation 
Eq. (\ref{eigenvector}) is transformed into an 
integral equation: 
\begin{equation} \label{int-eigenvalue}
\pi^s(\kappa) = \int W(\kappa,\lambda)\,\pi^s(\lambda)\,d\lambda\, \ \ \ 
\mbox{and} \ \ \ \int \pi^s(\kappa)\,d\kappa = 1\,.
\end{equation}
Numerical calculations show that the eigenvector $\pi^s(\kappa)$ 
takes the shape of two Gaussian peaks centered at symmetrical distances
$\pm \kappa_0$ from the origin (see Fig. \ref{STO-p04}).
The physical interpretation is that the majority switches
from one side to the other in every time step. Since approximately 
$(\kappa_0+1/2)p\,N$ agents switch sides every turn and the 
distance between the two peaks amounts to a number of 
$2\,\kappa_0\,N$ agents, we get $\kappa_0 = p/(4-2p)$. 

This reasoning can be made more precise, and also the width of the 
peaks for large but finite $N$ can be calculated by the following argument:
The well known normal approximation 
for the binomial coefficients in
Eq. (\ref{mat-binom}) leads to
\begin{eqnarray} \label{int-kernel}
W(\kappa,\lambda) = N\,W_{k\ell}  & \approx &
\frac{1}{\sqrt{2\,\pi}\,s(\lambda)} \exp\left[-\frac{1}{2}\frac{(\kappa
 - f(\lambda))^2}{s^2(\lambda)}\right]\,,  \nonumber \\ 
\mbox{where} \ \ 
f(\lambda) & = & (1-p)\,\lambda - {\rm sign}(\lambda)\frac{p}{2}
\nonumber \\
\mbox{and} \ \ 
s^2(\lambda) & = & \frac{p\,(1-p)\,(\frac{1}{2}+|\lambda|)}{N}\,.
\end{eqnarray}
A double Gaussian of the form
\begin{equation} \label{ansatz}
\pi^s(\kappa) = \frac{1}{2}\frac{1}{\sqrt{2\,\pi}\,b}
\left[\exp\left(\frac{(\kappa+\kappa_0)^2}{2\,b^2}\right) + 
      \exp\left(\frac{(\kappa-\kappa_0)^2}{2\,b^2}\right) \right]
\end{equation}
is transformed by the integral kernel (\ref{int-kernel}) into a double 
peak of the same type  under two approximations:

First, one must approximate  $s^2(\lambda)$ of 
(\ref{int-kernel}) by $s^2(\pm\kappa_0)$ in the integral
equation. This is justified if the peaks are narrow, i.e.,
$b^2 \ll 1 $.

Second, one must assume that the peaks are well localized on
each side of 0, i.e., the tail of the Gaussian on the
positive side does not have a significant contribution in 
the negative range and vice versa. The mathematical
condition for this is $b^2 \ll \kappa_0^2$. Also, the 
peaks should be well separated from $-1/2$ and $+1/2$.
If these conditions are fulfilled, one can extend the 
integral over each peak from $-\infty$ to $+\infty$.

By requiring $\pi^s(\kappa)$ from Eq. (\ref{ansatz}) to satisfy the
eigenvalue equation (\ref{int-eigenvalue}), one finds
\begin{equation}
\kappa_0 = \frac{p}{2(2-p)} \mbox{\ \ and \ }b^2=\frac{1-p}{(2-p)^2 N}.
\label{kappa-width}
\end{equation} 
The result for $\kappa_0$ confirms the simple argument given above, 
whereas the term for $b^2$ is slightly surprising: it does not
depend on $p$ in the leading order, i.e., it is not simply
the number of players who switch sides. Instead, the width 
of the peaks is the result of two conflicting mechanisms:
the variance in the number of players who switch sides 
makes the peak wider. It is counteracted by a
self-focusing mechanism: for example, if the system is at 
$\kappa = \kappa_0 +b$ at time $t$ , it will on average be at 
$-\kappa_0 + (1-p)\, b$ in the next time step, i.e., the
average distance from the center $-\kappa_0$ has decreased
by a factor $(1-p)$.

Eq. (\ref{kappa-width})
also allows to check whether the assumptions made for its 
derivation are true for a given $p$ and $N$, i.e., 
whether the approximation is self-consistent. For example, 
for $p = x/N$, $\kappa_0^2/b^2 \rightarrow 0$ for $N\rightarrow \infty$
according to (\ref{kappa-width}), so one cannot expect 
the formation of double peaks in this limit. The crossover from
single-peak to double-peak distribution occurs for $p \propto 1/\sqrt{N}$. 

\begin{figure}[h]
\epsfxsize= 0.65\textwidth
  \centerline{\epsffile{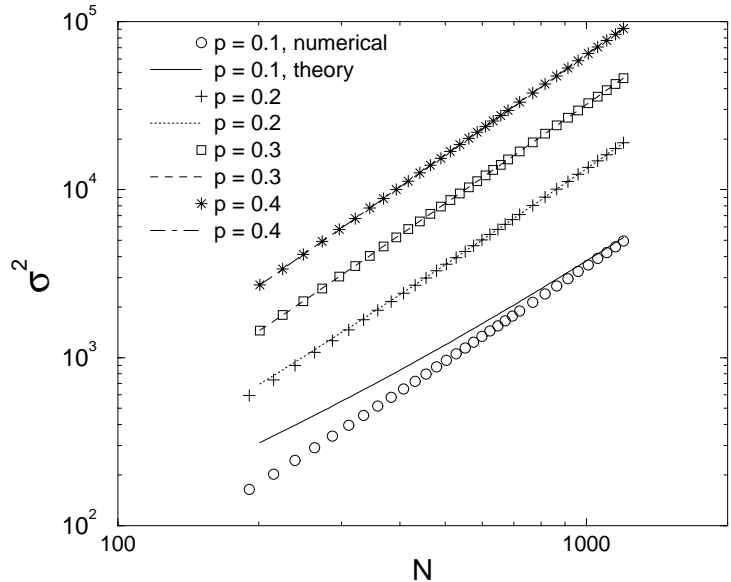}}
  \caption{$\sigma^2$ for several values of $p$ and $N$,
compared to predictions by Eq. (\ref{STO-meansm}).} 
  \label{STO-pfix}
\end{figure}

If the conditions are fulfilled, is easy to integrate over
the probability distribution Eq. (\ref{ansatz})
to get an expression for $\sigma^2$:
\begin{equation}
\sigma^2 = \frac{N}{(2-p)^2}(Np^2 + 4(1-p)). \label{STO-meansm}
\end{equation}
This agrees well with numerics well if the condition 
$\kappa_0 \gg b$ is fulfilled,
i.e., for sufficiently large $p$ and $N$, as seen in Fig.
\ref{STO-pfix}. 
Numerical evidence shows that $\sigma^2$ scales like $\sigma^2 \propto
N^2\,p^2 $ even when the condition $\kappa_0 \gg b$ is not
fulfilled, as
for $p \propto 1/\sqrt{N}$. In this case we found $\sigma^2
\propto N$, like
in the original game of Challet and Zhang and for random
guessing. However, the behavior of the system is different: the
distribution of $\kappa$ can have a  double peak structure
(depending on the proportionality constant) rather 
than a Gaussian shape, and the minority is 
still very likely to switch from one side to the other at consecutive
time steps. 

\subsection{Mixed strategy populations \& susceptibility to
  noise}
In Ch. Horn's diploma thesis \cite{Horn:Diplom}, computer 
experiments were presented in which populations with different strategies 
(decision tables, neural networks, the stochastic strategy etc.)
played in a common minority game, and the success of the
strategies depending on the fraction of players that use
them was studied. It turned out that in most cases,
each of the three mentioned strategies is robust to the
presence of a small number of players with other strategies
(meaning that the coordination is not completely disrupted 
by the presence of ``strangers'') and that populations that 
have a large majority of the total population  
usually score better than the small population using another
strategy -- the strategies are evolutionarily stable 
\cite{Maynard:Evolution}.

The susceptibility of a population of players using the
stochastic strategy to the presence of other agents 
can be calculated, provided that the output
of these other agents is not strongly correlated to the 
dynamics of the stochastic players -- in other words, the
collective output of the alternative-strategy agents
can be treated as a random number.

\begin{figure}[h]
\epsfxsize= 0.65\textwidth
 \centerline{ \epsffile{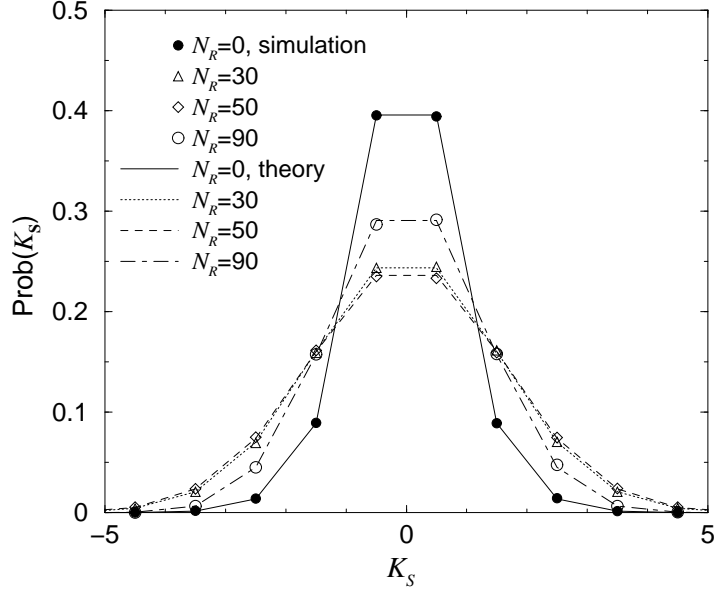}}
  \caption{Distribution of the output of $N_S$ stochastic
    players in the presence of $N_R = 101-N_S$ random
    players. Simulations are taken from
    Ref. \cite{Horn:Diplom},  numerical results are
    calculated using Eq.(\ref{mat-binomnoise}).} 
  \label{STO-noise}
\end{figure}

To be more precise, let us assume that among a total on $N$
players, a population of 
$N_S$ players using the described stochastic strategy
competes with $N_R = N-N_S$ players who are simply guessing
randomly. The output of guessers can be 
treated as a random variable $K_R$, analogous to 
Eq. (\ref{STO-defK}), with a probability distribution
\be
\mbox{Prob}(K_R=k)= \frac{1}{2^{N_R}} \binom{N_R}{k+N_R/2}.
\ee
These guessers only influence the stochastic players' 
coordination if they overrule their decision, that is, 
if $\s{K_S+K_R}\neq \s{K_S}$. The probability of 
that is
\be
p_o (K_S) = \frac{1}{2} \mbox{Prob}(|K_R|>|K_S|) = 
\frac{1}{2} \sum_{k>|K_S|}^{N_R/2} 2^{-N_R}
\binom{N_R}{k+N_R/2}.
\ee
If the decision is overruled, the stochastic players
who are on the minority side among their population 
roll the dice to see whether they change their opinion.
In terms of a random walk, there is now a probability
$p_o(k)$ that the step from position $k$ leads away
from the origin rather than towards it.
The entries of the transition matrix have to be
changed accordingly: 
 \begin{eqnarray}  \label{mat-binomnoise}
W_{k\ell} &=& 
  \binom{\frac{N_S}{2} + \ell }{\ell - k} p^{\ell - k}
  (1-p)^{\frac{N_S}{2}+k} (1-p_o(\ell)\,) \nonumber \\
&& ~~~~~~ + \binom{\frac{N_S}{2}-\ell}{k-\ell} p^{k-\ell}
(1-p)^{\frac{N_S}{2}-k} p_o(\ell)
  \mbox{\ \  for\ } \ell > 0\,, \nonumber \\
W_{k\ell} &=& 
  \binom{\frac{N_S}{2}-\ell}{k-\ell} p^{k-\ell}
  (1-p)^{\frac{N_S}{2}-k} (1- p_o(\ell)\,) \nonumber \\
&& ~~~~~~ + \binom{\frac{N_S}{2} + \ell }{\ell - k} p^{\ell - k}
  (1-p)^{\frac{N_S}{2}+k} p_o(\ell) 
\mbox{\ \  for\ } \ell < 0\,. 
\end{eqnarray}
The eigenvalues of this matrix can again be calculated 
numerically. The agreement with simulations confirms
the approach (see Fig. \ref{STO-noise}). 
In the limit where $p=2x/N$ and $N_S\rightarrow \infty$,
the distribution depends on the absolute number, not the
ratio, of random players.

\begin{figure}
\epsfxsize= 0.7\textwidth
 \centerline{ \epsffile{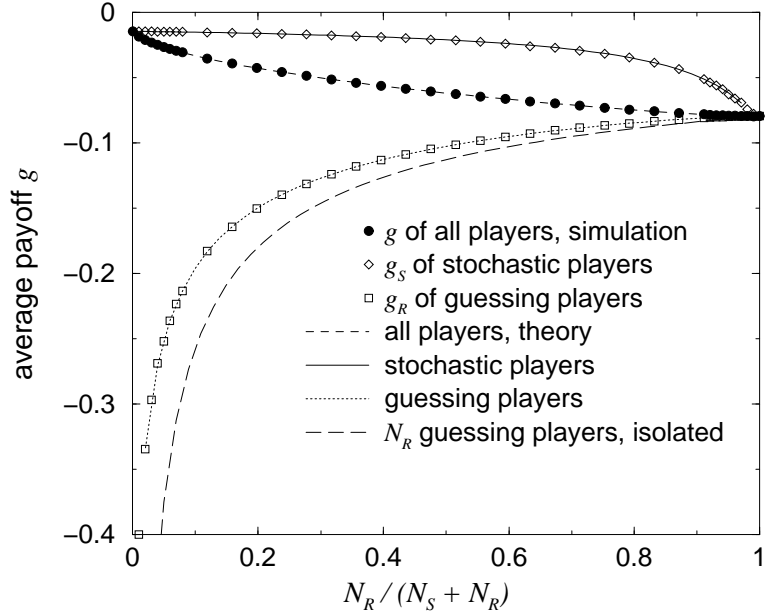}}
  \caption{Two populations of $N_S$ stochastic players
 with $p=1$ and
    $N_R = 101-N_S$ random players compete. The numerical 
results agree excellently with simulations taken from
Ref. \cite{Horn:Diplom}.} 
  \label{STO-mixedpop}
\end{figure}

Although the width of the distribution of $K_S$ increases
slightly from the presence of guessers, this drawback
does not have a drastic influence on the average gain
(or rather, loss) of the stochastic players: if
the guessers overrule the decision made by the 
stochastic players, that means that the majority of
those wins, and they actually increased their 
population's gain at the expense of the guessers.
This argument can easily be quantified: upon 
averaging the gain using the calculated probability
of $K_S$, the gain is $+2|K_S|$ with probability $p_o (K_S)$
and $-2|K_S|$ otherwise:
\be
\langle g_S \rangle  = 
\frac{1}{N} \sum_{K_S} \pi^s_{K_S} [ 2|K_S| p_o(K_S) -
2|K_S| (1-p_o(K_S))]. 
\ee
Likewise, the gain of the random players can be calculated
using the known probability distributions for $K_S$ and
$K_R$. Agreement with simulations is excellent, as seen in 
Fig. \ref{STO-mixedpop}.   

As pointed out in \cite{Horn:Diplom}, the guessers would be
better off in a pure population of $N$ guessers, but worse
off if the stochastic players were simply absent, and
the $N_R$ guessers made up the whole population.
The average loss of isolated random players is easy to 
calculate:
\be
\langle g_R \rangle = \frac{1}{N_R} 
     \sum_{K_R=-N_R/2}^{N_R/2} -2 |K_R| 2^{-N_R}
\binom{N_R}{N_R/2+ K_R},
\ee
and it is worse than that of the mixed population 
(see Fig. \ref{STO-mixedpop}).

\subsection{Correlations in the time series}
\label{SEC-STO-corr}
It was mentioned in Sec. \ref{SEC-STO-largep} that in 
the case of fixed $p$ and large $N$, the majority 
switches sides at every time step. On the other hand, 
it is obvious that for $p\ll 2/N$ no player will 
change his opinion during most time steps, leaving
the majority side unchanged. What happens between these
extremes? 

\begin{figure}

\epsfsize = 0.7 \textwidth
\centerline{\epsffile{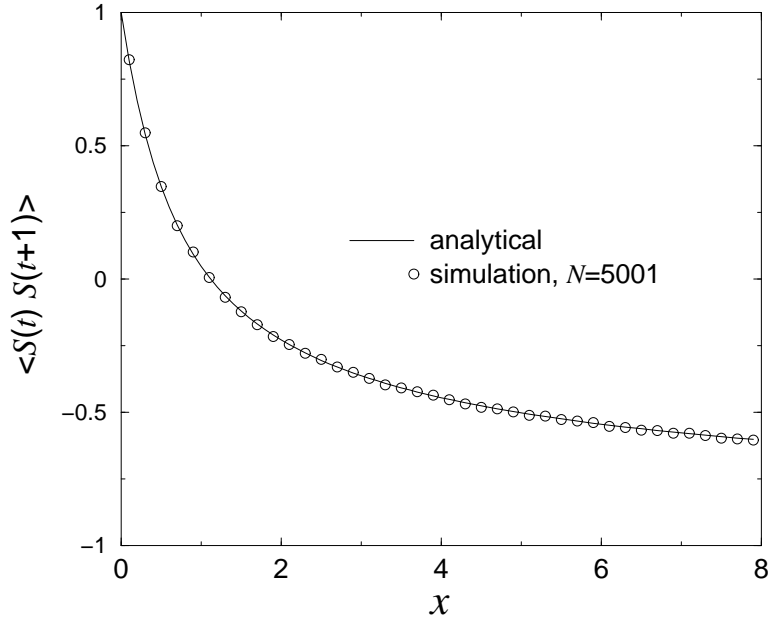}}
\caption{One-step autocorrelation of minority signs 
generated by the stochastic minority game in the
limit $p=2x/N$, $N\rightarrow \infty$.}
  \label{STO-1scorr}
\end{figure}

\begin{figure}

\epsfsize = 0.7 \textwidth
\centerline{\epsffile{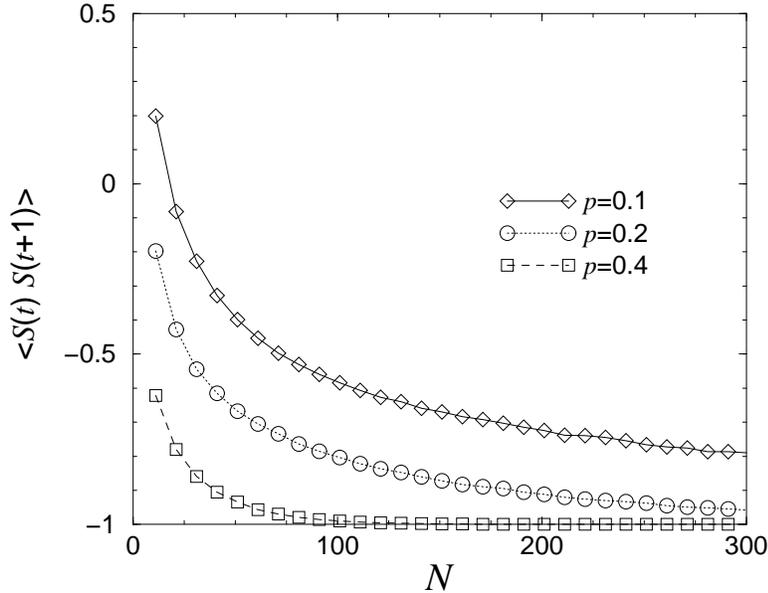}}
\caption{For fixed $p$ and increasing $N$, 
$\langle S(t) S(t+1) \rangle$ goes to zero as
the overlap between the peaks in $\pif^s$ 
decreases. See also Fig. \ref{STO-p04}.}
  \label{STO-1scorrN}
\end{figure}

To answer this question, one can calculate the
one-step autocorrelation function $\langle S(t)
S(t+1)\rangle$ of the minority sign $S(t)$ from 
the stationary probability distribution and the 
transition matrix:
\be
\langle S(t) S(t+1) \rangle = \sum_{k,l} \s{k}\s{l} \pi^s_k W_{k\ell}.
\ee
Otherwise, one can simply measure the autocorrelation in
a simulation. 
Figs. \ref{STO-1scorr} and \ref{STO-1scorrN} show
the autocorrelation function in the two limits
treated in Sec. \ref{SEC-STO-smallp} and
\ref{SEC-STO-largep}, respectively.
 In the limit
of $p=2x/N$, the correlation is positive for small $x$, 
as expected: if fewer than one player change sides on
average, the majority is not likely to switch. For
$x\approx 1.1$, $\langle S(t) S(t+1) \rangle =0$.
Beyond that point, the autocorrelation function is negative,
and it goes to $-1$ as $x\rightarrow \infty$.

For fixed $p$,  $\langle S(t) S(t+1) \rangle =0$ goes to 
$-1$ rapidly with increasing $N$. The interpretation in 
terms of the probability distribution is that an appreciable
overlap between the two Gaussian peaks in $\pif^s$
corresponds to a non-vanishing probability that the 
minority sign stays the same. As $N$ increases, the
peaks get narrower, and the overlap vanishes.

\subsection{Relation to the original MG}
The stochastic MG as it was presented here has only
a one-step memory. It is possible (although not necessary)
 to include a longer history of minority decisions,
analogous to the early papers on the evolutionary 
Minority Game \cite{Johnson:Segregation}.
In that case every player would keep
an individual decision table that tells him what to do if a given
sequence of minority decisions occurs, similar to the tables that the
players in the original Minority Game use. However, instead of changing to 
an entirely different table, each player changes individual entries in
his table with probability $p$ if he loses. It is easy to see -- and a
similar argument was given in Ref. \cite{Burgos:Selforg.} for the global
decision table in Johnson's variant -- that the entries for different
histories are completely decoupled. Each entry or row in the table
corresponds to a one-step Markov process as described above, influenced
only by the last time that the same history occurred.

In that sense, introducing a history changes the properties of the
time series generated by the decisions, but not the average loss of the
players in the stationary state. It might become relevant if
one mixes players with different strategies and different
memory length, who are susceptible to correlations on
different orders of the DeBruijn graph (see
Sec. \ref{AP-time} and \ref{AP-stoch}).
As Sec. \ref{SEC-STO-corr} showed, there is a crossover
from persistent to antipersistent behavior as the switching
probability is increased. If $p$ is constant and $N$ is
large enough, the ensemble of players can essentially
be replaced by one effective player, who generates
a time series with the properties studied in Sec. 
\ref{SEC-Decision}. 

The presented strategy shows similarities to the behavior of 
the original minority game in the limit $\alpha \rightarrow 0$, i.e.,
a small ratio of possible histories compared to the number of players
\cite{Heimel:Generating}. In the extreme case where each of the two decision
tables that each player keeps consists of only one entry
(i.e., only the last minority decision counts), the output 
of roughly $50\%$ of the players is set to either $+1$ or $-1$
independent of the history, whereas the other
$50\%$ can choose their output depending on their current
score. Out of those players, those who chose the minority side 
will repeat their decision, whereas the update of the scores
will cause some of the losers to switch sides. As mentioned 
before, it has been 
observed that $\sigma^2$ shows a crossover from $\sigma^2\propto N^2$
to $\sigma^2 \propto 1$ depending on the initial differences
in players' scores \cite{Garrahan:Continuous,Heimel:Generating}; 
these differences determine the typical number of players 
who switch sides when they lose. 
This situation is very much the same for players who use perceptrons
with only one input unit \cite{Metzler:Interact}, where 
a solution with $\sigma^2=1$ is reached if the learning
rate $\eta$ is smaller that the differences between the 
initial weights of the players.
However, in the absence of frozen disorder, the stochastic 
strategy obviously does not lead to a separation of agents 
into frozen and oscillating players.

\section{Multi-choice MGs}
\label{SEC-PMG}
A generalization of the MG that was suggested by I. Kanter 
and studied first by Ein-Dor, Metzler, Kinzel and Kanter in Ref. 
\cite{Ein-Dor:Multi-choice} was to lift the restriction to two 
alternatives and allowing for $Q$ choices, or options, or ``rooms'',
instead. The choice that is picked by the fewest players
wins; in case of a tie, one of the rooms with the lowest
attendance is picked at random and declared the winning option.
The motivation is clear, since many real-world problems offer
more than two choices: an investor has many different stocks
to choose from, more than two routes may connect two 
cities, and a predator may have more than two possible hunting
grounds. In a sense, the generalization from 
the binary MG to a multi-choice MG is analogous to 
the generalization from an Ising model \cite{Huang:Statistical} 
to a Potts model \cite{Wu:Potts}. However, this analogy does not
apply to the mathematical treatment of the standard MG using
Ising spin variables \cite{Challet:Analytical} -- there,
the spin variables represented the choice between the
two decision tables rather than the two actions. 

\subsection{Random Guessing}
\label{SEC-PMG_random}
The default strategy against which any more 
sophisticated approaches have to be measured is that
of randomly choosing one of the options. It is not
as obvious as in the binary MG what numbers come 
out, or even what quantities one should look at.
I will therefore devote a few words to that topic.
The following calculations were largely done by
Liat Ein-Dor.

The macrostate of the game can be described by the
set of numbers $\{N_q\}$ of players who choose each option
$q=1,\dots,Q$. 
However, the only quantity relevant for global gain or loss is
the number $N_{min}= \min_q(\{N_q\})$ 
of players in the winning room. In analogy to 
the binary MG, one can define a variance
\be
 \sigma_{min}^2 =  \langle N_{min} -N/Q \rangle_t.
\ee
Analogous to the probability distribution of $N_{+}$
in the binary MG, which, for random guessing, follows 
a binomial distribution, the joint probability 
distribution of the $N_q$s is given by the
multinomial distribution:
\be
P(\{N_q\})= Q^{-N} \frac{N!}{N_1!N_2!
\dots N_Q!} \delta_{N-\sum N_q}. \label{PMG-P_Nq}
\ee
Averages over $N_{min}$ can be evaluated by exploiting the
symmetry of the system: I assume that
$N_1$ is the smallest occupation number (which happens
with a probability of $1/Q$), include a factor of $Q$
to correct this assumption,  
 and sum over 
the other $N_q$ with appropriate boundaries. For example,
$\sigma_{min}^2$ can be calculated as follows:
\be
\sigma_{min}^2 = Q \sum_{N_1=0}^{N} \left(N_1 -\frac{N}{Q}
\right)^2 \sum_{N_2, N_3, \dots= N_1}^{N} P(\{N_q\}).
\ee
Numerics show that for large $N$, $\sigma_{min}^2$
is proportional to $N$, as expected. To calculate
the proportionality constant for $N\rightarrow
\infty$, one can
introduce new, properly scaled quantities $\epsilon_q$
to measure the deviations from equidistribution:
\be
N_q = N/Q + \epsilon_q \sqrt{N},
\ee 
and approximate the factorials in 
Eq. (\ref{PMG-P_Nq}) using the Stirling approximation
\be
N! \approx \left(\frac{N}{e} \right)^N \sqrt{2\pi N}.
\ee
The result 
\be
P(\{\epsilon_\rho\})\propto \exp\left({-\frac{Q}{ 2}\sum_{\rho=1}^{K}
 \epsilon_\rho^2 }\right )
\delta \left( \sum_{q=1}^Q \epsilon_q\right)
\ee
can be used to numerically integrate the continuous
expression for $\langle \epsilon_{min}^2 \rangle$:
\be
 \langle \epsilon_{min}^2 \rangle =\frac{
\int_{-\infty}^{0} d\epsilon_{1} 
{\epsilon_{1}}^2
\int_{\epsilon_{1}}^{\infty}
\left (\prod_{q \geq 2} d\epsilon_{q} \right) 
P(\{\epsilon_{\rho}\})}
{
\int_{-\infty}^{0}
d\epsilon_{1} 
\int_{\epsilon_{1}}^{\infty}
\left( \prod_{q \geq 2}  d\epsilon_{q}\right)
 P(\{\epsilon_{q}\}) }.  \label{PMG-epsmin}
\ee
Structurally, this is an integral over the square
of the smallest of $Q$ Gaussian random numbers. 
The only difficulty is the global constraint for the
sum of $\epsilon_q$.
Results of the numerical integration for $Q=3,4,5$ and
$6$ are $\sigma_{min}^2/N = \langle \epsilon^2_{min} \rangle \approx $ 
$0.313$, $0.322$, $0.320$,  and $0.309$, respectively.

Another quantity that is easier to calculate and often 
just as meaningful as $\sigma_{min}^2$ is 
\be 
\sigma^2 = \left \langle \frac{1}{Q} \sum_q \left (N_q -
  \frac{N}{Q}\right)^2 \right \rangle.
\ee
If there is no systematic preference for one of the 
options, it is sufficient to look at one of the
options, e.g. the option 1. For random guessing, this
quantity takes the form
\bea
\sigma^2  &=& \left \langle \left (\sum_{i=1}^N
    \delta_{\sigma_i,1} - \frac{N}{Q} \right)^2 \right \rangle
        \nonumber \\
&=& \left \langle \sum_{i,j} \delta_{\sigma_i,1}
  \delta_{\sigma_j,1} - 2 \frac{N}{Q} \sum_i
  \delta_{\sigma_i, 1} + \frac{N^2}{Q^2} \right \rangle
    \nonumber \\
&=& \left \langle \sum_i \delta_{\sigma_i, 1} + 
\sum_{i\neq j }\delta_{\sigma_i, 1} \delta_{\sigma_i,
  \sigma_j} - 2\frac{N}{Q} \sum_i \delta_{\sigma_i, 1} +
\frac{N^2}{Q^2} \right\rangle
\nonumber \\
&=& 
 \frac{N}{Q} + \frac{N (N-1)}{Q} \left \langle
 \delta_{\sigma_i,\sigma_j} \right \rangle - \frac{N^2}{Q^2} \label{PMG-eps1}\\
&=& N \left(\frac{Q-1}{Q^2}\right). \label{PMG-eps2}
\eea 
The form of the expression in Eq. (\ref{PMG-eps1}) will be 
useful later on, since it reduces the calculation of
$\sigma^2$ to the average probability that two agents agree
on their output.
 
\subsection{Neural Networks}
\label{SEC-PMG_NN}
\subsubsection{Network architecture}
The first scenario for the multi-choice minority game 
that was studied in detail \cite{Ein-Dor:Multi-choice}
included an ensemble of neural networks making their choice
based on an $M$-dimensional vector whose components are 
the most recent minority decisions. Since both the input and the output
of the networks are now integer numbers in the 
range between 1 and $Q$, the architecture of the networks
has to be able to handle this. A simplified version of 
a Potts perceptron \cite{Watkin:Multi-Class}  
 which is introduced in greater detail in Chapter \ref{CHAP-game}, 
seems appropriate. In this simplification, each 
player has a weight vector $\vw_i$, from which $Q$ 
hidden fields are calculated. Each hidden field $h_{iq}$
only gets contributions from components of the weight vector
if the corresponding component of the vector is
equal to $q$:
\be
h_{iq} = \sum_{j=1}^M w_{ij} \delta_{x_j,q}.
\ee
This architecture is limited compared to the
full Potts perceptron: first, there is no 
interaction between the different input values --
the appearance or non-appearance of some $q_1$
in the pattern does not influence the hidden field 
for any other $q$. Second, the options are
completely interchangeable: if one would interchange
all occurrences of $q_1$ and $q_2$ in the pattern,
one would interchange $h_{q_1}$ and $h_{q_2}$ as
well. This seems reasonable, given that the options
are completely equivalent \`{a} priori. 

Each agent determines his output $\sigma_i$ by choosing
the option that corresponds to his largest hidden field:
\be
\sigma_i = \{ k | h_{ik} = \max_q h_{iq}  \}
\ee 
From these individual outputs, the occupation numbers
$N_q = \sum_j \delta_{\sigma_j,q}$ are calculated,
and the winning option $S$ is determined:
\be
S = \{ q | N_q = \min_m N_m\}
\ee
If the learning rule is Hebbian, analogous to
Sec. \ref{SEC-MGNN}, an analytical approach is possible.
The update rule for the vectors is therefore
\be
w_{ij}^{t+1} = w_{ij}^t + \frac{\eta}{M} (Q \delta_{x_j,
  S}-1).\label{PMG-Hebb}
\ee
The rule does what one would expect: it increases the hidden
field for option $S$ and decreases the fields for the
other options. Also, the same update step is applied
to all weights, leaving relative vectors unchanged.

Just like in Sec. \ref{SEC-MGNN}, weight vectors can 
be split into a center of mass $\C = \sum \vw_i/N$
and relative vectors $\r_i = \vw_i-\C$. The initial
conditions are such that
vectors have a norm of $|\r_i|=1$ 
and symmetrical overlaps of $\sp{\r_i}{\r_j} =-1/(N-1)$
(the first condition is a choice of length scale, the
second is automatically fulfilled approximately if
vectors are chosen randomly with a large number $M$
of dimensions).
If one sets $C=|\C|=0$, this is indeed the optimal 
configuration that can be achieved without 
breaking the symmetry between the perceptrons.

For random patterns and large $N$, the hidden fields $h_{iq}$ are 
Gaussian random numbers. It is convenient to rescale them 
to variables $\hh_{iq}$ which  obey $\langle \hh_{iq}\hh_{ir}
\rangle = \delta_{q,r}$ and $\langle \hh_{iq} \hh_{jr} \rangle =$
$R\delta_{q,r}$ for $i \neq j$, where $R = \langle
\sp{\vw_i}{\vw_j}/(w_i w_j) \rangle_{i,j}$ is the overlap 
between different weight vectors. 

The interesting task is calculate $\sigma_{min}^2$ 
for non-vanishing $R$, i.e., for correlated weight vectors.
This is apparently only feasible for small correlations
$R =\mathcal{O}(1/N)$, and even there only in a rather
convoluted way. The following calculation was basically 
done by L. Ein-Dor and I. Kanter. However, its presentation
in Ref. \cite{Ein-Dor:Multi-choice} is somewhat cryptic
and slightly wrong,
and I will attempt to give an understandable and correct
account of it.

\subsubsection{Calculating $\sigma^2$ and $\sigma_{min}^2$}
The first step is to calculate the variance $\sigma^2$,
which can be broken down to an average over two 
players, as pointed out in Sec. \ref{SEC-PMG_random}.
The question is thus: what is the probability that 
two perceptrons $i$ and $j$ with a small overlap $R$ give the
same output, let us say, $\sigma_i=\sigma_j=1$? 
Equivalently, what is the probability that 
$\hh^i_1 \geq \hh^i_q \wedge \hh^j_1 \geq \hh^j_q $ for all
$q$? The hidden fields can be combined into one vector
$\hat{\v{h}}= (\hh^i_1, \dots,\hh^i_Q, \hh^j_1, \dots, \hh^j_Q)$ of
Gaussian variables with the correlation matrix
\be
\Cm = \left(\begin{array}{cccccc} 
1 & \cdots & 0 & R & \cdots & 0 \\
\vdots  & \ddots  & \vdots & \vdots & \ddots &\vdots \\
0 & \cdots &1 &0 &\cdots&R \\
R&\cdots & 0 & 1& \cdots &0  \\
\vdots & \ddots & \vdots & \vdots & \ddots & \vdots \\
0 & \cdots & R & 0 & \cdots &1 
\end{array}\right).
\ee
The joint probability distribution of these variables
follows 
\be
P(\v{\hh}) = \frac{1}{\sqrt{(2\pi)^{2Q}
    \mbox{Det}(\Cm)}} \exp\left(-\frac{1}{2} \hat{\v{h}}^T \Cm^{-1}
  \hat{\v{h}} \right).  \label{PMG-pdist1}
\ee
For general $R$ and $Q$, this expression is fairly 
complicated. We simplify it by approximating
$\Cm^{-1}$ to the first order in $R$. It turns out that
\be
\Cm^{-1}  \approx \Cm^{-1}_{app}=\left(\begin{array}{cccccc} 
1 & \cdots & 0 & -R & \cdots & 0 \\
\vdots  & \ddots  & \vdots & \vdots & \ddots &\vdots \\
0 & \cdots &1 &0 &\cdots&-R \\
-R&\cdots & 0 & 1& \cdots &0  \\
\vdots & \ddots & \vdots & \vdots & \ddots & \vdots \\
0 & \cdots & -R & 0 & \cdots &1 
\end{array}\right),
\ee
which can be verified by checking that $\Cm\cdot \Cm^{-1}_{app}$
yields the identity matrix multiplied with $1-R^2$.

In that approximation, the probability distribution from 
Eq. (\ref{PMG-pdist1}) looks more amiable:
\be
P(\hat{\v{h}}) \approx \frac{1}{\sqrt{(2\pi)^{2Q}}}
\exp \left(-\frac{1}{2} \sum_q \left ((\hh_q^i)^2 +
    (\hh_q^j)^2 -2 R \hh_q^i \hh_q^j \right ) \right ). 
\ee
Now the probability that both networks give an output of
$1$ can be calculated, again in a consequent 
approximation to first order in $R$:
\bea
& & \mbox{Prob}(\sigma_i = \sigma_j =1)  \nonumber \\
&=&
\int_{-\infty}^{\infty} d\hh_1^i 
\int_{-\infty}^{\infty} d\hh_1^j
\int_{-\infty}^{\hh_1^i} \prod_{q=2}^Q d\hh_q^i  
\int_{-\infty}^{\hh_1^j} \prod_{q=2}^Q d\hh_q^j \times  \nonumber \\
& &~~~~\frac{1}{(2 \pi)^Q} \exp\left( -\frac{1}{2}
  \sum_{q'}(q{\hh_{q'}^i} ^2 + {\hh_{q'}^j}^2 -2 R \hh_{q'}^i
  \hh_{q'}^j ) \right) \nonumber \\
&=& \int_{-\infty}^{\infty} d\hh_1^i 
\int_{-\infty}^{\infty} d\hh_1^j \frac{1}{2\pi} 
\exp \left (-\frac{1}{2} ({\hh_{1}^i}^2 + {\hh^j_{1}}^2 -2 R
\hh_{1}^i \hh_{1}^j) \right ) \times \nonumber \\
& & ~~~~ \left [
\int_{-\infty}^{\hh_1^i} \int_{-\infty}^{\hh_1^i} d\hh^i d\hh^j
\frac{1}{2\pi} \exp \left (-\frac{1}{2} ({\hh^i}^2 + {\hh^j}^2 -2 R
\hh^i \hh^j) \right )
\right ]^{Q-1} \nonumber \\
&=& \int \dots \left [ \int_{-\infty}^{\hh_1^i}
\frac{1}{\sqrt{2 \pi}} \exp \left( -
  \frac{-{\hh^i}^2}{2}(1-R^2) \right)\Phi(\hh_1^j - R\hh^i)
\right]^{Q-1} \nonumber \\
&\approx& \int \dots  \left [\int_{-\infty}^{\hh_1^i}
\frac{1}{\sqrt{2 \pi}} \exp \left(-\frac{{\hh^i}^2}{2}
\right)\left(\Phi(\hh_1^j) - \frac{R\hh^i}{\sqrt{2\pi}}
\exp  \left(-\frac{{\hh^j}^2}{2}\right) \right ) \right ]^{Q-1}
 \nonumber 
\eea
\bea
&=& \int \dots \left[ \Phi(\hh_1^j) \Phi(\hh_1^j) +
  \frac{R}{2\pi}
\exp \left( -\frac{1}{2}( {\hh_1^i}^2 + {\hh_1^j}^2 \right )
\right]^{Q-1} \nonumber \\
&\approx&  \int_{-\infty}^{\infty} d\hh_1^i 
\int_{-\infty}^{\infty} d\hh_1^j \frac{1}{2\pi} 
\exp \left (-\frac{1}{2} ({\hh_{1}^i}^2 + {\hh^j_{1}}^2)
\right)
\left (1 +2 R
\hh_{1}^i \hh_{1}^j \right ) \times \nonumber \\
& & ~~~~\left [ (\phi(\hh_1^i)\phi(\hh_1^i))^{Q-1} +
  \frac{R(Q-1)}{2\pi}\exp \left( -\frac{1}{2}( {\hh_1^i}^2 +
    {\hh_1^j}^2) \right ) (\phi(\hh_1^i)\phi(\hh_1^i))^{Q-2}
\right ]
\nonumber \\
&\approx& \frac{1}{Q^2} + R \left( \left[\int_{-\infty}^{\infty} dh 
\frac{h}{\sqrt{2\pi}} \exp(-h^2/2) \Phi(h)^{Q-1}\right ]^2
\right. \nonumber \\
&&
 + \left. (Q-1) \left [
    \int_{-\infty}^{\infty} 
 dh \frac{1}{2\pi} \exp(-h^2) \Phi(h)^{Q-2} \right ]^2
\right).
\label{PMG-si_sj}
\eea
The first term in Eq. (\ref{PMG-si_sj}) is the default result
for uncorrelated vectors. The second term is due to the 
correlations between $h_1^i$ and $h_1^j$, and the third
term represents correlations between  $h_q^i$ and 
$h_q^j$ for $q\geq 2$. For some reason, Ref. 
\cite{Ein-Dor:Multi-choice} only lists the first
and third term. 

Using this result, $\sigma^2$ can be calculated 
according to Eq. (\ref{PMG-eps1}):
\be
\sigma^2(R)/N = \frac{Q-1}{Q^2} + (N-1) R I, 
\label{PMG-eq_s_R}
\ee
where $I$ stands for the two integrals in 
Eq. (\ref{PMG-si_sj}). Plugging in the
optimal value $R = -1/(N-1)$, one gets an optimal 
$\sigma^2/N$ which is independent of $N$, just 
like in the binary-choice MG. 

\begin{figure}
\epsfxsize= 0.7\textwidth
 \centerline{ \epsffile{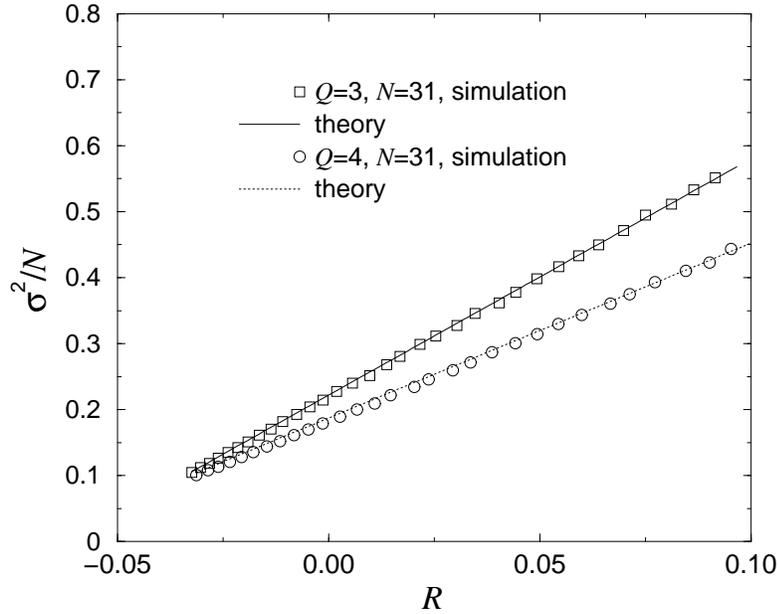}}
  \caption{Variance $\sigma^2/N$ as a function of $R$ 
    in the multi-choice minority game with neural networks.
    Eq. (\ref{PMG-eq_s_R}) agrees well with simulations
    using $M=200$. All simulations in this section use
    random patterns.} 
  \label{PMG-s_R}
\end{figure}

Considering that Eq. (\ref{PMG-eq_s_R}) is a first-order 
approximation,
it agrees surprisingly well with simulations, 
as seen in Fig. \ref{PMG-s_R}.

What is still left to do is to deduce $\sigma_{min}^2$ 
from $\sigma^2$. This can be done in a surprisingly
easy way: for large $N$, the occupation numbers $N_q$ are still
Gaussian random numbers whose variance is now given by 
Eq. (\ref{PMG-eq_s_R}) instead of (\ref{PMG-eps2}).  
Therefore, the integral
necessary to calculate $\sigma^2_{min}/N = \langle
\epsilon_{min}\rangle$ has the exact form of
Eq. (\ref{PMG-epsmin}), and if $\epsilon$ is rescaled 
properly, one can reuse the integrals evaluated to
get $\sigma_{min}^2$  in the random case:
\be
\frac{\sigma^2(R)}{\sigma^2_{min}(R)} = 
\frac{\sigma^2(R=0)}{ \sigma^2_{min}(R=0)}.
\ee 
The values on the right hand side can be taken from 
Eqs. (\ref{PMG-eps2}) and (\ref{PMG-epsmin}), while the
enumerator on the left hand side is given by Eq. (\ref{PMG-eq_s_R}).

\begin{figure}
\epsfxsize= 0.7\textwidth
 \centerline{ \epsffile{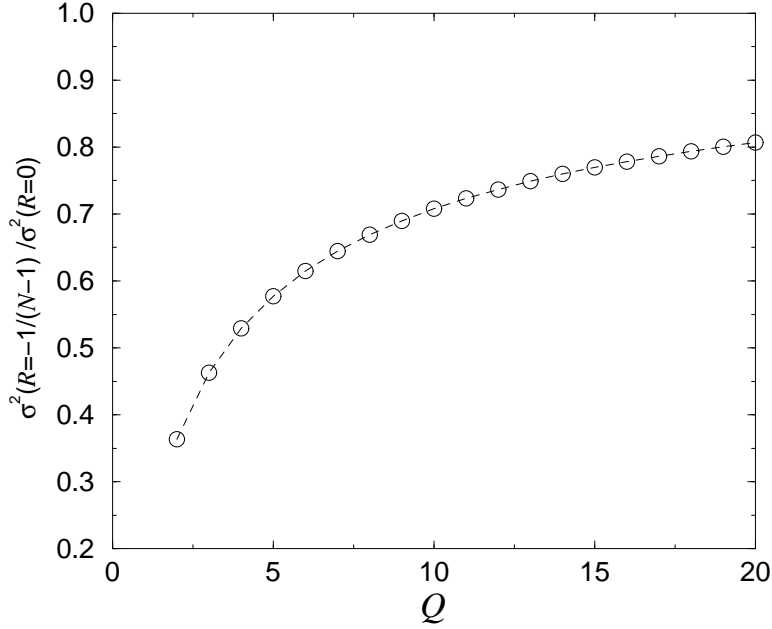}}
  \caption{Improvement by coordination:
    $\sigma^2(R=-1/(N-1))/\sigma^2(R=0)$
    as calculated from Eq. (\ref{PMG-eq_s_R}). For larger $Q$,
    the benefit gained from learning decreases.} 
  \label{PMG-sigratio}
\end{figure}

Just how good is perfect coordination ($R=-1/(N-1)$) compared to 
random guessing? As the calculations have shown, this 
depends only on $Q$. The ratio  $\sigma^2(R=-1/(N-1))/\sigma^2(R=0)$ 
is shown in Fig. \ref{PMG-sigratio}. For $Q=2$, one gets
the ratio $1-2/\pi$, as calculated in Sec. \ref{SEC-MGNN}.
For larger $Q$, the ration slowly converges to 1:
the benefit from coordination decreases with $Q$.

\subsubsection{Learning Dynamics}
The preceding paragraphs showed how $\sigma^2$ depends
on $R$, and that $R$ depends on the norm of the
center-of-mass vector $\C$. To finish the argument, one
can show that the proposed learning rules leads to
$C\rightarrow 0$, and thus optimal coordination, 
for $\eta \rightarrow 0$. The calculation, which
is again by L. Ein-Dor, is 
roughly analogous to that for the binary-choice MG,
but takes a shortcut at a convenient spot.

To find $C$ in dependence on $\eta$, $K$, and $Q$,
one starts with the update rule for $\C$, which
is analogous to Eq. (\ref{PMG-Hebb}):
\be
C_j^{t+1}= C_j^t + \frac{\eta}{M} (Q \delta_{x_j,S} -1),
\ee
takes the square of it, averages over random patterns, 
and sums over $j$:
\be
(C^{t+1})^2 = (C^t)^2 + 2 Q \frac{\eta}{M} \left \langle
 \sum_j^M C_j \delta_{x_j,S} \right \rangle 
+ \frac{\eta^2}{M^2} \sum_j^M \langle (Q  \delta_{x_j,S}
-1)^2 \rangle. 
\ee
For $M\rightarrow \infty$, this turns into a 
differential equation for $C$, where one 
infinitesimal time step $d\alpha=1/M$ corresponds to one
learning step:
\be
\dd{C^2}{\alpha} =  2\eta Q\left \langle \sum_j C_j
  \delta_{x_j,S} \right \rangle + \eta^2 (K-1).
\label{PMG-NNdiff}
\ee
At this point, Ref. \cite{Ein-Dor:Multi-choice} 
makes a convenient approximation: it is assumed that
the output of the ensemble of networks always 
follows the output of the center-of-mass. While 
possibly true
for large $N$ and large $\eta$, this is certainly 
not correct for small $\eta$ -- the range we are interested 
in! However, I will first try to explain the approximation,
and then follow it to the end to see if it makes a difference
in the final result.

The hidden fields $h_{iq}$ of each player can again be 
split into two contributions, one from the random vector and
one from the center-of-mass:
\be
h_{iq} = \sum_m^M \delta_{x_m,q} {r_i}_m + \sum_m^M \delta_{x_m,q} C_m.
\ee
The fields induced by the center-of-mass,
$h_q^C =  \sum_m \delta_{x_m,q} C_m$, thus act as 
a bias on the random decision of the relative
vectors. If $C$ is of order 1, the center-of-mass
heavily influences the decision of each individual. However, to 
influence the decision of the minority, it only has
to tip the decision of $\mathcal{O}(\sqrt{N}/Q)$ players.

\begin{figure}[h]
\epsfxsize= 0.7\textwidth
\centerline{  \epsffile{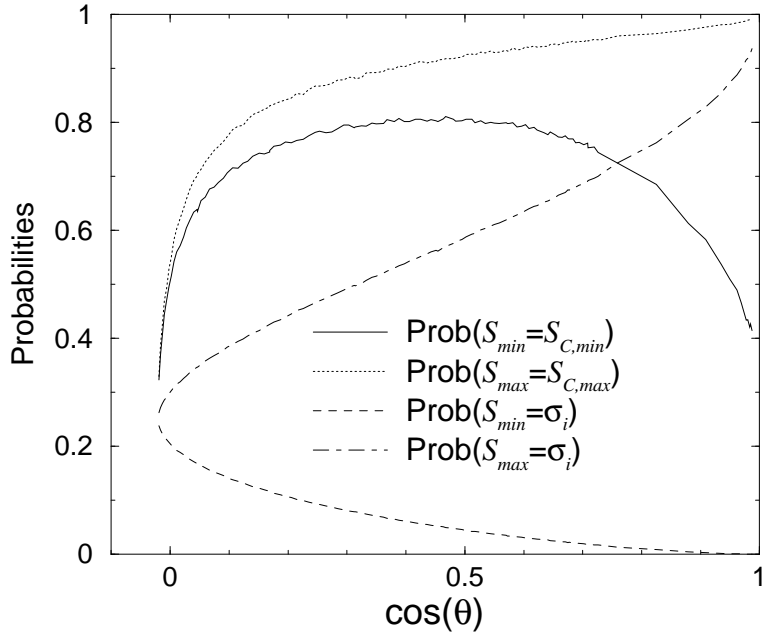}}
  \caption{Perceptrons in the multi-choice MG:
    probabilities that the least occupied room $S_{min}$
    corresponds to the option
    $S_{C,min}$ with the smallest bias $h_q^C$, that the
    most occupied room $S_{max}$ corresponds to the
    largest bias $S_{C,max}$, and that the individual
    decisions $\sigma_i$ correspond to the respective 
    largest and smallest bias, as a function of the
    weight vectors' mutual angle $\cos(\theta)$. 
    Simulations used $Q=4$, 
    $N=41$ and $M=100$.}
  \label{PMG-probs}
\end{figure}

It is messy to calculate the probability that
the output of a single player follows the smallest
hidden field of the center-of-mass, not to mention 
the probability that the ensemble joins this 
decision. The probability can be measured in 
simulations, though, and results are shown in
Fig. \ref{PMG-probs} for $Q=4$. As expected, 
the probability that the room with the largest
occupation corresponds to the largest $h^C_q$
goes to 1 as the overlap between weight vectors
increases. However, the probability that the 
smallest occupation is given by the smallest hidden
field takes high values for intermediate $\cos(\theta)$
and the decreases again. The reason for this is not yet clear.

If one nevertheless assumes that the option with the smallest 
$h_q^C$ is the minority decision $S$ of the ensemble,
$\sum_j C_j
  \delta_{x_j,S}$ is simply the smallest of $Q$
uncorrelated Gaussian random numbers with zero 
mean and variance $C^2/Q$, and the average over
it can be calculated. Again, let us assume that 
the winning option is 1, and correct the assumption by a
factor of $Q$:
\bea
\left \langle \sum_j C_j
  \delta_{x_j,S} \right \rangle &\approx& 
Q  \int_{-\infty}^{\infty} dh^C_1 \frac{h^C_1}{\sqrt{2 \pi C^2/Q}}
 \exp\left(-\frac{{h_1^C}^2}{2C^2/Q}\right) \times \nonumber
 \\
& &~~~~~ \int_{h_1^C}^{\infty}
\prod_{q\geq 2} dh^C_q \frac{1}{\sqrt{2 \pi C^2/Q}} 
\exp\left(-\frac{{h_q^C}^2}{2C^2/Q}\right) \nonumber \\
&=& C \sqrt{Q} \int_{\infty}^{\infty}
dh\frac{h}{\sqrt{2\pi}}  \exp(-h^2/2) (1-\Phi(h))^{Q-1}.
\label{PMG-hc_avg}
\eea
This is proportional to $C$, and together with Eq. 
(\ref{PMG-NNdiff}) one gets for the fixed point
\be
C_{fix} \approx \eta \frac{Q-1}{2\sqrt{Q}^3  J}, 
\ee
where $J$ is the integral expression in 
Eq. (\ref{PMG-hc_avg}).

As mentioned before, this approximation is 
never truly valid. It is the same approximation that replaces the
center-of-mass vector in the binary-choice 
perceptron with a single confused perceptron
(see Sec. \ref{SEC-MGHebb}), but the existence
of more than two alternatives complicates the situation.

Nevertheless, simulations indicate that $C$ indeed goes to 0 
as $\eta\rightarrow \infty$  for small $N$,
although not linearly. This is again analogous to the
binary-choice-perceptron, where $C\propto \sqrt{\eta}$ 
was found for small $\eta$. It does not change
the conclusion that $\sigma^2$ takes its optimal
value for very small learning rates.

\subsection{Decision tables}
The possibility of applying other rules of behavior (such 
as Challet and Zhang's decision tables) to the multi-choice 
minority game was only hinted at in
Ref. \cite{Ein-Dor:Multi-choice}. I will present a
straightforward generalization of the standard MG 
and some simulations which indicate that the 
behavior of the multi-choice MG is very similar to
the two-choice MG.
A more thorough analysis based on a slightly 
modified generalization was presented by Chow and Chau 
in Ref. \cite{Chow:Multiple}.

The simplest generalization from the standard MG
\cite{Challet:Emerg.,Challet:Analytical,Challet:Theory} 
is to give each player two decision 
tables with entries $a^{\mu} \in \{1,\dots,Q\}$ 
for each of the $Q^M$ possible histories $\mu$. 
Scoring does not have to be modified: a table receives a 
point if it would have predicted the correct minority 
room, and loses a point for an
incorrect prediction. The table with the highest score is used.

Since the number of table entries $Q^M$ increases rather
drastically, it simplifies simulations to substitute the
time series/ history scheme by a number of $p$ possible
patterns $\mu$ that are picked at random at each time step.
In the conventional MG, this alters results 
only very little \cite{Cavagna:Memory}.

\begin{figure}[h]
\epsfxsize= 0.65\textwidth
\centerline{  \epsffile{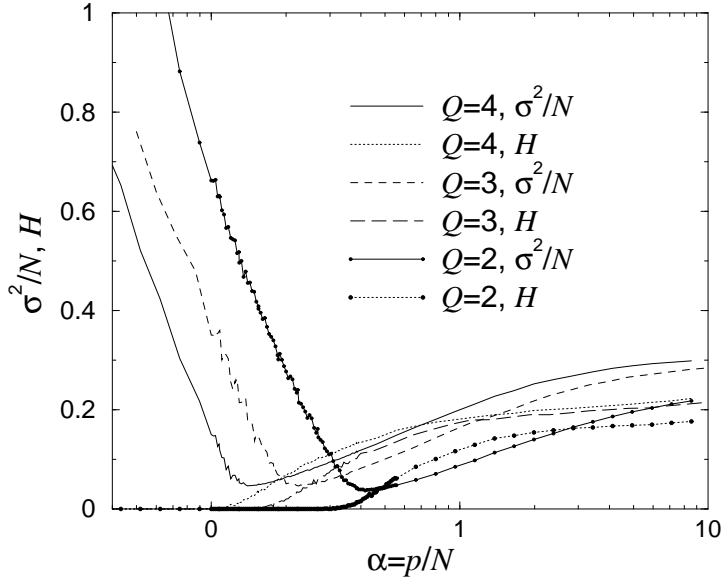}}
  \caption{Standard Minority Game with $Q\ge 2$ options.
    Curves for $\sigma^2/N$ and $H$ look qualitatively
    similar, whereas the value $\a_c$ of the phase
    transition depends on $Q$. Simulations use $N=101$ and $N=201$.}
  \label{MG-CZpotts}
\end{figure}

Result can be seen in Fig. \ref{MG-CZpotts}: curves for 
$\sigma^2/N$ look qualitatively similar for $Q=2$ (the
binary case, apart from a factor of 4 that comes from
different definitions of $\sigma^2$), $Q=3$ and $Q=4$.
For $\alpha \rightarrow \infty$, the values of $\sigma^2$
approach those of random guessing ($\sigma^2/N=(Q-1)/Q^2$), 
whereas at small $\alpha$ there is a crowded phase.

As in the case of $Q=2$, a preference for a certain response
to each pattern $\mu$ appears at $\alpha_c$ and increases 
with $\alpha$. This preference can be expressed as 
an information $H$ in the following way: if $c_q^{\mu}$ is
the probability that following a given 
pattern $\mu$, the minority chooses option $q$, one defines
\be
H = \frac{1}{p} \sum_{\mu=1}^p \sum_{q=1}^Q \left
  (c_q^{\mu} -\frac{1}{Q} \right)^2.
\ee 
The phase transition can be located by estimating where
$H$ becomes 0 for $N\rightarrow \infty$. As seen in Fig.
\ref{MG-CZpotts},
the critical value $\alpha_c$ of the phase transition
decreases with $Q$.

\subsection{Johnson's evolutionary MG}
\label{PMG-evol}
The evolutionary Minority Game suggested in 
Ref. \cite{Johnson:Segregation} and briefly described in 
Sec. \ref{SEC-EvolMG} has several possible generalizations
to $Q$ options. 
In the first, players have a probability $p$ to choose the
option that was successful the last time. Otherwise, they 
choose one of the remaining options at random. 
In the second, players  $i$ have a set of probabilities $\{p_i^q\}$ 
to visit each of the possible rooms $q=1, \dots,Q$. If their
score drops below a certain threshold, they discard their
probabilities and choose a new combination of $p_i^q$.

Let us first consider the first generalization. It is fairly
obvious that this prescription can do very little to 
improve coordination. Even in the extreme case where 
exactly $N/Q$ players choose the last winning option 
with probability 1 and $N(1-1/Q)$ roll the dice with
probability 1, the standard deviation is that of 
$N(1-1/Q)$ players choosing between $Q-1$ options. 
The best thing that can happen is that the number of 
guessers is effectively reduced by a factor of $1-1/Q$.

\begin{figure}
\centerline{
\epsfxsize= 0.7\textwidth
  \epsffile{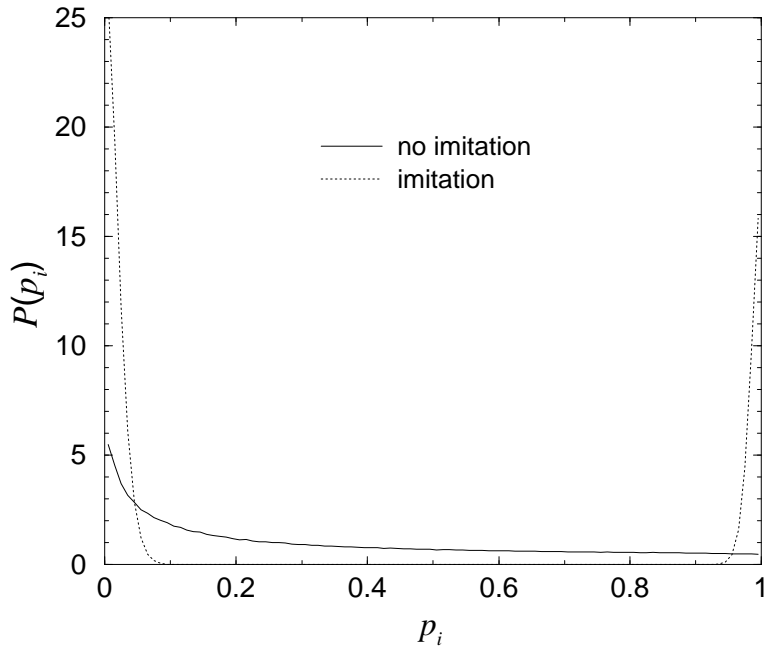}}
  \caption{Johnson's Evolutionary Minority Game with
    $Q=3$ options and
    one probability $p_i$ of choosing the last winning
    option: if new players pick their $p_i$ at random, 
    specialization is weak (solid line). If they imitate
    an existing player, the population segregates into 
    two crowds (dotted line). 
    }
  \label{EMG-sj}
\end{figure}

Simulations show that, compared to the case of $Q=2$,
the self-organization is less pronounced for $Q>2$. 
The probability density
distribution $P(p_i)$ decreases monotonically  towards
higher $p_i$; there is no double-peak structure 
(see Fig. \ref{EMG-sj}). 
There is, in the stationary state, a chance of $1/Q$ for
the most recent option to win again, i.e., probabilities
$p_i$ distribute themselves to arbitrage away any 
systematic advantage:
\be
\int_0^1 P(p_i)p_i\, dp_i = 1/Q.
\ee
Furthermore, coordination is slightly improved, compared
to completely random guessing.

What is the cause of the relatively bad coordination in 
this evolutionary game?
As pointed out in Ref. \cite{Horn:Diplom}, 
the stationary distribution $P(p_i)$ is determined by two 
processes: the removal of agents with a given $p_i$, i.e.,
the mean lifetime of an agent with $p_i$ in an 
environment with a given distribution $P(p_i)$, and the
addition of new agents, i.e., the rules according to which
the $p_i$s of removed players are replaced. In a dynamics
in which new $p$s are chosen at random, the stationary 
distribution $P(p_i)$ is proportional to the mean lifetime
$L(p_i)$ of a player with probability $p_i$.
In an environment with a wide distribution $P(p_i)$, 
having an extreme position apparently does not provide
enough advantage to bring about a strong polarization
of opinions.

This can be remedied by a small change in the dynamics:
a removed player is replaced by a (possibly slightly
modified) copy of one of the other players. This
ensures, in the spirit of evolutionary biology,
that ``fitter'' strategies spread more quickly than
less successful ones. In simulations, it proves essential
to add a small ``mutation'' (a random number) to the
value of $p$ that is being copied, to allow the population
to reach advantageous $p$-values even if the agents that held them
initially died out due to fluctuations. In
Fig. \ref{EMG-sj}, this mutation is a Gaussian random number
with mean 0 and $\sigma =0.005$ that is added to $p_i$ with 
reflecting boundary conditions.
 
For $Q=2$ (binary decision), this leads to a 
split of the population into two groups: one
with $p\approx 0$ and roughly $N/2$ members,
the other with $p\approx 1$ and again $N/2$
members. 
This even leads to $\sigma^2 = \mathcal{O}(1)$.
One can argue that this situation is analogous to 
Reents' and Metzler's stochastic strategy, with 
evaluation of one's strategy stretched out over 
a longer time: let us exploit the symmetry of the
problem again and say that $p_i$ is the probability of
choosing $+1$.  If the population is 
already separated into two peaks at $p\approx 1$ and 
$p\approx 0$, these correspond to different populations 
that say $+1$ or $-1$, respectively.
Players who have lost a certain amount pick a 
new $p_i$ -- either they rejoin their previous side,
or they switch sides. If few enough players die per round,
this can lead to $\sigma^2\approx 1$.

For $Q>2$, the population again segregates into two
factions with extreme opinions ($p_i \approx 0$ or $1$).
As explained above, even this split 
cannot do more than reduce the number of 
guessers, and hence $\sigma^2/N$, by a factor of $(Q-1)/Q$.
Numerical evidence shows that this is exactly what happens.

The second generalization, in which each player $i$ has an 
individual preference $p_i^q$ for each room $q$, offers 
(at least in theory) a 
configuration of perfect coordination: for each option,
there is a fraction of $1/Q$ players who choose it with
certainty. Analogous to the case of binary decisions,
where probabilities close to $0$ or $1$ are favorable,
the preferred probability combinations would be in the
edges of the simplex. A suitable order parameter
is therefore the average self-overlap of agents' 
strategy vectors:
\be
R = \left \langle \sum_q {p_i^q}^2 \right \rangle_i,
\ee
which takes values from $1/Q$ if each option is 
equally likely, to 1 if there is complete specialization.
The average of this quantity can be determined 
analytically for random vectors. The simplest 
way to generate a random vector on a simplex is to 
take $q$ random number $r_q$, equidistributed between  
$0$ and $1$, and normalize them $p_k = r_k/\sum_q^Q r_q$.
The average of $R$ is then
\be
\langle R_Q \rangle = \int_0^1 \prod_q dr_q \frac{\sum_k
  r_k^2} {\left (\sum_k r_k \right)^2}.
\ee
This yields $R_2= 2 + 12 \log(2)  - 9 \log(3)  \approx
0.43026$, $R_3 = 2- 88 \log(2) + 54 \log(3) \approx 0.32812$
etc.. These results, which can be well approximated
by $R_Q \approx 1.31/Q$, are the baseline against which a
coordination by evolution has to be measured.

Simulations of the evolution process show that if a 
deceased player is replaced by one with a randomly
updated strategy vector, coordination is and 
stays bad. The probability distribution of $R$ is
almost indistinguishable from that of random 
vectors, and $\langle R_3 \rangle$ merely increases
to $\approx 0.445$, compared to $0.43026$ for
random vectors. In a noisy environment, players who 
attempt to specialize in one of the options to 
not have a significant advantage.

Again, the picture changes if new players are 
allowed to copy (with modifications) the strategy
of a randomly chosen other player. The imitation mechanism
allows players to explore areas of parameter space that 
have a slim probability of being chosen at random,
like the edges of the simplex. In fact, a 
strong specialization takes place: in the
stationary state, each player has a decided 
preference for one of the rooms, leading to high 
average $R$. 

\begin{figure}
\begin{minipage}{0.49 \textwidth}
\epsfxsize= 0.95\textwidth
  \epsffile{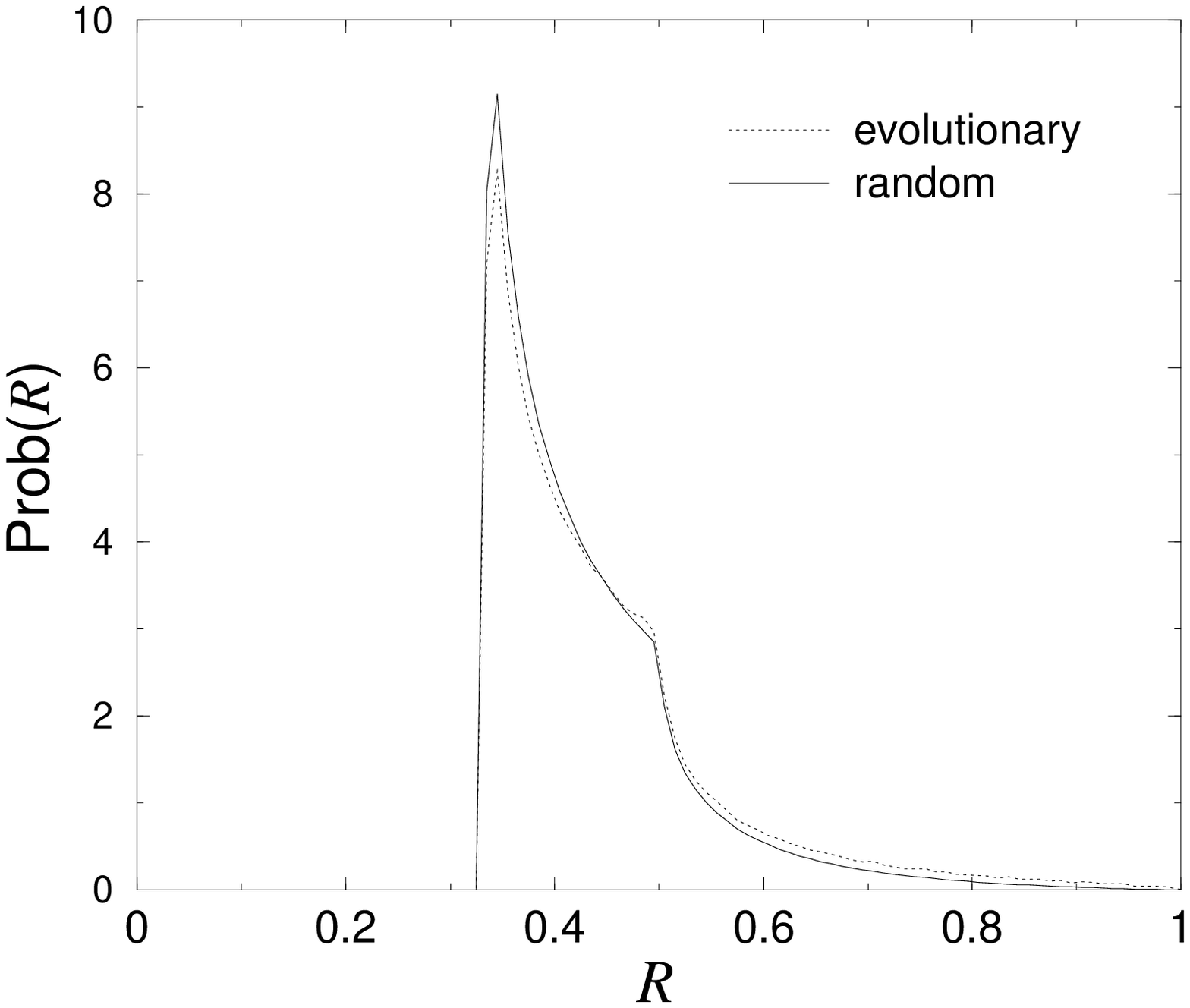}
\centerline{(a)}
\end{minipage}
\begin{minipage}{0.49 \textwidth}
\epsfxsize= 0.95\textwidth
  \epsffile{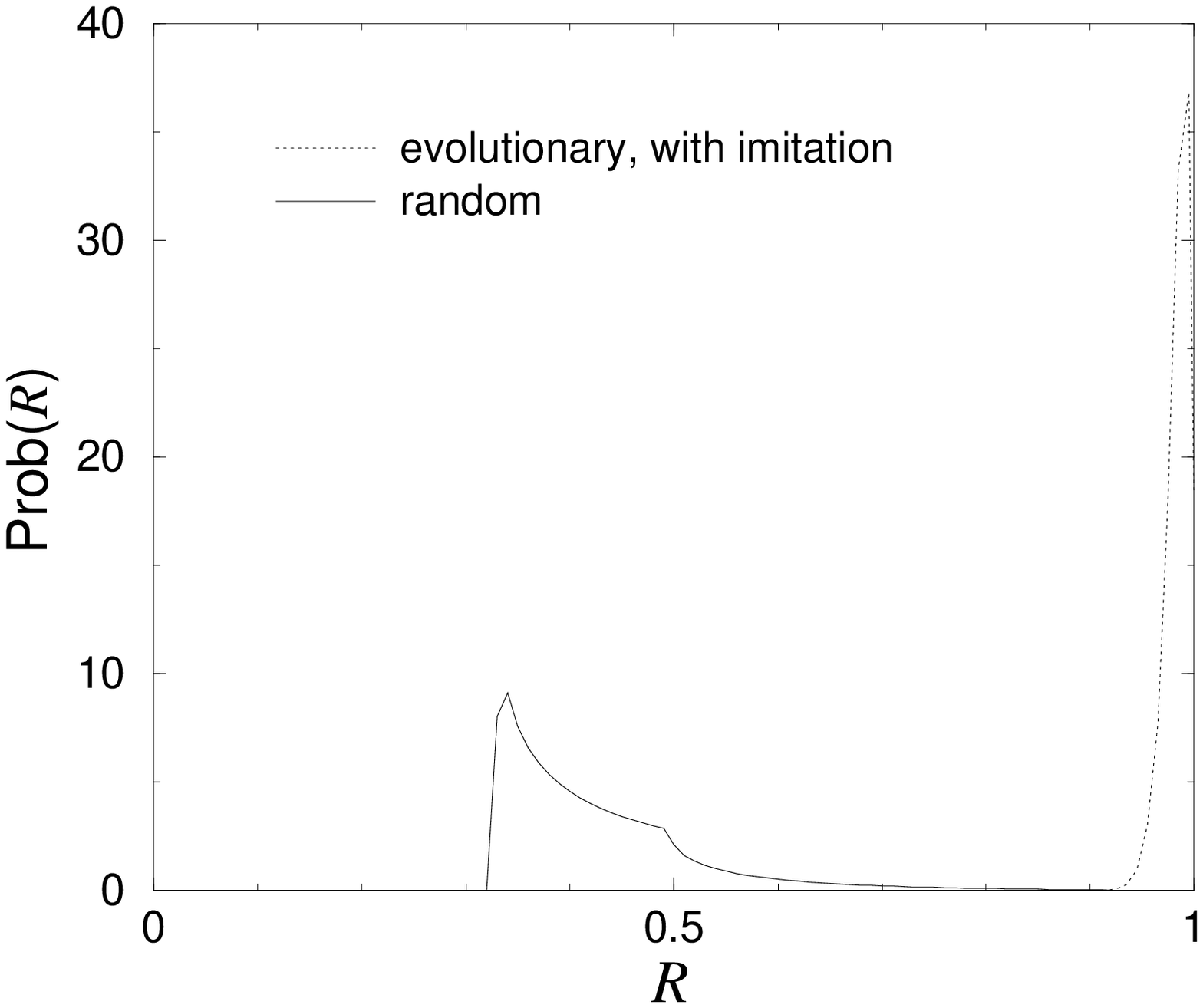}
\centerline{(b)}
\end{minipage}
  \caption{Johnson's Evolutionary Minority Game with $Q=3$
    options and individual preferences: if new players' 
    strategies are picked at random
    (a), the stationary distribution of self-overlaps $R$ 
    shows no specialization. This changes if new players
    imitate existing players with slight modifications (b).
    Simulations used $N=1001$.}
  \label{EMG-R_3}
\end{figure}

The exact distribution of $R$ depend on the details
and parameters of the mutation mechanism. In 
Fig. \ref{EMG-R_3}, the following mode was chosen: a
Gaussian random number with $\sigma=0.005$ is
added to each component, with reflective boundary 
conditions at 0 and 1. The resulting vector is 
normalized to $\sum p_i^q =1$. 

As I have shown, the mechanism of imitation improves
coordination among agents, at least for the instances
discussed in the last paragraphs. However, strictly
speaking, it violates the assumptions made about the
rules of the game: Although copying another player's 
strategy is not the same thing as making an
explicit contract stating ``Each round, I play this and
that, and you play such and such'', it amounts to the
same thing if a player's strategy has a strong preference
for one of the options.

Interestingly, the ``contracts'' made by the agents 
by imitation do not follow the lines of ``I play $+1$ each time
you play $-1$, and one of us wins for sure'' as one
would expect. To the contrary, players ``agree'' to
make the same statement most of the time! That this still
leads to good coordination is a consequence of the
slow adaption brought about by the evolutionary process
with sufficiently large thresholds and correspondingly
small death rates. If the threshold is very small,
such that each round a significant proportion of
the population dies, one might expect something 
like in Reents' and Metzler's stochastic strategy to 
happen.

However, even if the threshold is set to $-0.5$, i.e.,
new players are replaced immediately if they 
make a mistake, only a small fraction of the population
dies each round. This is because those players that survive
their first few round typically accumulate $\approx 100$
points. As outlined in \cite{Horn:Diplom}, the score
of each player is a random walk with a drift that is
proportional to the loss of the player. If a player is
very firm in his opinion and gives the same output 
every time, and there is no systematic preference for either
output, the mean loss of that player is 0, and the average
time for that player to die is actually infinite (this is 
a well-known result for random walks with one absorbing
edge). Since this limit is not realized completely, players
do die eventually, but at a very small rate. 
Simulations show that the time series generated by 
the evolutionary MG for $Q=2$ displays very weak
antipersistence ($p_{AP}\approx 0.52$) for a threshold
of $-0.5$, and significant persistence ($p_{AP}\approx 0.35$)
for large negative thresholds.

\subsection{Reents' and Metzler's stochastic strategy}
\label{SEC-RMSQ}
The stochastic strategy for the MG described in 
Sec. \ref{SEC-sto} can be adapted to accommodate 
several options as well. Again, two generalizations
suggest themselves:
\begin{list}{}{\leftmargin6mm}
\item[1)] Players are informed about which option 
won the last round. If they won, they repeat their
decision; otherwise they move to the last winning 
room with probability $p$.
\item[2)] Players only know if they lost or won. If they
lost, they choose, with probability $p$, to try one
of the other options at random.
\end{list}
In both cases, the system can still be considered a one-step 
Markov process. However, the state of the system must now be
characterized by $Q-1$ values $N_1, \dots, N_{Q-1}$,
which give number of players in each room (the
remaining value $N_Q$ can be calculated from normalization
constraints: $\sum_q N_q =N$), and the joint probability
distribution is a $Q-1$-dimensional tensor, or a function
living in $\mathbb{R}^{Q-1}$ if one wants to go to continuous  
variables. Transition probabilities look even worse, taking the
form of $2(Q-1)$-dimensional tensors or integral kernels.
Put briefly, this problem is only accessible to simulations
and very crude approximations. 

\subsubsection{Large $p$}
In the limit of $p = \mathcal{O}(1)$, $N\rightarrow \infty$,
the behavior is analogous that detailed in
Sec. \ref{SEC-STO-largep}: finite fractions of players
move from room to room, and the minority option changes 
at every time step. Suitable variables are $n_i= N_i/N$, 
the fractions of players who chose option $i$.
Occupation probabilities $\pi^s(n_i)$  
are a superposition of $Q$ Gaussian peaks whose widths
decrease with $N$. Self-consistent values for the centers 
of the peaks can be found analytically, as the following
example will show:

Let us take $Q=3$, and the players are not informed about
which option was successful, i.e., if they lose and decide
to switch, they choose one of the remaining options at
random. At any given step, there are three occupation
numbers, which we order $n_1 < n_2< n_3$.
Room 1 will now receive players from rooms 2 and 3, whereas
rooms 2 and 3 gain players from the respective other room 
and lose players to all other rooms. Neglecting
fluctuations, the rate equations for the 
occupations $n_i^+$ at the next time step look like this:
\bea
n_1^+ &=& n_1 + (p/2) (n_2 + n_3); \nonumber \\
n_2^+ &=& n_2  -p n_2 + (p/2) n_3; \nonumber \\
n_3^+ &=& n_3  -p n_3 + (p/2) n_2. \label{STO-q3rates}
\eea
If one can find a permutation of $n_i$ such that each 
$n_i$ is equal to $n_j^+$ with some $j\neq i$, one 
has a solution. In the present case, the solution 
for Eq. (\ref{STO-q3rates}) is 
$n_1^+ = n_3$, $n_2^+ = n_1$ 
and $n_3^+ = n_2$ for $0<p\leq 2/3$,
and  $n_1^+ = n_3$, $n_2^+ = n_2$ and $n_3^+ = n_1$
for $2/3 \leq p <1$. The corresponding equations are
\bea 
n_1 &=& \frac{4-6p+3p^2}{3(2-p)^2} \mbox{ for } p\leq2/3, 
   1- \frac{6}{10-3p} \mbox { for } p>2/3; \nonumber \\
n_2 &=& \frac{4(1-p)}{3(2-p)^2} \mbox{ for } p\leq2/3, 
   \frac{2}{10-3p}    \mbox { for } p>2/3;  \nonumber \\
n_3 &=& \frac{2}{3(2-p)} \mbox{ for } p\leq2/3, 
    \frac{4}{10-3p}  
    \mbox { for } p>2/3.
\label{STO-q3sol3}
\eea

\begin{figure}[h]
\epsfxsize= 0.7\textwidth
\centerline{
   \epsffile{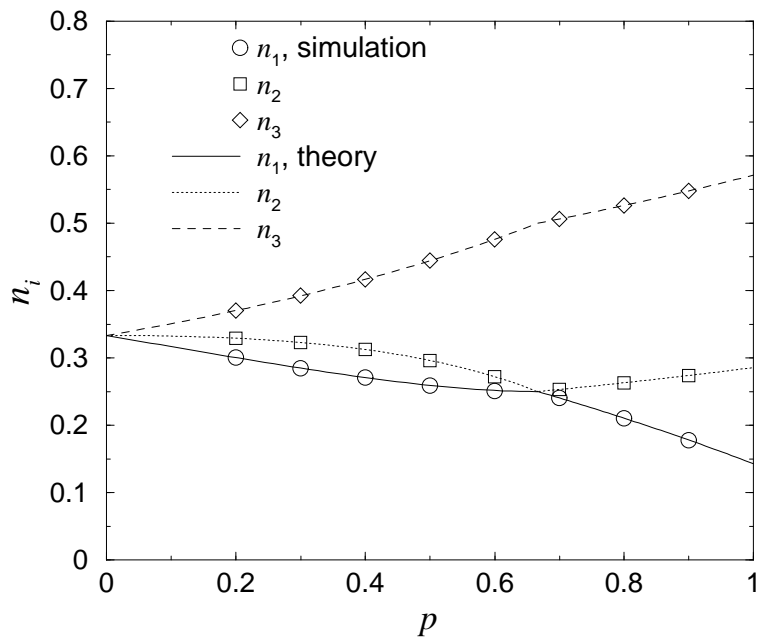}}
  \caption{Centers of the peaks of $\pi(n)$ in the
    stochastic MG with $Q=3$. Simulations with $N=4001$ 
    agree well with Eqs. (\ref{STO-q3sol3}).}
  \label{STO-pottsq3ns}
\end{figure}

Fig. \ref{STO-pottsq3ns} shows the results of
Eqs. (\ref{STO-q3sol3})
compared to simulations. Data collection from the
simulations is tedious, because the centers of each 
peak have to be fitted individually for each $p$;
however, even the few points give a good impression of
the accuracy. 

\subsubsection{Small $p$}
The case of small $p \propto 1/N$ is even less accessible to
analytics. I will therefore only present results from
simulations. The probability
distribution for the deviation from the optimal
configuration, $\pif^s(N_i-N/Q)$, depends less and 
less on the order of magnitude of $N$  
as $N \rightarrow \infty$. However, it does depend on 
$N \bmod Q$ -- this is less relevant for larger $p$, but
it becomes clear that very close to the optimal state, 
it makes a difference whether the optimal state has a 
net gain of $0$, $-1/Q$, or $-(Q-1)/Q$.

It also turns out that, if players are not informed about
which option won, the probability distribution does not 
converge to an optimal state, regardless how small $p$ 
is chosen: even if at most one agent per time step 
switches sides, there is a chance of $(Q-2)/(Q-1)$
of choosing the wrong room, making matters worse.

A different mechanism prevents perfect coordination if
agents are informed about the last minority option,
even if $N \bmod Q =0$, i.e., equidistribution among the
rooms is possible: according to the rules outlined above,
only one room is declared winner. Even if perfect
coordination is achieved at some point, agents from the
rooms that are tied with the winner but lost the coin toss
will move to the winning room, ruining equidistribution.

The symmetry between $N_i-N/Q <0$
and $N_i-N/Q >0$ that was still present for $Q=2$ is broken,
resulting in an increasingly skewed distribution for 
increasing $Q$. 

Although details are more complex, the broad features
of the stochastic MG remain the same: for large $p$, 
the ensemble of players can be treated as $Q$ effective
players who switch rooms, and strong herd behavior is
observed. For $p=\mathcal{O}(1/N)$, there is 
good, although not perfect, coordination.

\section{Summary and remarks on the Minority Game}
This chapter has provided an overview over different 
strategies for the Minority Game, with different 
approaches: the standard MG provides the players with
elaborate strategies, but little freedom to change them.
The quenched randomness inherent in the model is the
source of most of the interesting behavior.

The neural networks approach presented in
Sec. \ref{SEC-MGNN} also includes quenched
randomness in the differences between weight vectors.
However, this randomness is irrelevant for all practical
purposes, and a calculation that assumes a perfectly
symmetrical state gives good results. The idea
here is to give players a fairly sophisticated 
prediction algorithm and a learning algorithm that
can, in principle, completely alter the parameters
of this algorithm. The rate at which this modification
takes place determines whether the agents can
find an anti-correlated state or whether they
over-compensate and show herd behavior. Adaption
has to be slower with increasing population size,
and there is a smooth crossover from negative
to positive correlations between players.

Johnson's evolutionary MG (Sec. \ref{SEC-EvolMG}) 
models each player in a very simplistic way, 
with only one parameter, and an evolutionary
learning algorithm. Nevertheless, it is almost 
intractable analytically. The evolutionary
algorithm suggested in the original publications
induces significant self-organization. However,
if newly generated players are allowed to copy 
the strategy from existing players, almost total
coordination results (see Sec. \ref{PMG-evol}). 

Reents' and Metzler's stochastic strategy is 
also very simple. In the treatment given in 
Sec. \ref{SEC-sto}, the whole population is 
described by a single parameter! Still, my
feeling is that the suggested strategy is 
close to what people do in real-life situations
where little information is available. 
It would be interesting to check this in  
psychological experiments. 

As in the case of the neural networks, coordination
becomes better as learning becomes slower, and 
the rate of change has to be proportional to $1/N$
to achieve optimal coordination. Between this 
regime and the regime of herding behavior ($p=$const), 
there is again a smooth crossover. 

It can be shown in simulations and calculations
that the stochastic strategy (at least in the
small-$p$ regime) is robust to the presence of
noise (i.e., other players playing different
strategies), and evolutionarily stable.   

The section on the multi-choice Minority Game  
showed that one can find suitable generalizations
of all presented strategies to scenarios where players
have to choose between more than two different options.
With the exception of neural networks, an analytical
solution seems too involved to be a realistic option; 
however, simulations show that the general behavior of the strategies
is similar to that for $Q=2$, with respect to 
the presence or absence of phase transitions, the
scaling regimes for parameters, and so forth. The
evolutionary MG is an exception: the original update
rule suggested by Johnson et al. yields no
specialization; however, a strategy-copying mechanism 
again leads to self-organization.
 
Of the presented strategies, only the standard MG 
shows a phase transition. This has drawn significant
attraction \cite{Challet:Phase}; however, I attribute this
more to the general fondness of statistical physicists
for phase transitions than to good evidence that 
minority-like settings in real life have a phase 
transition. Admittedly, it is an established fact that
time series from financial markets display power-law
probability distributions typical for critical 
behavior \cite{Farmer:Physicists}. With sufficient tweaking,
it is possible to find a variation of the standard
MG (the so-called Grand-canonical MG 
\cite{Challet:Stylized,Jefferies:Market})
which replicates these properties, or ``stylized facts''.
It might be worthwhile to explore if this is 
possible with any of the other strategies as well.  

This chapter has highlighted some connections 
between the Minority Game and the concept of 
anti-predictability. All presented strategies 
show (very pronounced in some cases, and weakly
in others) a transition to 
antipredictability in the time series generated 
by the ensemble of players.

If players adapt slowly, they can find a well-coordinated
state where the next minority decision is either
very random (neural networks) or so narrow that 
even though it can be predicted, it cannot be
exploited (stochastic strategy, evolutionary MG with
imitation). If quenched randomness is present, however,
even slow adaption is sometimes not sufficient to 
remove preferences for certain outputs (standard MG).

If players adapt quickly (large learning rates in
neural networks, large $p$ in the stochastic
strategy, small initial differences in the 
crowded regime of the standard MG), they overcompensate:
large numbers of players change their opinions 
at the same time, defeating their own prediction.
The Minority Game is thus a well-motivated framework in
which antipredictability emerges naturally in some
limits, and it allows to study the circumstances that
lead to failure of prediction algorithms.

  \cleardoublepage
\chapter{Neural Networks in Two-Player Games}
\label{CHAP-game} 
The Minority Game that was introduced in Chapter \ref{CHAP-MG}
and related to the concept of antipredictability 
is only one example out of a large number of
game-theoretical scenarios. Neural networks 
have been applied to some of these scenarios;
however, there seems to be little work on 
very basic, concrete questions concerning simple 
neral networks playing simple two-player games.

This chapter will
present a short overview over the different 
areas of game theory, introduce some basic concepts,
and then go into details on how and to 
what extent a simple neural network can be used
to learn strategies in two-player games.

\section{An overview over game theory}
The field of game theory can be divided into several
distinct areas. Although the mathematical treatment 
of pure games of chance triggered important research on
probability theory \cite{Bernoulli:Ars,Girenzer:Empire}, game theory in 
the modern sense often does not include randomness as
a necessary ingredient. The usual scenario consists of
two or more players who have a choice between a set of 
{\em options} or {\em strategies}.
\footnote{The usual convention in game theory is to 
call each possible choice a strategy. However, to avoid
confusion between strategies, pure strategies, and mixed 
strategies, I will continue to call them options, and
use the term strategy for the set of probabilities of choosing each 
option.} 
 The payoff that
each player receives depends on the choices he makes and
the choices of the other player(s), which he cannot influence.
The aim of each player is to maximize his personal 
payoff \footnote{Newcomers to game theory often feel that 
  this approach oversimplifies the motivations and emotions
  of players. Game theorists usually answer that any 
  motivation can, in principle, be quantified and accounted
  for by modifying the payoffs. Whether this is indeed the
  case can be argued; however, to allow for an analytical
  approach, a quantification of preferences seems necessary.}. 
The difference between the 
fields lies in the rules of the game, and 
in the way that the players change or don't change during 
repetitions of the game:
   
\subsection{Combinatorial game theory}
In {\em combinatorial game theory}, players 
take alternating turns and usually have full information 
about the state of the game and all possible future 
combinations of moves. A simple example of this type
of game is Nim: the game starts with several heaps of
matches, and each player can take an arbitrary number
$\geq 1$ of matches from one of the heaps. The last
player to take a match wins. Many combinatorial 
games can be reduced to Nim and studied using a 
generalization of rational numbers called Nimbers 
(see ``Winning Ways'' by Berlekamp et al.  
\cite{Berlekamp:Winning} for details on Nim,
an introduction to Nimbers, and 
lots of information on other combinatorial games). 

More complex cases of combinatorial games are 
Four-in-a-Row, Dots-and-Boxes, Go and Chess. Although 
it is in principle possible to tell in 
advance how the game will end if all players play 
optimally and which moves are optimal in a given 
situation, one can argue that all ``interesting''
real-life games require too much computational 
effort for a human player to go through the 
complete analysis (in fact, 
some games have been proven to be NP-complete,
i.e., the computational effort increases exponentially with
the size of the board \cite{Demaine:Playing}). A case in 
point is TicTacToe, which is simple enough to grasp completely and
considered boring by most people beyond the age of 10.
On the other hand, a complete analysis of chess is 
beyond the capacities of even the most advanced 
computers.
 
Thus, players of such games rely on heuristics (or
experience and ``intuition'', which amounts to the same 
thing) to evaluate the quality of moves in games like Chess
\cite{King:Wie}. If moves seem equivalent to
players, they pick among them more or less randomly, 
which allows for stochastic treatments of games that are,
in principle, deterministic (see e.g. Ref. \cite{Metzler:Jamming}).

\subsection{Economic game theory}
In {\em economic game theory} \cite{Jianhua:Games}, 
uncertainty is an essential
ingredient, since players make their choices simultaneously,
and none of the players know what choices the other(s) will 
make. The simplest scenario considered in economic game
theory is that of two-player zero-sum games
\cite{Neumann:Theory}. 
Here, each of the
two players has a finite number of options. 
Each combination
of options is assigned a number (an entry in the payoff
matrix $\mathcal{C}$) which the first player has to pay to the second if
that combination is chosen, such that the gain of one
player is always the loss of the other. A popular example 
is Rock-Paper-Scissors. Zero-sum games always have a unique
combination of strategies for the two players where 
the strategy of each player is optimal if the respective
other player does not deviate from his strategy.
Such a combination is called a {\em Nash equilibrium}.
More details on zero-sum games, 
as well as a formal explanation of Nash equilibria, will follow
in Sec. \ref{SEC-zero-sum}.

In contrast, non-zero-sum games can have different payoff tables
for each player. This makes the situation more complicated,
allowing for cases in which there are several equilibria
or in which the solution that maximizes ``global good''
(the sum of average gains) is not stable because
options exist that increase one player's payoff, but
decrease the sum of payoffs. The most infamous scenario
is the Prisoner's Dilemma \cite{Poundstone:Prisoners}.

Some non-zero-sum games involve a large number of 
players and a fairly simple payoff scheme. For example,
the Minority Game described in Chap. \ref{CHAP-MG} 
is a negative-sum game with an arbitrarily 
large number of participants.

Games are also classified as complete information games
if all players know all entries of all payoff matrices,
and as incomplete information games if each participant only
knows part of the payoffs (usually at least his own).

\subsection{Evolutionary game theory} 
This branch may be considered a spin-off
of economic game theory which is concerned with the
mechanisms by which equilibria can be reached
\cite{Maynard:Evolution,Axelrod:Evolution}. Typical 
models include populations of players with individual
strategies that repeatedly play against each other.
Successful players generate offspring with the same
or at least similar strategies, whereas less successful
strategies can become extinct. 

Evolutionary game theory has become a testing ground for
simple models of interpersonal interactions, allowing 
to study under which circumstances cooperation among
players can exist, when cheating is to be expected, and
what ways of behavior favor which response. As the name
suggests, it has become extremely important in evolutionary
biology, where it is used to model animal behavior and
has shed light on many aspects of mating behavior, 
cooperation within herds, and so forth.

Interestingly, many popular games do not easily fit into the
categories of either theory. For example, card games 
like Poker, Romm\'{e}e or Magic:The Gathering\textregistered
\ and board games like Monopoly\textregistered  \ or Siedler von
Catan\textregistered \  mix combinatorial elements with 
a strong stochastic component (rolling the dice, drawing
cards) and/or 
incomplete information (what cards are the other players 
holding?). In principle, they can be formulated as matrix
games if the realization of randomness is included as the
choice of a ``dummy player'', and picking an option actually
means deciding, in advance, what move to make for each
conceivable situation during the course of the game. The 
number of different options is uasually exponential in the
number of cards and/or die rolls, 
and the formulation as a matrix game 
serves purposes of theoretical analysis rather than 
practical guidelines how to play.

\section{Zero-sum games}
\label{SEC-zero-sum}
Zero-sum games, in which the gain of one player always
corresponds an equivalent loss of the other player, 
can be treated with a fairly elegant 
mathematical apparatus introduced by John von Neumann
\cite{Neumann:Theory}. 
Let us call one player A and the other B. In each round,
A chooses one among $K$ options. In the following analysis,
the option that is actually chosen will be denoted by $i$,
$1 \leq i \leq K$. At the same time, B picks an option $j$ from
his repertoire of $L$ options. They then consult the 
entry $c_{ij}$ of the payoff matrix $\Cm$, and A receives
$c_{ij}$ from B. Accordingly, A will try to play strategies
where large $c_{ij}$ are picked, and B will try to keep 
$c_{ij}$ small (negative, if possible). 

In some realizations of $C$, it may be optimal for
A to keep choosing one option which is better than 
the alternatives, no matter what B plays, like in this
example:
\be
\Cm= \left( \begin{array}{rrr}
    10 & 2 & 5 \\ 
    3  & -2 & 1 \\
    -5 &  1 & 4  \end{array}\right).
\ee
Here, all entries in row 1 are higher than the
corresponding entries in rows 2 and 3 -- row 1 is said
to {\em strictly dominate} rows 2 and 3, and picking
it is a no-brainer, so to speak. Player A is said to 
play a {\em pure strategy} if he chooses the same option 
all the time.
If B knows that A is going to repeatedly pick row 1, all that is left 
to do is minimize the losses and pick column 2 
every time as well. 

For random payoff matrices, the probability that one 
pure strategy dominates all other 
strategies decreases exponentially with increasing $K$.  
Even for small payoff matrices, the interesting cases
are the ones without dominating strategies, such as
Penny-Matching:
\be
\Cm= \left( \begin{array}{rr}
    1 & -1  \\ 
    -1  & 1  \\
     \end{array}\right).
\ee
In this game, player A tries to play the same option 
as player B, who in turn tries to play the opposite
option of player A. The only way for a player to
avoid being exploited is to choose at random 
between the two options -- in this case, with 50\%
probability for each.

Generally, the behavior of players A and B at a given time 
can be defined by the vectors $\va$ and $\vb$ of 
probabilities $a_k$ and $b_l$ for playing options $l$ and
$m$, respectively. These vectors follow the normalization
constraint $\sum_k a_k=1$ and $\sum_l b_l=1$, i.e., they
lie on the $K$- and $L$-dimensional simplex. If a player 
is playing a pure strategy, his strategy vector only has
one nonvanishing component.

The expected payoff $\l_k^A$ for A of one of his strategies $k$ 
depends only on B's strategy:
\be
\l_k^A = \sum_l b_l c_{kl}.
\ee
Player A will prefer strategies with higher $\l_k^A$. In the language
of game theory, a {\em strictly rational} agent will {\em
  always} prefer the option with the highest $\l_k$, thus
playing a pure strategy every time, unless the the 
expected payoffs of several strategies are exactly tied.
This almost never happens in models where the player 
estimates the opponent's strategy from experience, rather 
than calculating the opponent's equilibrium strategy and
assuming he will play it \cite{Foster:Impossibility}.
A more tolerant approach, and one that emerges automatically
in the neural network learning model discussed later,
is to give higher probabilities to options with higher
payoffs, and gradually eliminate those that promise
significantly lower payoffs completely.

\subsection{Nash equilibria}
\label{SEC-Nash}
A {\em Nash equilibrium} is a combination of strategies
$\vas$ and $\vbs$ such that no player can improve his 
average payoff by unilaterally deviating from his 
equilibrium strategy. 
The average payoff for player A is 
\be
\l^A = \sum_{i,j} a_i c_{ij} b_j.
\ee
He will try to maximize the payoff he receives if 
B makes life as difficult as possible for him, and will
choose
\be
\max_{\va} \min_{\vb} \sum_{i,j} a_i c_{ij} b_j.
\ee
On the other hand, player B, who is trying to 
minimize the amount he pays to A, will choose
\be
\min_{\vb} \max_{\va} \sum_{i,j} a_i c_{ij} b_j.
\ee
The Nash equilibrium fulfills both conditions
at the same time.
As von Neumann proved \cite{Neumann:Theory},
in zero-sum games, there is exactly one Nash equilibrium. 
If it is a mixed strategy,
all options that are played with non-zero probability
(the {\em support} of the strategy) yield the exact
same average payoff, also called the value of the game
$\nu$.
The necessary and sufficient conditions
for $\vas$ and $\vbs$  can be written in a very concise way:
\bea
\l_i^A &=& \sum_j c_{ij} \bs_j \leq \nu  \mbox{~~~for all
  }i; \label{NE-cond1} \\
\l_j^B &=& \sum_i \as_i c_{ij} \geq \nu \mbox{~~~for all }j. 
 \label{NE-cond2}
\eea
The minimax theorem guarantees that the smallest value
for which a solution to Eq. (\ref{NE-cond1}) can 
be found is the largest value that yields a solution to
Eq. (\ref{NE-cond2}).
For a given payoff matrix, the equilibrium strategies
can be calculated using linear optimization. For details,
consult Ref. \cite{Kinzel:Physics} and  \cite{Press:Numrec}. 

The interpretation of the Nash equilibrium is
not very intuitive and worth spending a few words on.
If player A plays the ``optimal'' strategy $\vas$, 
he is guaranteed the average payoff $\nu$. 
There is no way for B to exploit A. However, there
are usually several options for B that are precisely
equivalent -- each one makes B pay $\nu$ to A on average.
In that sense, it is not a big challenge to play against
a player who sticks to $\vas$ and does not respond to his
opponent's actions -- B has to find one of the
strategies in the support of $\vbs$ and play it. 
\footnote{This can be readily verified using the
  corresponding program in Ref. \cite{Kinzel:Physics}.}
Doing this opens up player B to exploitation from A,
which in turn would allow B to exploit A. 
On the other hand, it means that a learning algorithm
of player B cannot be expected to converge to the 
minimal-optimal solution $\vbs$ if A always plays $\vas$.

In a game with a finite number of options,
a small deviation from $\vas$ or $\vbs$ only
causes a small difference in the $\l^A_i$, which can 
only be empirically detected through many repetitions
of the game. A pair of learning algorithms that rely 
only on observing the opponent's actions will therefore 
converge to equilibrium very slowly (if they converge at all).

\section{Multi-Choice Perceptrons}
\label{SEC-MC}
Economic game theory often assumes  perfectly 
rational players with unlimited computational resources
and a perfect grasp of the situation: both players
study the game thoroughly before they start playing,
find the Nash equilibrium and play the corresponding
strategy, thus fulfilling the expectations of the
opponent, who likewise plays his optimal strategy.
Since this rarely
applies to real-world situations, the question of
{\em learning} in games has received much attention
(see Ref.\cite{Fudenberg:Learning} and references herein):
how can a player who is not familiar with the behavior
of the opponent find a strategy, based on his knowledge
of the payoff matrix and the observed behavior of 
the opponent?

Even in learning scenarios, there is a concept of
a {\em rational} player: it means a player who,
based on some prediction algorithm, calculates the
expected payoff for each of his options and {\em always}
chooses the one that promises the highest payoff.
However, it was shown in Ref. \cite{Foster:Impossibility}
that under rather general circumstances, a rational 
player has no chance of learning to play the correct
equilibrium strategy.

It is therefore interesting to search for strategies that
are not rational, but that will converge to the Nash 
equilibrium, at least under some circumstances.
One obvious candidate are neural networks, whose most
interesting property is the ability to learn from examples. 
If each player has only two choices,
a simple perceptron could be used to make the decision.
This has been studied in some detail in
Ref. \cite{Metzler:Interact} for some simple 
games like ``Matching Pennies''. That case was also 
examined in Ref. \cite{Samengo:Competing}. 
However, even in the case of $K=2$, a more complex
network may be appropriate: a simple perceptron with
random unbiased patterns will pick each option 
with $50\%$ probability -- this is generally not 
the optimal mixed strategy. 
If a larger number of options is available, a different
architecture is needed in any case. The obvious choice
is the Potts perceptron, also known as Multi-Class
perceptron \cite{Watkin:Multi-Class}.

A simplified version of the Potts perceptron
was introduced in Section \ref{SEC-PMG_NN}; here, we 
need the full architecture. This is still the simplest
neural network that can handle inputs and outputs of
the required form -- quite possible, more elaborate networks
could give better results; however, their behavior becomes
mathematically less tractable.

To be able to choose one of $K$ output options, the
network of player A consists of $K$ hidden units
that generate $K$ hidden fields $h_i$. 

The output $\sigma$ is usually taken to be the option with the highest
hidden field:
\be
\sigma=\{ i | h_i = \max_l (h_l) \} \label{WTAP-rule1}.
\ee
This rule is the reason why this architecture is
also called ``Winner-takes-all perceptron'' (WTAP,
for the sake of brevity).
For reasons explained below, it can be helpful to
take as output the option with the highest absolute
value of the hidden field:
\be
\sigma=\{ i | |h_i| = \max_l |h_l| \} \label{WTAP-rule2}.
\ee

For real-valued
input vectors $\x$, each hidden unit is a linear 
perceptron: 
\be
h_l =  \sp{\vw_l}{\x}. 
\ee
If the input vectors consist of integers between 1 and $Q$
(for example, a time series of the opponent's decision),
a more elaborate scheme is needed: each hidden unit has a 
set of $Q$ weight vectors $\vw_{l}^q$ with entries 
${w_{l}^q}_n$. The hidden field for output $l$ is
\be
h_l = \sum_q^Q \sum_n^N {w_l^q}_n (Q\delta_{x_n,q} -1).
\ee
This can be written as
\be
h_l = \sum_q^Q h^q_l \mbox{~with~~} h^q_l = \sp{\vw_{l}^q}{\vx^q},
\ee
where the $\vx^q$ are vectors with components 
$x^q_n = Q\delta_{x_n,q}-1$.
Thus each of the possible input values generates a
separate contribution to the total hidden field.

In the following sections, I will assume real-valued
inputs, which are easier to treat and to imagine. 
As usual, the variance of one component of the 
input is taken to be $\langle x_n^2\rangle =1$.
Later on, the learning rules will be generalized to 
multi-value inputs.

What can and should we expect the neural network to do?
If the inputs are completely random, the network 
should at least be able to use these patterns as
seeds to generate a probability distribution of
outputs and to adapt this distribution such that it
gradually improves the perceptron's average payoff.
If the inputs indeed contain some information on 
the opponent's action, the network might be able to 
capitalize on this. However, if the opponent is
also a network of similar capabilities, there is
no a priori reason why one should be able to outplay
the other.

It turns out that learning has to follow different
principles depending on whether the input pattern 
has a bias (a preferred direction) or whether it 
is an isotropic random vector. Note that it is 
not important whether this bias carries any information
about the opponent or not. If the aim were to 
create a good, applicable learning algorithm,
one could artificially add a bias vector. However,
my first aim is to see what happens for a rather na\"{\i}ve
approach to learning in different scenarios.

\section{Learning from unbiased patterns}
\subsection{Generating a probability distribution}
As mentioned in Section \ref{SEC-MC}, one of the
feats that a neural network in a two-player game 
should be capable of is to generate any probability
distribution of outputs when fed with random patterns.
The first step is therefore to check whether this
is feasible. 

If the weight vectors of the network's hidden units are
mutually perpendicular ($\sp{\vw_i}{\vw_j} =w_i^2
\delta_{ij}$), the hidden fields are uncorrelated
Gaussian random numbers with  variances  $\langle
h_i^2\rangle= w_i^2$. \footnote{The assumption of
  perpendicular vectors simplifies things immensely. Of
  course, it must be justified later, when learning
  algorithms are considered.}
The probability $a_i$ that option $i$ is chosen is then
\bea
a_i &=& \int_{-\infty}^{\infty} \frac{d h_i}{\sqrt{2 \pi w_i^2}}
  \exp\left(-\frac{1}{2}\frac{h_i^2}{w_i^2} \right )
  \prod_{j\neq i}^{K} \int_{-\infty}^{h_i} 
   \frac{d h_j}{\sqrt{2 \pi w_j^2}} 
   \exp \left(-\frac{1}{2} \frac{h_j^2}{w_j^2}\right ) \nonumber \\
&=& \int_{-\infty}^{\infty} \frac{d \hh_i}{\sqrt{2 \pi}}
    \exp \left(-\frac{\hh_i^2}{2} \right ) 
      \prod_{j\neq i}^{K} \Phi\left(\frac{\hh_i w_i}{w_j}\right).
      \label{WTA-biasex}
\eea
It is clear that under these circumstances, no probability 
larger than $1/2$ can be
achieved: even if all other weights are set to 0
(and correspondingly their hidden fields are 0),
there is only a 50\% chance that the hidden field of the
nonzero vector is larger than 0.

It is also clear that no probability can be smaller 
than $2^{-K+1}$: if one weight is set to 0 while all
others have a finite value, there is still a chance
of $2^{-K+1}$ that all others are negative, and thus
the field with a value of 0 wins. The function 
(\ref{WTA-biasex}) interpolates between these two 
extremes. 

A small variation of the decision rule removes these
limitations: if the output is determined by the 
largest absolute field, as suggested in
Eq. (\ref{WTAP-rule2}),
$a_i $ is given by
\bea
a_i & =& \int_0^{\infty} \frac{2 dh_i}{\sqrt{2\pi w_i^2}} 
  \exp \left( -\frac{1}{2}\frac{h_i^2}{w_i^2} \right)
  \prod_{j\neq i} \int_0^{h_i} \frac{2 dh_j}{\sqrt{2 \pi w_j^2}}
    \exp \left ( -\frac{1}{2} \frac{h_j^2}{w_j^2}\right) \nonumber \\
 &=& \int_0^{\infty} \frac{2 dh_i}{\sqrt{2\pi}} 
  \exp ( - h_i^2/2) \prod_{j\neq i} 
    \hat{H} \left(\frac{w_i h_i}{w_j} \right), \mbox{  where } \nonumber \\
   \hat{H}(x) &=& \int_0^x  \frac{2 dh}{\sqrt{2\pi}} \exp ( - h^2/2).
\label{WTA-biasabs}
\eea
This rule allows for probabilities $a_i$ between 0 and 1
 -- however, the interpretation
of Eq. (\ref{WTAP-rule2}) is admittedly dubious if the
pattern has any meaning beyond being a random number seed:
The invariance of Eq. (\ref{WTAP-rule2}) to the
sign of $h$ essentially states, ``If the
situation is such-and-such, do this, and if the situation
is exactly the opposite, do the same.''

In the simplest case of $K=2$, Equation (\ref{WTA-biasabs}) 
can be evaluated analytically, and we obtain
\be
a_1 = (2/\pi)\arctan (w_1/w_2) \mbox{ and } a_2 = 1-a_1.
\label{WTA-K2}
\ee

\subsection{Learning rule}
If we assume that the patterns are random and the
strategy is only determined by the set of weight norms
$\{ w_i\}$, we need a learning rule that adapts 
the norms accordingly. The following rule   
 employs the mechanisms of the Confused
Bit Generator: if a perceptron learns the opposite 
of its own output, its norm converges to a finite 
value. If it reinforces its output, the norm grows 
linearly with $\alpha$. A plausible first attempt at a learning 
rule is therefore
\be
\vw_l^{t+1} = \vw_l^{t} + \frac{\eta}{M} \vx \s{\sp{\vx}{\vw_l}} c_{lj}.
\label{MatG-Imp1}
\ee
The row player updates each of its weights using the
column $j$ of the payoff matrix that his opponent
chose. Weights with a high payoff get positive 
feedback, while weights with a negative payoff
are suppressed. Averaging over many outputs $j$
and patterns $\vx$,
Eq. (\ref{MatG-Imp1}) becomes
\be
\langle \vw_{l}^{t+1} -\vw_{l}^t \rangle_{j,\vx} = 
 \frac{\eta}{M} \langle \vx
 \s{\sp{\vx}{\vw_{l}}}\langle_{\vx}
 \sum_j b_j c_{lj}.  
\ee
That means that the feedback term on the right hand side
is proportional to the expected payoff $\l^A_l = \sum_j b_j c_{lj}$
for that option. A positive value of $\l^A_l$ leads to 
linear growth, a negative value leads to a finite value,
and a vanishing value $\l^A_l=0$ gives $w_l \propto \sqrt{\alpha}$.

This may be problematic since the value
of the game $\nu$ may be different from 0, leading
to a suppression of all options for $\nu<0$ 
or enhancement of non-optimal options for $\nu>0$.
An even better learning rule incorporates this and
estimates the value of the game by subtracting
the current payoff $c_{ij}$ from the potential payoff
of the updated option, $c_{lj}$:
\be
\vw_{l}^{t+1} = \vw_{l}^{t} + \frac{\eta}{M} \vx \s{\sp{\vx}{\vw_{l}}} 
 (c_{lj}-c_{ij}).
\label{MatG-Imp2}
\ee
This rule is somewhat more difficult to treat,
since the update for each option explicitly depends 
on what option was chosen by the player, i.e., it 
depends on all  other $w_i$ as well. For large $K$,
one can ignore this effect, but for small $K$ it turns
out to be important. Thus I will first look closer
at the textbook case of $K=L=2$ for specific
matrices (called generalized
penny-matching in the literature) and then go to 
the limit of large $K$ for random matrix entries.

\subsection{The case of $K=2$}
\begin{figure}
  \epsfxsize= 0.7\columnwidth
  \centerline{\epsffile{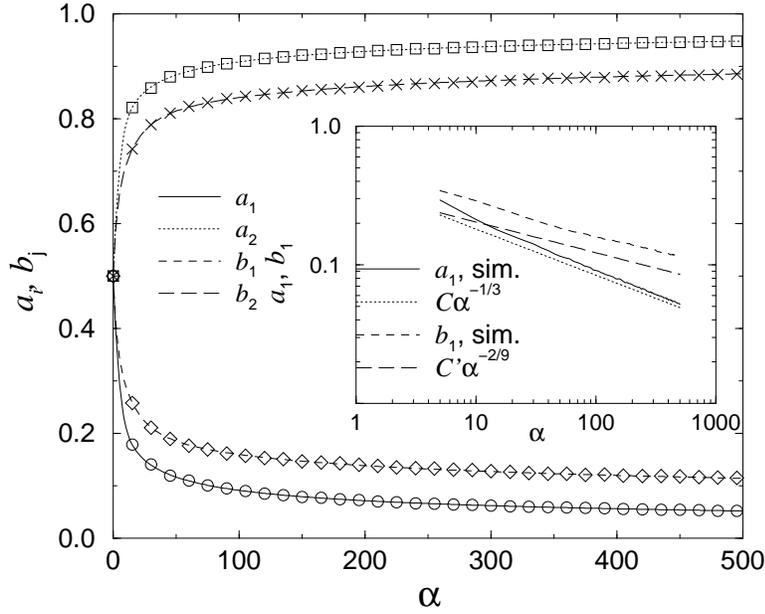}}
  \caption{Development of strategies in a zero-sum game with
payoffs according to Eq. (\ref{ZSG-K2payoff}), under learning
rule Eq. (\ref{MatG-Imp2}). Symbols in the large plot denote
simulations with $M=2000$.} 
  \label{ZSG-K2}
\end{figure}

As mentioned above, for small $K$, the output of the
WTAP has to follow the largest absolute value
of the inner field $|h_i|=|\sp{\vw_i}{\vx}|$ --
otherwise both options would be chosen with 
$50\%$ probability each. Using absolute
fields, the strategy for two options is given by 
Eq. (\ref{WTA-K2}). In the spirit of on-line 
learning, we can take the square of Eq. (\ref{MatG-Imp2})
and use the limit $M\rightarrow \infty$ to transform
the update equations it into a set of differential equations
for the $w_i$. For example, for option 1 of player A, 
one obtains (with $\sp{\vx}{\vx}=M$):
\be
{w_1^{t+1}}^2 = {w_1^t}^2 + \frac{\eta}{M} |h_1| (c_{1j}-c_{ij})
 + \frac{\eta^2}{M} (c_{1j}-c_{ij})^2.
\ee
The second and third term on the 
right hand side vanishes if $|h_1| > |h_2|$, which has to 
be taken into account when averaging:
\bea
\langle |h_1| (c_{1j}-c_{ij})\rangle_{h_1,h_2} &=& 
\left (\int_0^\infty Dh_1 2 w_1 h_1 \int_0^{h_1 w_1/w_2}2 Dh_2 \right ) 
(\l^A_1-\l^A_2) \nonumber \\
&=& \sqrt{\frac{2}{\pi}} w_1 
  \left(1-\frac{w_1}{\sqrt{w_1^2 +w_2^2}} \right)(\l^A_1-\l^A_2); \\
\langle (c_{1j}-c_{ij})^2\rangle &=& x_2 (b_1 (c_{11}-c_{21})^2 + 
b_2 (c_{12}-c_{22})^2).
\eea
In this special case, $\l^A_i$ is an abbreviation for 
$b_1 c_{i1} + b_2 c_{i2}$.
Using the chain rule, a differential equation for $w_1$ can be derived:
 \bea
\dd{w_1}{\a} &=& \sqrt{\frac{2}{\pi}} \eta 
 \left(1-\frac{w_1}{\sqrt{w_1^2 +w_2^2}} \right)(\l^A_1-\l^A_2)
 + \nonumber \\
& &~~~~ \eta^2 a_2 \left[b_1 (c_{11}-c_{21})^2 + 
b_2 (c_{12}-c_{22})^2\right]/(2w_1). \label{K2-eq1}
\eea
A similar equation can be derived for $w_2$. 
If player B is using a neural network with the
same learning rule (whose vectors I will denote
as $\vv_j$ with norms $v_j$ to avoid confusion), 
the differential equations for its 
weight vectors norms $v_1$ and $v_2$ can be written 
down as well:
\bea
\dd{w_2}{\a} &=& \sqrt{\frac{2}{\pi}} \eta 
 \left(1-\frac{w_2}{\sqrt{w_1^2 +w_2^2}} \right)(\l^A_2-\l^A_1)
 + \nonumber \\
& &~~~~  \eta^2 a_1 \left [ b_1 (c_{11}-c_{21})^2 + 
b_2 (c_{12}-c_{22})^2\right ]/(2w_2).  \\
\dd{v_1}{\a} &=& \sqrt{\frac{2}{\pi}} \eta 
 \left(1-\frac{v_1}{\sqrt{v_1^2 +v_2^2}}
 \right)(\l^B_2-\l^B_1) +
\nonumber \\
& &~~~~ \eta^2 x_2 \left[ b_1 (c_{11}-c_{12})^2 + 
a_2 (c_{21}-c_{22})^2 \right]/(2 _1).  \\
\dd{v_2}{\a} &=& \sqrt{\frac{2}{\pi}} \eta 
 \left(1-\frac{v_2}{\sqrt{v_1^2 +v_2^2}}
 \right)(\l^B_1-\l^B_2) +
\nonumber \\
& & ~~~~ \eta^2 b_1 \left[ a_1 (c_{11}-c_{12})^2 + 
a_2 (c_{21}-c_{22})^2\right]/(2v_2).  \label{K2-eq4}
\eea
The symbol $\l_j^B$ stands for the expected
payoffs of B's options $a_1 c_{1j}+ a_2 c_{2j}$, and
player B favors strategies $j$ with minimal $\l^B_j$.
This set of ODEs can now be solved numerically for any 
$2\times 2$ payoff matrix. 
In simple cases, the asymptotics can be calculated analytically.
For example, take the payoff matrix
\bea
\mathcal{C} &=& \left (  \begin{array}{cc}
    1 & 0 \\
    1 & 1 \end{array} 
\right ).\label{ZSG-K2payoff}
\eea
In this case,
$\l^A_1 = b_1$, $\l^A_2 = 1$, $\l_1^B = 1$, and $\l_2^B =a_2$.
One can assume that for large $\a$, $w_1\ll w_2$ and $v_1 \ll v_2$,
and correspondingly make the expansions $a_1 \approx (2/\pi) w_1/w_2$,
$b_1 \approx (2/\pi) v_1/v_2$, $w_2 /\sqrt{w_1^2 + w_2^2} \approx
1 - (w_1/w_2)^2/2$, and $v_2/\sqrt{v_1^2 + v_2^2} \approx 1- (v_1/v_2)^2$,
a short calculation leads to the asymptotic value for $w_1$ and $v_2$:  
$w_1, v_2 \rightarrow \sqrt{\pi/8}\eta$; for $w_2$ one gets
$w_2 \propto \alpha^{1/3}$, corresponding to $a_1\propto \alpha^{-1/3}$.
With this result, the exponent for $v_2$ and $b_1$ can be
calculated; it is $v_2 \propto \alpha^{2/9}$ and $b_1 \propto \alpha^{-2/9}$.
The complete numerical solution of Eqs. (\ref{K2-eq1}--\ref{K2-eq4})
agrees excellently with simulations and also confirms the 
calculated asymptotic exponents, as seen in Fig. \ref{ZSG-K2}.

It turns out that the minimax-optimal solution for this problem 
is $\vas = (0,1)$, $\vbs = (1,0)$; i.e., $\vb$ converges to 
the wrong solution. However, if $\va = \vas$, 
the payoff is indifferent to the choice of $\vb$, and as long 
as $\va \neq \vas$, $\vb = (0,1)$ is the best response of
player B.

In this example, the proposed learning algorithm
converges to a strategy that gives the same payoff
as the equilibrium strategy, but the rate of
convergence is slow (power-law convergence with small negative
exponents) and depends on the realization 
of the matrix. A look at Eqs. (\ref{K2-eq1}--\ref{K2-eq4})
shows that the Nash equilibrium (which is characterized by
$\l^A_1= \l^A_2$ and $\l_1^B = \l_2^B$) is a stationary point of
the update equations only in the long-time limit 
where $1/w_i$ and $1/v_j$ are negligible. This is better 
than no convergence at all, but can hardly be considered
good in a situation where a few round should be enough 
to clarify what is the best thing to do. However, a learning
algorithm, even a slow one, may be more useful in si\-tu\-at\-ions where the optimal
behavior is not as obvious, such as for large payoff matrices.

\subsection{Large $K$}
\label{ZSG-largeK}
Let us now turn to the case of large payoff matrices.
To make statements that are valid for fairly 
generic cases, I will also assume that the 
entries are random and uncorrelated -- to be specific,
they are random Gaussian numbers of variance $\langle
c_{ij}^2 \rangle = K$. 
The statistics of the  minimax-optimal solution for 
this scenario were studied by J. Berg and A. Engel 
\cite{Berg:Matrix,Berg:Statistical,Berg:Diss}.
Combining these results with simulations and calculations
that I did, the following picture emerges: 

If player A picks a random strategy $\va_r$, the 
expected payoffs $\l^B_j$ of player B's options
are again Gaussian random numbers with zero mean
and a variance of roughly 1 (the exact value depends 
on the self-overlap of the random strategy vector).

If both players are playing the minimax-optimal
strategies $\vas$ and $\vbs$, and $K=L$, on average
50\% of each player's options are in the
support of this strategy, i.e., they are played with 
non-zero probability. Each of these yields an 
average payoff of $\nu$, which is 0 for $K\rightarrow
\infty$ and a Gaussian random number of mean 0 and 
variance $\langle \nu^2 \rangle \propto 1/K$ for finite $K$. 
The strategies which are not played give a 
payoff that is lower than $\nu$ and follows
a Gaussian probability distribution.

In order to see how a WTAP that learns according to
Eq. (\ref{MatG-Imp2}) behaves, it is necessary to assume
that the entries
in the payoff matrix are not correlated to 
the strategies of the players. This may seem 
counter-intuitive at first: after all, player
A ought to prefer strategies with above-average
entries. However, if player B can avoid being 
exploited completely (by a learning algorithm or
other clever strategy), player A will have to put up 
with taking strategies with below-average entries
as well, at least for large $K$, where dominating 
strategies are exceedingly unlikely.

 The mentioned assumption leads
to $\langle c^2_{lj}\rangle_{i,j} = \langle c^2_{ij}\rangle_{i,j}
 K$ and $\langle c_{lj} c_{ij} \rangle_{i,j} = K a_l$.
It is easy to calculate that 
$\langle \sp{\vx}{\vw_l} \s{\sp{\vx}{\vw_l}} \rangle =
\sqrt{2/\pi} \w{l}$, and to combine these results into
\be 
\dd{w_l^2}{\alpha} = \sqrt{8/\pi} \eta \left (\l_l - \sum_i
  a_i \l_i
\right ) w_l + 
  2 \eta^2 K(1-a_l).
\label{MatG-wsq}
\ee
The second term on the right hand side is a random walk term. 
The factor $(1-a_l)$ can be 
explained by a look at Eq. (\ref{MatG-Imp2}): the weight 
that belongs to the currently chosen option is not updated.
More important, however, is the first term, which causes
the weights with above-average success to grow and those
with below-average success to shrink. This term is proportional
to $w_l$, so it will dominate for large $\alpha$ and 
correspondingly large $w_l$. 

If both players play their optimal strategies $\vas$ and $\vbs$,
$\sum_i a_i \l_i $ is equal to the value of the game $\nu$, 
and the first term in Eq. (\ref{MatG-wsq}) goes to 0 
for those options $l$ that are part of the optimal
strategy, and stays negative for those that are not.
Neglecting the second term, this would be an indication
that the optimal strategies are a fixed point of
the considered learning rule.

To be sure that the assumption of perpendicular
vectors underlying Eq. (\ref{WTA-biasex}) holds,
we must examine the time development, and hence the fixed point,
of angles $\theta_{lm} =$ $ \arccos (\sp{\vw_l}{\vw_m}/(w_l w_m))$
between the vectors. A calculation similar to the one above
yields for the unnormalized overlap $R_{lm} =\sp{\vw_l}{\vw_m}$ :
\bea
\dd{R_{lm}}{\a} &=& \eta \sqrt{\frac{2}{\pi}} \cos(\theta_{lm}) 
 \left [w_l \left (l_m - \sum_i \l_i a_i \right ) + w_m
   \left (\l_l - \sum_i \l_i
 a_i\right )\right] \nonumber \\
& &~~~~ + \eta^2 \left (1- \frac{2 \theta_{lm}}{\pi}\right) K (1-a_l -a_m).
\eea
Using the chain rule, an expression for the development of
$\c_{lm}$ can be derived. Interestingly, all terms proportional
to $\eta$ vanish:
\bea
\dd{\cos(\theta_{lm})}{\a} &=& \eta^2 K \left ( 
  \frac{1 - 2 \theta_{lm}/\pi}{w_l w_m}(1- a_l -a_m)
\right. \nonumber \\
&&~~~~ 
  \left. - \frac{\cos(\theta_{lm})}{w_l^2} (1-a_l) 
  -\frac{\cos(\theta_{lm})}{w_m^2} (1-a_m) \right ) . \label{MatG-dcda}
\eea
Of the remaining terms, the ones proportional to 
$\cos(\theta)_{lm}$ dominate: the prefactor $1/(w_l w_m)$ 
is smaller or equal to $1/w_l^2 + 1/w_m^2$, 
$1-a_l-a_m$ is smaller or equal to $1-a_l$ and $1-a_m$,
and near the fixed point $\theta = \pi/2 + \Delta\theta$, 
an expansion gives $1-2\theta/\pi \approx -2\Delta\theta/\pi$,
whereas $\cos (\pi/2 +  \Delta\theta) \approx - \Delta\theta$.
Hence, for small $\Delta\theta$ the right hand side is 
proportional to $-\Delta\theta$, and $\theta = \pi/2$ is 
a stable fixed point. This is reassuring and tells us that
the use of Eq. (\ref{WTA-biasex}) is at least a good 
approximation. 

The approximation can be tested in simulations, which show
that angles between different weight vectors fluctuate 
around 0, with a variance that depends on $M$. Therefore,
Eq. (\ref{WTA-biasex}) agrees well ($\sim 1\%$ accuracy 
for $M=100$) with the output probabilities measured in the
simulations. This allows to calculate the mixed strategy,
and hence the learning success, at any point in the
simulations with reasonable computational effort.
 
\subsubsection{Playing against an opponent with a fixed strategy}
The first test case is playing a simplified
matrix game against an opponent who follows
a pre-determined strategy, i.e. chooses 
columns $j$ with  constant probabilities $b_j$.

If this fixed strategy is chosen at random,
the distribution of expected average payoffs $\l_i$
will be Gaussian with a variance near 1, as mentioned above. 
Without loss of generality, we can rename options such that
$\l_1> \l_2 > \dots \l_K$. \footnote{The case that
two expected payoffs are exactly equal occurs with 
probability 0.} The best strategy for 
player A is then $a_i = \delta_{i, 1}$. 
A WTAP that uses Eq. (\ref{WTAP-rule2})
and is updated according to Eq. (\ref{MatG-Imp2}) 
should converge to the optimal solution:
The weight belonging to the most profitable option
always grows at least like $\sqrt{\alpha}$:
\bea
\dd{w_1}{\a}&=& \frac{1}{2w_1} \left [ \sqrt{\frac{8}{\pi}}
    \eta \left (\l_1- \sum_i a_i \l_i \right ) w_1 + 2
    \eta^2 K (1-a_1) \right ] \nonumber \\
&> & \sqrt{\frac{2}{\pi}}\eta (\l_1 - \l_2) (1-a_1).
\eea
 
\begin{figure}
\centerline{
\begin{minipage}{0.49 \textwidth}
\epsfxsize= 0.95\textwidth
  \epsffile{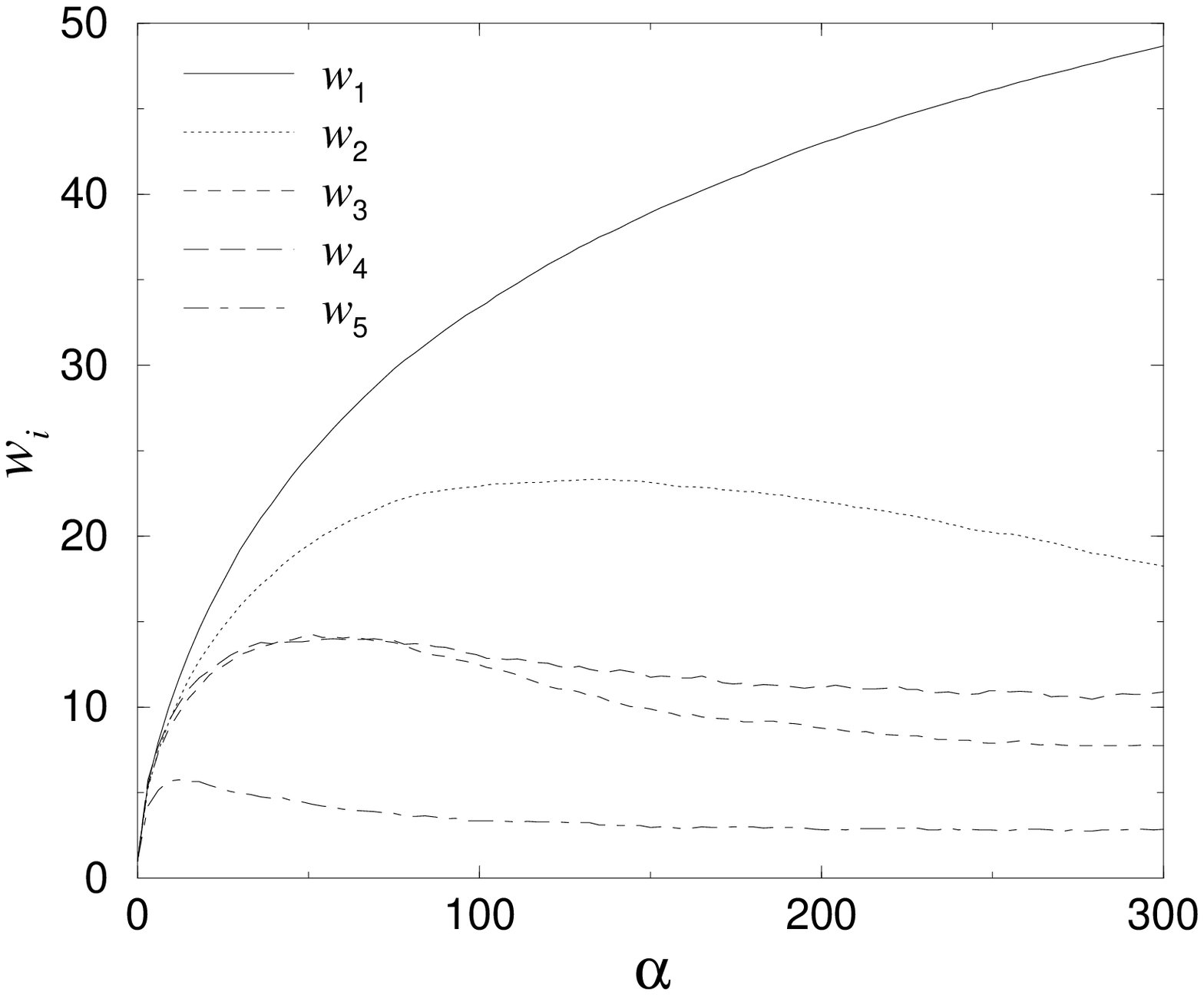}
\centerline{(a)}
\end{minipage}
\begin{minipage}{0.49 \textwidth}
\epsfxsize= 0.95\textwidth
  \epsffile{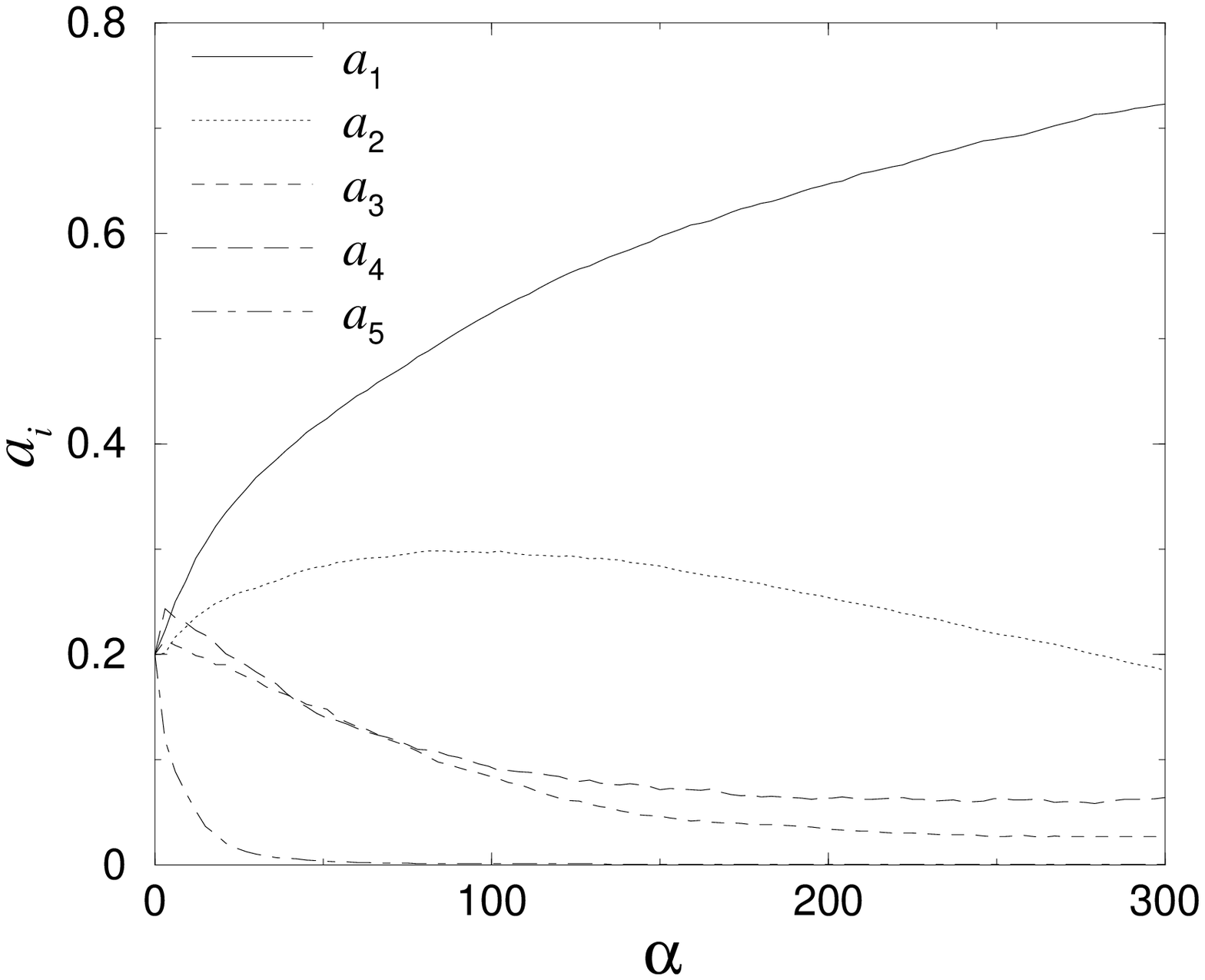}
\centerline{(b)}
\end{minipage}}
  \caption{Playing against a non-adaptive opponent with a 
    random strategy: the weight of the most advantageous
    strategy 1 grows fastest, whereas the weights of below-average
    strategies shrink (Plot (a)). Correspondingly, the most
    profitable strategy is used increasingly (Plot (b)). 
    Simulations used a random matrix with $K=L=5$ and $M=100$.}
  \label{ZSG-vsrand}
\end{figure}

Pure strategies that give below-average payoff (compared to
the current average, i.e., to $\sum a_i \l_i$) have a 
negative growth term in their update equation, i.e.,
their weight stays bounded, and their probability $a_i$
of being played goes to zero, removing them 
from the competition. Gradually, the average payoff 
increases, eliminating one non-optimal choice after 
the other. 
This is what happens in simulations as well, as seen in 
Fig. \ref{ZSG-vsrand}.

\begin{figure}
  \epsfxsize= 0.65\columnwidth
  \centerline{\epsffile{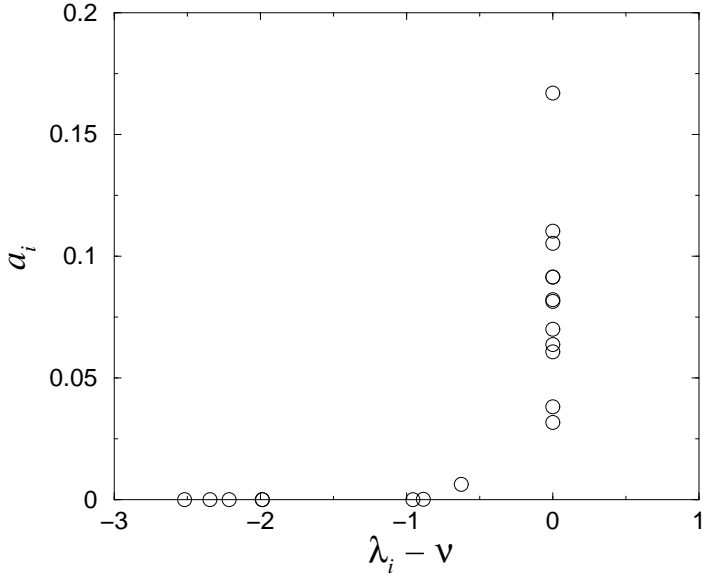}}
  \caption{Components $a_i$ of the strategy vector versus 
  expected payoff $\l_i$, against an opponent using the 
  equilibrium strategy $\vbs$.
  The figure shows the state after $10^4$ learning steps with
    $K=20$, $M=100$.} 
  \label{ZSG-vsopt}
\end{figure}

As pointed out previously, if B is constantly playing the optimal
strategy $\vbs$, the best thing that A can do is to 
eliminate the strategies with $\l_i<\nu$ that are not in 
the support of $\vas$. As Fig. \ref{ZSG-vsopt}
shows, the neural network succeeds in doing this: after
a learning time of $\a=100$, the suboptimal options
are strongly suppressed (only the option with the payoff 
closest to $\nu$ still has an appreciable probability),
whereas the strategies in the support of $\vas$ are 
played with random positive probabilities.

\subsubsection{Playing against adaptive opponents - Fictitious Play}
One of the standard learning algorithms in game theory
is {\em Fictitious Play} \cite{Fudenberg:Learning}: 
the adaptive player (in this 
case, player B) keeps a histogram of his opponent's 
decisions $i^{\tau}$ up to a time $t$, 
which he uses to estimate A's strategy $\va$ by
$\vat$:
\be
\at_k = \frac{1}{t} \sum_{\tau} \delta_{k,i^{\tau}}.
\ee 
Player B then estimates the average payoffs of his
available strategies assuming that A will continue
playing as he did before: 
\be
\lt_j^B = \sum_k c_{kj} \at_k,
\ee
and chooses the output with minimal $\lt_j^B$.
This updating rule has been proven to converge
to the optimal strategy if both players use it,
in the sense that the empirical estimate
$\vat$ converges to $\vas$, and $\vbt$ to $\vbs$, as $t
\rightarrow \infty$
\cite{Fudenberg:Learning}.

However, in a different sense, it does not 
converge: at any given time, the output of
the ``Fictitious Play''-player is deterministic,
and he often repeats the same output for 
several time steps, so he never plays the optimal 
mixed strategy.

\begin{figure}
  \epsfxsize= 0.65\columnwidth
  \centerline{\epsffile{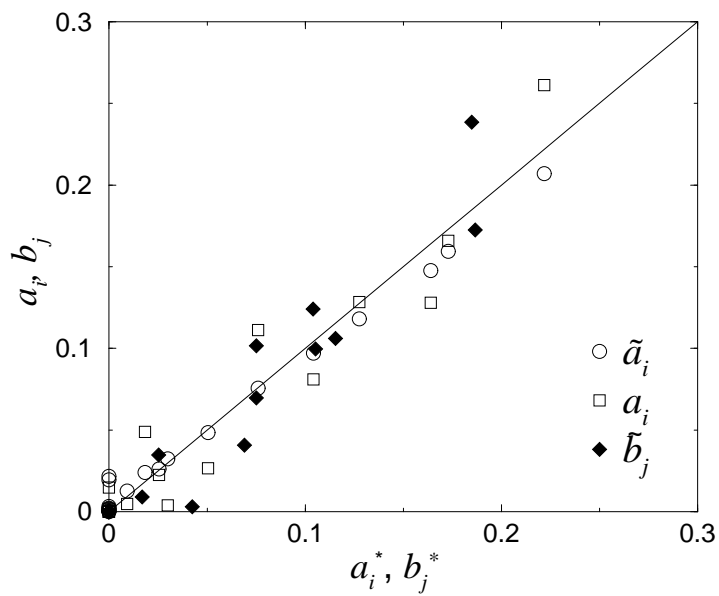}}
  \caption{Averaged empirical and current strategies 
    $\vat$ and $\va$ of
    a neural network playing against Fictitious Play
    (empty symbols) and empirical strategy  $\vbt$ of the opponent
    (filled symbols) after $10^6$ learning steps with
    $K=20$, $M=100$, compared to the equilibrium strategies
    $\vas$ and $\vbs$. Perfect agreement with the
    equilibrium strategies would result in all points lying
    on the diagonal.} 
  \label{ZSG-FictFig}
\end{figure}

Simulations show that in the long run, when Fictitious Play and
the proposed neural network algorithm play against each
other, they converge to 
the Nash equilibrium in the long-time average sense.
At any given moment, the output distribution of A
is significantly different from $\vas$. However,
just as the histogram of outputs of two players doing Ficticious Play  
converges to $\vas$ and $\vbs$, so does the output
of the Neural Network player. Fig. \ref{ZSG-FictFig}
shows the components of $\vat$ and $\vbt$ (the empirical
long-time averages) versus the components of $\vas$ and
$\vbs$ after $10^6$ steps in a simulation with
$K=20$. Perfect convergence to equilibrium would result
in all points lying on the diagonal. The figure also shows
the current strategy $\va$ of the Neural Network at the
moment when the simulation ended. One can see that 
the deviation of the current $\va$ (empty squares) 
from the equilibrium strategy is larger than that
of $\vat$ (empty circles), which agrees reasonably well
with $\vas$. 

\begin{figure}
  \epsfxsize= 0.7\columnwidth
  \centerline{\epsffile{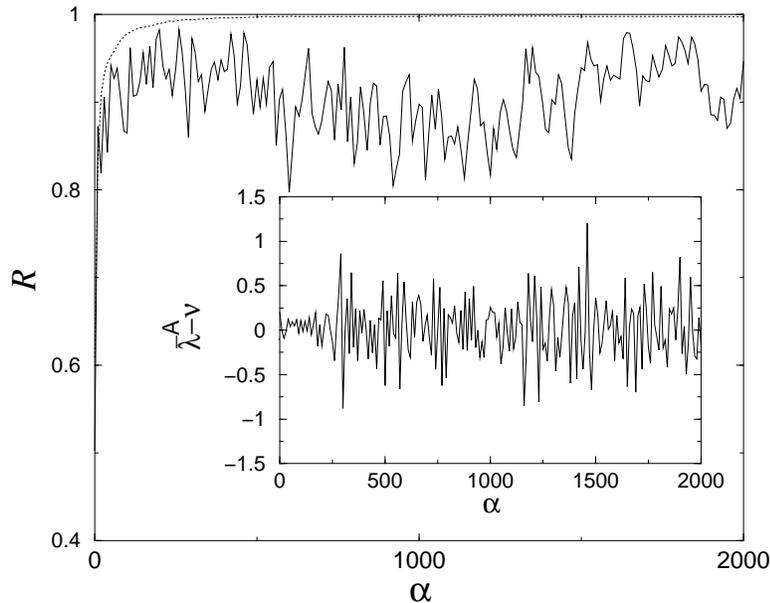}}
  \caption{Dynamics of the overlap $R$ between 
    $\va$ and $\vas$ (solid line) and between 
    $\vat$ and $\vas$  (dotted line) of a neural
    network playing against Fictitious Play.
    The inset shows the payoff $\bar{\l}^A$ for A, 
    averaged over 1000 round of play. The simulation used
    $M=100$ and $K=20$. 
    } 
  \label{ZSG-FictFluct}
\end{figure}

The fluctuations of $\va$ around the optimal strategy 
can be quantified by introducing an overlap $R=
\sp{\va}{\vas}/|\va| |\vas|$ and an expected payoff
$\bar{\l}^A=\langle \sum_i \l_i^A \rangle $ averaged over 
times short compared to the total learning process. 
$R$ is observed to fluctuate around high values 
$\approx 0.9$, but never converges to 1, whereas 
$\bar{\l}^A$ fluctuates around the value of the game,
as seen in Fig. \ref{ZSG-FictFluct}. The figure also
shows the overlap of $\vat$ with $\vas$ (dotted line),
which seems to converge to 1. However, the rate
of convergence depends on the realization of the payoff
matrix.

\subsubsection{Playing against an equivalent opponent}
One rather obvious question is how the proposed learning
algorithm fares against an opponent who is using the
same architecture and the same update rule.
The answer is, quite good, with some exceptions.
Generally, the principles of the algorithm work fine:
options that give above-average payoffs are strengthened,
the others are suppressed. If equilibrium is reached, 
the weight vectors continue to grow slowly due to the
random walk term in Eq. (\ref{MatG-wsq}), which 
gradually slows down the dynamics. Simulations show
that the strategies of the two players usually reach
a high overlap with the Nash equilibrium. Fluctuations
around the optimal strategy decrease with increasing $M$
and decreasing $\eta$.

The mentioned exceptions concern two possible problems:  
for one, there is again the possibility that an option
that is not in the support of the optimal strategy
has a payoff that is only negligibly below the
value of the game. This option is very hard to tell
apart from a ``good'' one, i.e., one that is in the
support, by looking at average payoffs, and may 
stay in the strategy mix for a long time.
From the point of view of optimization, such an
option corresponds to a very shallow valley
whose minimum has to be found. 

The second problem can occur if the networks of
both players are fed with the same pattern.
If two weight vectors of the two networks have
an angle other than $\pi/2$, the output of
the players is mutually correlated.
Although a calculation similar to that leading to 
Eq. (\ref{MatG-dcda}) shows that the overlaps between
the different weight vectors of the two players,
$\sp{\vw_l}{\vv_k}/(w_l v_k)$, should have a 
stable fixed point at $0$, simulations show that
a significant overlap (on the order of $0.9$) can 
develop. If one network can predict the output of
the other beyond the point of knowing the probability
distribution $\va$, the concept of an equilibrium 
mixed strategy breaks down.
It appears, however, that this irregularity appears
less frequently with increasing $M$.

\begin{figure}
  \epsfxsize= 0.7\columnwidth
  \centerline{\epsffile{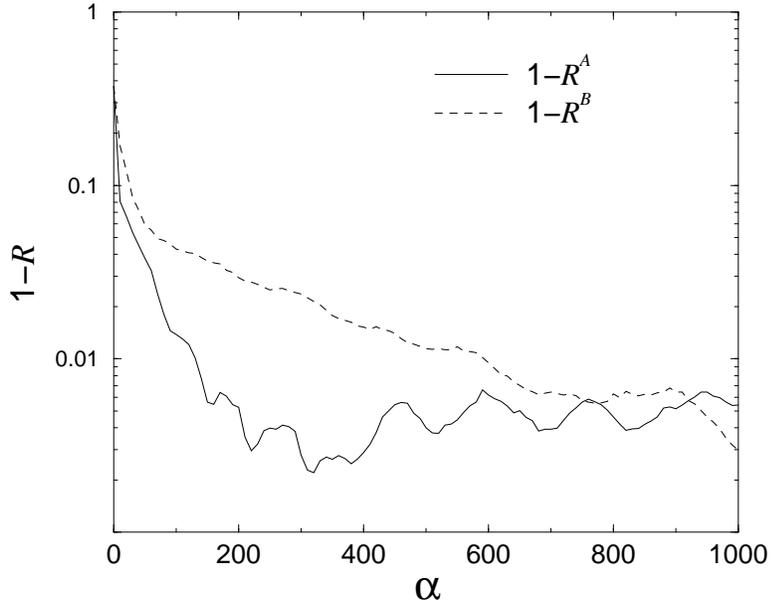}}
  \caption{Two neural networks with $M=1000$ playing
    a zero-sum game: the plot shows how the overlaps with the
    equilibrium strategy increase during learning, then
    saturate at some value that depends on $M$ and $\eta$.} 
  \label{ZSG-eqR}
\end{figure}

\subsubsection{A remark on efficiency and more general schemes}
The proposed learning algorithm essentially
uses the patterns as random number seeds, without
drawing any further information from them. The
only relevant property of the weight vectors 
was their norm. If the task is to find an
algorithm that learns to play a two-player game,
would it not be more efficient to reduce the
problem from updating and multiplying vectors
to rolling Gaussian random numbers and updating
their variance according to Eq. (\ref{MatG-wsq})? 
Indeed it would. This abstraction also opens
the way to all sorts of fiddling with the
learning rule: the noise term (the term proportional to
$\eta^2$ in Eq. (\ref{MatG-wsq}) can be made larger or
smaller, the learning rate can be made time-dependent,
the variance can be increased linearly, polynomially
or exponentially\dots
possibilities are limitless.  

This opens the door to a more general class of 
learning algorithms for two-player games:
adjusting the variances of random numbers,
the largest of which determines the output,
is one way of updating the probability 
vector. However, it is not trivial to 
find one that even comes close to the Nash
equilibrium. 


\subsubsection{A remark on time scales}
The analysis of the learning algorithm tacitly
assumed the usual limit $M\rightarrow \infty$,
which allows to replace $\langle c_{lj} \rangle$
with $\sum_j b_j c_{lj}$ -- as long as $K$ is 
finite, each option $j$ that is played at all is  
played an infinite number of times during a
short time interval $\Delta \a$. In reality, 
$M$ is always finite -- and as the remark 
above showed, $M$ is not really relevant as
the dimensionality of a vector space. 
However, a certain number of rounds have to be
played to get a good estimate of the 
expected payoff of an option. Let us assume that
strategies are fixed for a moment. If a large number $t$ 
of rounds are played, player A played option $i$  $t (a_i+\Delta_i)$
times, where $\Delta_i$ is a Gaussian random number of mean
0 and variance $\langle \Delta_i^2 \rangle = a_i(1-a_i)/t$.
The mean square deviation between the observed and the
expected payoff for one of B's options $j$ is then 
\bea
\langle (\l_j^B -\lt^B_j)^2 \rangle &=& \left \langle \left(
    \sum_i \Delta_i c_{ij} \right )^2\right \rangle
\nonumber\\
&=& \left \langle \sum_i \Delta_i^2  \right \rangle + 
\left  \langle \sum_{i, l\neq i} \Delta_i \Delta_l c_{ij}
  c_{lj} \right \rangle  \nonumber \\
&=& \sum_i c_{ij}^2 a_i (1-a_i)/t \nonumber \\
&\sim & \frac{K}{t}. 
\eea
To decide which one of two options with payoffs $\l_1^B$ and
$\l^B_2$ is better, player $B$ has to wait roughly
$t>K/(\l^B_1 - \l^B_2)^2 $ time steps.
 
Learning in two-payer games is, as is often the
case in learning scenarios, a tradeoff between 
accuracy (exact judgment which option is the best)
and speed (to avoid being exploited by insisting on a 
suboptimal strategy for a long time). Things become
even trickier near equilibrium, since the difference
in expected payoffs decreases the closer the opponent
comes to his optimal strategy, and vanish if he 
reaches it. At that point, any drift term in a 
learning algorithm vanishes. Since the
noise term remains, fluctuations are inevitable.

An optimized learning algorithm might take this 
into account and adapt quickly as long as 
differences in the expected payoffs are 
large, while it reduces both the rate of adaption and,
if possible, additional noise terms when it 
is near an equilibrium. However, it is not obvious 
how proximity to equilibrium can be detected 
unambiguously.

\subsubsection{Learning from time series}
So far, the patterns that the neural networks
used had no meaning in themselves and contained
no information on the opponent's actions. This could
be considered the worst case, and it is good to 
know that the networks responded reasonably if the
update rule was chosen accordingly. If the patterns
now represent real information -- e.g., they contain
the time series of the opponent's output -- should not
the networks perform much better? 

To check this, we have to leave the simplicity of 
real-valued patterns behind and use $Q$-state patterns, as
described in Sec. \ref{SEC-MC}: the components of 
A's patterns can take $Q^A= L$ values, whereas 
B's patterns consist of $Q^B=K$ values. The first 
update rule to try is the generalization of
Eq. (\ref{MatG-Imp2}): 
\be
{\vw_l^q}^{t+1} =  {\vw_l^q}^{t} + \frac{\eta}{M} \vx^q
\s{h_l} (c_{lj}-c_{ij}). \label{ZSG-PottsUp1}
\ee
The output is determined by the hidden field with the 
highest absolute value, and the patterns that each
network sees are the time series of the other player's
output. 

The outcome is disillusioning: the output probabilities
have no similarity with the optimal strategy. A closer
look reveals that both perceptrons lock into long 
stretches of repeated outputs, similar to a single 
Bit Generator with fixed weights. Since weights are
not exactly fixed in this case, cycles do no continue 
forever, but nevertheless, no convergence to an 
equilibrium is observed. 

Could the network possibly draw useful information from
the patterns and at the same time avoid falling into
cycles if the components of the patterns had no temporal
correlation? For example, one can feed the networks 
an artificial time series of an opponent playing his optimal
strategy with true random numbers. The results are again
surprising: while the strategy of the networks at any given time 
is not very close to equilibrium, the long-time 
average of outputs converges to equilibrium fairly quickly,
similar to two players playing Ficticious Play against each
other.

One more detail shows that some new mechanisms are at
work: if the output is determined by the largest hidden
field $h_i$ instead of $|h_i|$, there is no convergence
to equilibrium in any sense. That points to the
fact that the sign of $h_i$ is no longer random, 
which has its reason in an inherent bias in the patterns.

\section{Learning from biased patterns}

A bias in the patterns can easily lead to a 
bias in the weights (which are, after all, 
a linear combination of the initial weights and 
the patterns that are shown during the learning
process), which can in turn cause a bias
in the hidden fields. To understand these effects
systematically, it is easier to go back to 
real-valued patterns with a bias, and later 
make the generalization of the terms introduced there
to multi-valued inputs.

To be specific, a biased pattern vector consists of 
an unbiased random vector $\vxt$ and a constant
vector $\vy$ of norm $y$. The weight vectors can
also be split into components parallel and
perpendicular to the bias:
\bea
\vwb_i &=& \frac{\sp{\vw_i}{\vy}}{y^2} \vy; \nonumber \\
\vwt_i &=& \vw_i - \vwb_i.
\eea
The appropriate order parameters are $\wb_i = \sp{\vw_i}{\vy}/y$,
which can take positive and negative values, and
$\wt_i = \sqrt{\sp{\vwt_i}{\vwt_i}}$, which is
positive or zero. 

For any pattern $\vx$, the hidden field $h_i$ 
of strategy $i$ now has three components:
\begin{itemize}
\item a stochastic component $\htt_i = \sp{\vwt_i}{\vx}$
that is assumed to be independent from the $\htt_l$ of the other
options $l$;
\item a constant component $\sp{\vwb_i}{\vy}= \wb_i y$; 
\item and a stochastic component $\sp{\vwb_i}{\vxt} = h_c \wb_i$ 
where the factor $h_c$ is a Gaussian variable of 
variance 1 that takes the same values for all strategies $l$
for a given $x$.
\end{itemize}
The chance that $h_i$ for a given $i$ is the largest now reads
(analogous to \ref{WTA-biasex}):
\be
a_i = \int_{-\infty}^{\infty} D\hh_i \int_{-\infty}^{\infty} Dh_c  
\frac{1}{2 \pi} \prod_{l\neq i} \Phi \left[ \frac{1}{\wt_l} 
\left ( \wt_i \hh_i + (h_c+y) (\wb_i - \wb_l) \right )
\right ].
\label{WTA-biasxr}
\ee
The integrals can possibly be eliminated by linear transformation
of the variables; however, the transformed expression neither becomes
prettier nor gives more insight.

If the bias dominates the output, a second look at the
update rule is necessary. If Eq. (\ref{MatG-Imp2}) is used
and the weight of an option has a large negative bias
$\wb_i$, the hidden field will be negative, and the update
will tend to make it more negative if the option is
favorable. This is not a problem if the output follows the
maximal $|h_i|$; however, it is now also possible to 
let the output follow the  maximal $h_i$
(Eq. (\ref{WTAP-rule1})) and  drop the $\s{h_l}$-term 
in the update rule:
\be
\vw_{l}^{t+1} = \vw_{l}^t + \frac{\eta}{M} \vx c_{lj}.
\label{WTA-HebbUpdate}
\ee
The normalizing term proportional to $-c_{ij}$ is not needed
anymore either -- it would result in adding the same update
vector to all $\vw_{l}$, thus shifting all hidden fields by the
same amount for a given pattern vector. This no longer has
an influence on the decision of the network.

The development of the two order parameters $w_l$ and $\wb_l=
\hb_l/y$
is straightforward:
\bea
\dd{w_l}{\a} &=& \frac{\eta^2}{2 w_l} \sum_j b_j c_{lj}^2; \\
\dd{\wb_l}{\a} &=& \eta y \sum_j b_j c_{lj} = \eta y\l^A_l.
\label{WTA-HebbODE}
\eea

\begin{figure}
\centerline{
\begin{minipage}{0.49 \textwidth}
\epsfxsize= 0.95\textwidth
  \epsffile{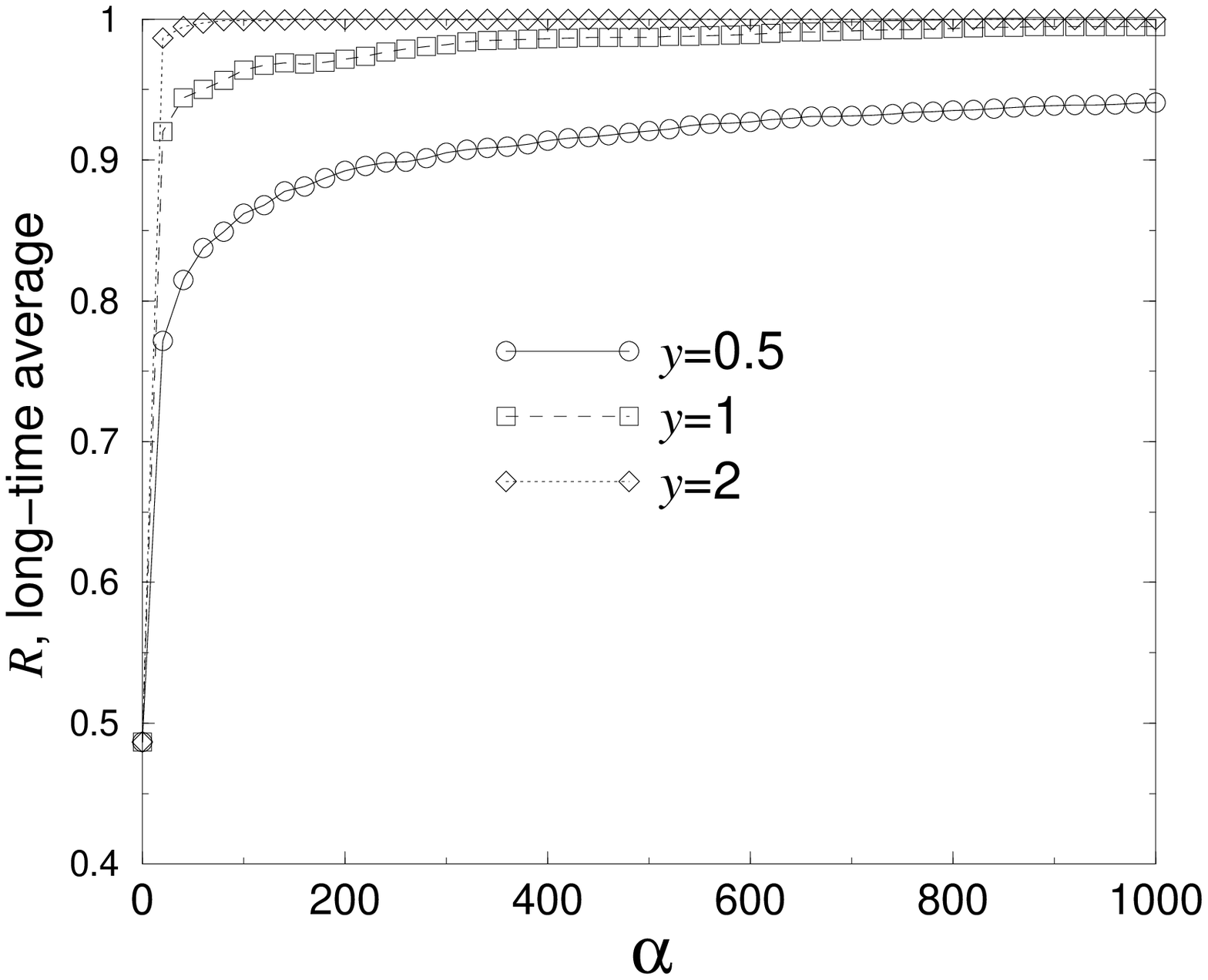}
\centerline{(a)}
\end{minipage}
\begin{minipage}{0.49 \textwidth}
\epsfxsize= 0.95\textwidth
  \epsffile{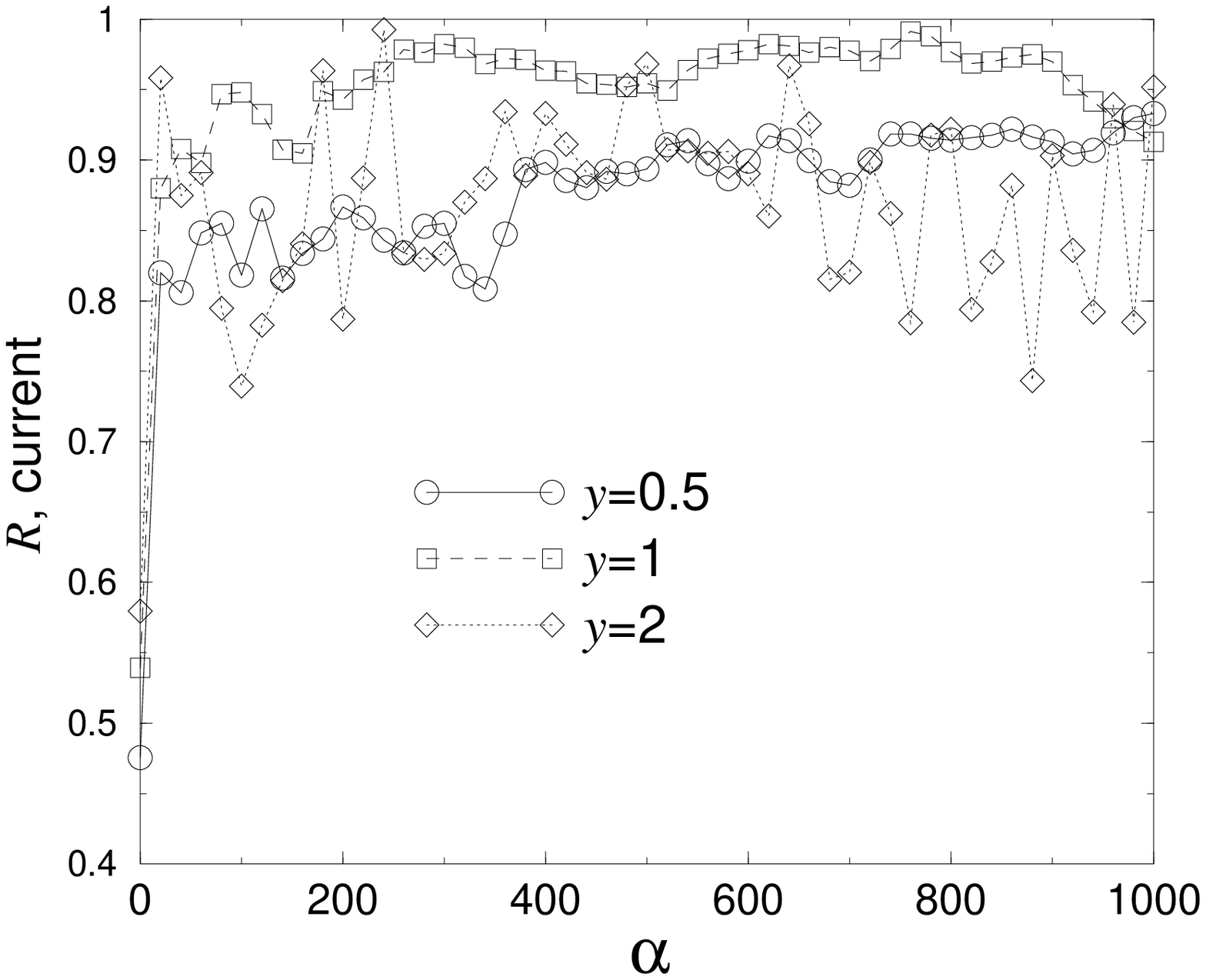}
\centerline{(b)}
\end{minipage}}
  \caption{Learning from biased patterns: while 
    the long-time averaged strategy converges to 
    $\vas$ quickly for large bias $y$ (as in Fictitious
    Play), the current strategy comes closest to $\vas$
    for intermediate $y$.}
  \label{ZSG-biasfig}
\end{figure}

Interestingly, rule (\ref{WTA-HebbUpdate}) implements a 
``soft'' fictitious play rule. This can be seen in the
limit $y\rightarrow \infty$, where the stochastic component
can be neglected completely. At time $t$, the strategy 
$i^{t}$ with the largest
$\wb^{t}_i$ is played, while the opponent plays $j^t$; 
$\wb^t_l$ at time $t$ can be written as 
\be
\wb_l^t = \sum_{\tau=0}^t \frac{\eta}{N} y\,_{lj^t} = 
\frac{\eta}{N} 
y \sum_j c_{lj} t \tilde{b}_j = \frac{\eta}{N} y t \lt^A_l,
\ee  
where $\tilde{b}_j$ and $\lt^A_l$ are analogous to Section
\ref{ZSG-largeK}. By adjusting the range of $y$ 
between $0$ and $\infty$, it is possible to interpolate
between random guessing and purely deterministic fictitious
play. If $y$ takes intermediate values, there is enough
stochasticity to get, at any given time, a mixed strategy
that is reasonably close to $\vas$; however, the convergence
to $\vas$ in the long-time average is slower than for
larger bias $y$. This is shown in Fig. \ref{ZSG-biasfig}.

\subsubsection{Multi-state patterns}
The same ideas can be generalized to $Q$-state input patterns
and weight vectors. It will become clear how 
unequal probabilities of the $Q$ inputs amount to a bias.
 As mentioned, each output $l$ now has a 
set of $Q$ weight vectors $\vw_{l}^q$ with entries 
${w_{l}^q}_n$. The hidden field for output $l$ is
\be
h_l = \sum_q^Q \sum_n^N {w_l^q}_n (Q\delta_{\xi_n,q} -1)
= \sum_q^Q h^q_l \mbox{~with~~} h^q_l=  \sp{\vw_{l}^q}{\vx^q},
\ee
where the $\vx^q$ are vectors with components $Q \delta_{\xi_n,q}-1$.
One can then assume that the bias in the pattern is the same
for all of its components (this is necessarily the case
with time series patterns, since each value is shifted 
through the entire weight vector) and is thus completely
defined by the probabilities of encountering the
possible inputs $q$. We then use the deviations $\delta q$ 
from the uniform distribution to describe this distribution:
\be
\mbox{prob}(x_n = q) = \frac{1}{Q}(1 + \delta q).
\ee
Note that $\sum_q \delta q =0$. We then proceed by 
splitting $\vx^q$ into an average and a random
component:
\be
\vy^q = \left( \begin{array}{c}
            \delta q \\
            \vdots \\
            \delta q \end{array} \right); \vxt^q = \vx^q - \vy^q.
\ee 
The hidden field $h_q^l$ can be split into three parts
analogous to the case of real-valued input patterns:
\be
h_l^q = \hb_l^q + \wb_l^q h_c^q + \htt_l^q,
\ee
where  $\hb_l^q = \sp{\vwb_l^q}{\vy^q} = \wb_l^q y^q$
is the constant part, $h_c^q = \sp{\vwb_l^q}{\vxt^q}/\wb_l^q$
is a random variable that is the same for all $l$, and
$\htt_l^q = \sp{\vwt_l^q}{\vxt^q}$ is random and different
for each $l$ (the $\htt_i^q$ are uncorrelated if the 
$\vwt_i$ are pairwise orthogonal, which I assume for
simplicity's sake).

The variance $\langle {h_c^q}^2 \rangle$ of $h_c^q$ can be 
calculated in a straightforward manner; the result is 
\be
\langle {h_c^q}^2 \rangle = Q (1 - (1 + \delta q)/Q) (1 + \delta q).
\ee
Similarly, the variance of $\htt_l^q$ is 
\be
\langle {\mbox{${\htt}_l^q$}} ^2 \rangle = 
\mbox{$\wt_l^q$}^2 Q (1 - (1 + \delta q)/Q) 
(1 + \delta q).
\ee
The addition of the fields $h^q$ belonging to different
$q$ is almost, but not quite straightforward. The total
field can be written like this:
\be
h_l = \hb_l + z_l h_{ct} + \htt_l.
\ee
$\hb_l$ is simply the sum of the constant components
$\sum_q \hb_l^q$. The second term is based on the assumption
that $\wb_l^q$ is approximately proportional to $y^q$ for all $q$,
with a coefficient characteristic for $l$: 
$\wb_l^q = z_l y^q$. This assumption will be 
justified later. The sum $\sum_q \wb_l^q h_c^q$ can 
then be written as $z_l h_{ct}$, where $h_{ct}$ is 
a Gaussian variable with a variance of $\langle h_{ct}^2 \rangle =
\sum_q {y^q}^2 Q (1- (1+ \delta q)/Q)(1+ \delta Q)$.
Again, $h_{ct}$ is the same for all $l$.

The components of the strategy vector $\va$ can now be 
calculated in analogy to Eq. (\ref{WTA-biasxr}):
\be
a_i = \int_{-\infty}^{\infty} D\hh_i \int_{-\infty}^{\infty}
Dh_c \prod_{l\neq i} \Phi \left [ \frac{1}{\wt_l }
(\wt_i \hh_i + (y^2+ h_c)(z_i -z_l) \right ].
\label{ZSG-MCstrat}
\ee
Three questions remain: what is $z_l$? What is all this
splitting of the hidden fields good for? And what 
relevance does this have for time-series patterns? 

The first question can be answered by a look at the 
learning rule, which, in analogy to Eqs. (\ref{ZSG-PottsUp1}) 
and (\ref{WTA-HebbUpdate}), can be taken as
\be
{\vw_l^{q}}^{\ t+1} = {\vw_l^q}^{\ t} + \frac{\eta}{M} \vx^q c_{lj},
\ee
where $j$ is the opponent's action at time $t$.
The average update of the weight component parallel to 
the bias is then
\be
\langle \mbox{$\vwb_l^q$}^{~t+1} \rangle_{\vx} = \mbox{$\vwb_l^q$}^{~t} +
\left \langle \frac{\eta}{M} \vy^q c_{lj} \right \rangle_j.
\ee
The order parameter $\wb^q_l$ then follows the differential
equation
\be
\dd{\wb^q_l}{\a} = \eta y^q \sum_j b_j c_{lj}= \eta y^q
\l^A_l \mbox{~~for all~}q, 
\ee
which can be solved:
\be
\wb^q_l = \mbox{$\wb^q_l$}_0 + y^q \left ( \eta \int_{0}^{\alpha}
  \l^A_l(\tau) d\tau \right). 
\ee
Neglecting the initial weight $\mbox{$\wb^q_l$}_0$,
the characteristic quantities $z_l$ are therefore
proportional to the
average payoffs of option $l$, integrated from the
beginning of the game.

Regarding the second question, what is this good for?
Eq. (\ref{ZSG-MCstrat}) states the following: The bias in the pattern
creates a strong preference for those options that have
accumulated higher $z_l$, i.e., higher expected payoffs
in the course of the game. This preference can be 
overruled by either a fluctuation in the pattern (if the
total pattern $\vx = \vxt + \vy$ has a negative overlap with the bias
$\vy$, the bias accumulated in the weights works in the
opposite direction) or by the contribution from 
the weight and pattern perpendicular to the bias.
For sufficiently large bias, either case becomes 
unlikely, and the learning process again reduces to 
Fictitious Play. 

As for time-series patterns, the relevance is this:
if the strategy played by the opponent 
has a significant bias (for example, if the 
opponent is playing $\vbs$, where roughly half of
the options do not appear at all), some of the
$\delta q$ are of order 1. Since the norm of the bias vectors 
is $y^q = \sqrt{M}|\delta q|$, that means that there is
indeed a strong bias (of order $\sqrt{M}$ rather than
1), and nothing significantly different from 
Fictitious play can be expected -- if the opponent's
output is properly randomized. 
If two networks with equivalent architecture use
each other's outputs as input patters, both lock 
into short cycles for many repetitions, never
coming close to anything like equilibrium.

\section{Summary}
The scenarios presented in this chapter have shown that 
a slightly modified Winner-Takes-All Perceptron 
with a learning algorithm inspired by the Confused
Bit Generator can be used to learn a good strategy
in zero-sum games: it concentrates on the optimal
pure strategy if there is one, and finds a good 
mixed strategy even against adaptive opponents. 
The ingredients for success are a drift term 
which strengthens options whose payoff is better than
the current one, and weakens the others, and a 
noise term which gradually increases the norms of
the weight vectors, decreasing the impact that a 
single learning step has.

This learning algorithm can be expected to converge to 
Nash equilibrium in the limit of infinitely slow
adaption ($M\rightarrow \infty$) and infinite 
learning times ($\a \rightarrow \infty$), and 
gives a reasonable approximation of equilibrium
for finite $M$ and $\a$. It is not ``rational'' --
on the other hand, it generates a truly mixed
strategy, as opposed to algorithms like
Fictitious Play, which generate mixed strategies
only in the long-time average. 

Finding the equilibrium mixed strategy $\vas$ in a zero-sum
game is a nontrivial problem of stochastic optimization:
the ``energy landscape'' becomes increasingly flat as
one's opponent comes closer to his equilibrium strategy
$\vbs$;
however, only then is $\vas$ the best response.
Considering this difficulty, the success of he 
presented learning algorithm is a pleasant surprise,
even if learning is slow.

A different learning algorithm is required if
there is a bias in the patterns. This bias 
leads to a bias in the weight vectors, and
the combination of these causes a bias in the
hidden fields of the neural network. This can 
be exploited: it is now the overlap of the weight vector 
with the bias in the patterns, rather than the norm of
the weight vector, which gives the preference for one
or the other option. Depending on the strength of
the bias, one gets pure guessing in the limit of small
bias, fictitious play in the limit of infinite 
bias, and ``soft'' fictitious play in intermediate
regimes.   

So far, no scenario has been found where the neural 
network could explicitly make use of any actual
information encoded in the patterns. If time series 
of the opponent's output are used for patterns,
and both players use networks, they lock into short 
cycles. More work is necessary to decide whether 
this is a systematic problem of the used network 
architecture, and whether it can be remedied with a
different learning algorithm.

  \cleardoublepage
\chapter{Conclusion and outlook}

The projects presented in this dissertation tried to 
make some connections between the fields named in its
title: neural networks, game theory, and time series 
generation. Neural networks played games, time series
were generated by games, networks learned and generated
time series, etc..  The concept that connects these
fields is that of prediction: neural networks can 
be used as prediction algorithms, time series generation
and prediction are so closely related that they are
almost interchangeable, and, last but not least,
economic game theory deals with the attempt to
predict an opponent's actions as accurately as
possible and not allow him to predict and outplay
oneself. 

The aim of Chap. \ref{CHAP-antipredictable} was to 
study the properties of time series that are 
antipredictable for three selected prediction algorithms,
and to find common aspects between them, such as 
a tendency towards longer cycles than for perfectly 
predictable series, and a suppression of
the features that the algorithm is sensitive to.

Chap. \ref{CHAP-MG} gave an overview over some of the
variations of the Minority Game, some of which were 
introduced first by our research group. 
It was demonstrated that
this game naturally gives rise to antipredictability 
in the time series of minority decision if agents 
try to adapt rapidly. I also demonstrated that,
with minor adaptions, all conventional strategies
can also be generalized to more than two options.

In Chap. \ref{CHAP-game}, a neural network tried to
find a suitable strategy for a zero-sum game by 
learning from experience. A learning algorithm was
found that uses random input patterns to generate 
a mixed strategy that can come close to the equilibrium
strategy. This algorithm probably can still be optimized by
a more systematic analysis what information can be 
extracted from watching the opponent's play. However,
it is not yet obvious under what circumstances a 
learning algorithm should play a mixed strategy at all. 

By its conception, this work is interdisciplinary,
and a reader with a background in physics
may have wondered, ``Is this physics?''. The answer, depending
on one's point of view, is no, yes, yes, and maybe.

This dissertation does not deal with the  classical subjects
of physics, namely, the properties of inanimate matter.
In that sense, it is not physics. 

A different definition
might be, ``physics is what physicists do''. In that sense,
it is physics: all of the co-workers who collaborated on
the various projects were trained as physicists, as are
the majority of scientists who have published on the 
statistical properties of neural networks and on the
Minority Game. 

The third answer concerns the methods: most of the methods
used to achieve results in this work, like averaging over
disorder in online learning, the application of Markov
chains in the stochastic MG, the analysis of the
nonlinear properties of the CSG, and the extensive
use of Monte Carlo simulations to check analytical
results  (or replace them where they are not available), 
are well-rooted in statistical physics and nonlinear dynamics. 

The fourth answer should really be, ``who cares, as long as 
it is interesting?'' If methods from one field can help
to get insights in other fields, they should be 
applied, no matter how the result has to be classified.

I hope that this work has shed some light on the 
problems it treated from unusual angles. What must 
be kept in mind is that all models presented here
are toy problems that may or may not have a strong
tie to real life. Many of these toy problems develop
a life of their own, intriguing researchers with 
mathematical subtleties and technical challenges. 
I have tried to avoid this tendency by presenting
and comparing several algorithms and strategies and
finding global aspects that connect them.

What remains to be done is to establish which
of these models has what relevance to ``real life'',
i.e., the behavior of humans, or animals, or
other entity: what strategies do stock brokers,
car drivers, predatory animals use when they
make their decisions in Minority Game-like
settings? How do humans adapt their strategy in
games? This will necessarily require
much know-how on sociology, ecology or psychology
and much experimental work, such as that begin in
Ref. \cite{Ruch:Interactive}. It may thus not
be the answer that a theoretical physicist
is well-equipped to answer.

 \cleardoublepage

\appendix 
\chapter{Notation}
In a work that deals with a rather large number of 
projects, concepts and mathematical objects, a certain
overloading of symbols like $\alpha$, $\lambda$ and $N$ 
is inevitable. This chart might be helpful to find out
what meaning a given symbol has in each chapter.

\begin{tabbing}

 \hspace{1 cm} 
\=A, B\hspace{2cm} \=players in a zero-sum game \\
\>$\va$, $\vb$ \> strategy vectors of players A and B \\
\>$\vas$, $\vbs$ \> optimal (Nash-equilibrium) strategies \\
\> $C$ \> Chap. \ref{CHAP-antipredictable}: autocorrelation
function; \\
\> \> Chap. \ref{CHAP-game}: norm of center-of-mass of perceptrons.\\
\> $\Cm$ \> Chap. \ref{CHAP-MG}: correlation matrix for
                 multi-choice perceptron; \\ 
\> \>  Chap. \ref{CHAP-game}: 
                 payoff matrix. \\

\> $h$ \> hidden fields of neural networks.\\
\> $H$ \> Chap. \ref{CHAP-antipredictable}: energy in the
Bernasconi Model;\\
\> \> Chap. \ref{CHAP-MG}: information generated in the standard MG. \\

\>$K$\> Sec. \ref{SEC-sto}: state of the stochastic MG; \\
\> $K$, $L$ \> Chap. \ref{CHAP-game}: 
number of options available to A and B; \\
\> $l$ \> Chap. \ref{CHAP-antipredictable}: 
                   cycle length. \\
\> $M$ \> memory length of prediction algorithm.\\
\> $\mathbf{M}$ \> Sec. \ref{SEC-lyapunov}: Jacobi matrix of
the CSG; \\
\> \> Sec. \ref{AP-propcyc}: transition probabilities on
DeBruijn graph.\\
\> $N$ \> Chap. \ref{CHAP-antipredictable}: number of nodes
                       in a graph;\\
\> \>  Chap. \ref{CHAP-MG}: 
                    number of players in the MG. \\
\> $p$ \> Sec. \ref{SEC-sto}: probability of changing
sides.\\
 \> $p_{AP}$ \> antipersistence parameter (see
 Sec. \ref{AP-stoch})\\
\> $p_h$, $P_h$ \> Sec. \ref{SEC-static}: probability of
hitting a cycle. \\
\>  $Q$ \> number of options in multi-choice MG; \\
\> \> number of different inputs/outputs in multi-choice
perceptrons \\
\> $\r$ \> Sec. \ref{SEC-MGNN}: relative vectors.\\
\>$R$\> Sec. \ref{SEC-MGNN}, \ref{SEC-PMG_NN}: overlap of weight
vectors;\\
\> \> Sec. \ref{PMG-evol}: self-overlap of strategy vector;
\\
\> \> Chap. \ref{CHAP-game}: normalized overlap between
$\va$ and $\vas$.\\
\> $s$\> Sec. \ref{AP-stoch}: success rate of prediction algorithm;\\
\> $S$ \> Sec. \ref{SEC-CSG-meanfield}: r.m.s. output of the
CSG. \\
\> \> Chap. \ref{CHAP-MG}: decision of the minority.\\
\> $\vw^t$ \> parameters of a prediction algorithm at time
$t$.\\
\> \> specifically: weight vector of a perceptron. \\
\> $\mathbf{W}$, $W$\> Sec. \ref{SEC-sto}: transition
matrix/ kernel. \\  
\> $x$\> Sec. \ref{SEC-sto}: proportionality constant $p=2x/N$.\\
 \> $x^t$  \> value of a time series.\\
 \> $\vx^t$ \> vector with components $x^{t-M-1},\dots
 x^{t-1}$.\\
\> $\alpha$ \> neural networks: rescaled learning time: $\a
=t/M$. \\
\> \> Bernasconi problem: ratio $p/M$  between length of sequence
and memory;\\
\> \> standard MG: ratio $p/N$ between length of decision
table.\\
\> \>and number of players. \\
\> $\beta$ \> amplification of continuous perceptron.\\
\> $\gamma$ \> rescaled amplification $\beta \eta$ of
continuous perceptron. \\
\> $\eta$ \> learning rate. \\
\> $\lambda$ \> Chap. \ref{CHAP-antipredictable}: Lyapunov
exponents of the CSG; \\
\> \> Chap. \ref{CHAP-game}: expected payoff.\\
\> $\pif$, $\pi$ \> Sec. \ref{SEC-sto}: state vector/ function. 

\end{tabbing}



\newpage
\pagestyle{empty}
\ \newpage
\section*{Acknowledgment}

At this point, I wish to express my gratitude
to the people and institutions who contributed to this work
directly or indirectly. I will start with the
Chair for Computational Physics in W\"{u}rzburg.

First of all, I want to thank my supervisor, Prof. Wolfgang
Kinzel, for the ideas and concepts that got this project
started, and for his supportive supervision. Special thanks
also go to Georg Reents and Christian Horn for the
collaboration on the stochastic Minority Game, and to 
Christoph Bunzmann and Michael Biehl for helpful advice
on various problems. Constructive work would have been
impossible without the system administrators, especially 
Ansgar Freking, Andreas Klein, and Alexander Wagner,
and the secretaries Uschi Eitelwein, Brigitte Wehner, and
Christine Schmeisser. 

The second group that contributed greatly to this work
is that of Prof. Andreas Engel at the university of
Madgeburg, and his students and post-docs Johannes Berg
and Stephan Mertens. There, I learned about the basics of
game theory and replica calculations and got valuable 
ideas and constructive criticism.

I am also grateful to Prof. Ido Kanter of Bar-Ilan
University, Ramat Gan, Israel, and his 
students Avner Priel, Liat Ein-Dor,  Michal Rosen-Zwi
and Hanan Rosemarin
for countless ideas and inspirations and for the
collaboration on the multi-choice Minority Game.
I greatly enjoyed the stays in Israel that Prof. Kanter
invited me to.

I acknowledge financial support by the German-Israeli 
Foundation for Scientific Research and Development
(GIF). My graduate student position was financed by
the GIF grant No. I -521-8.14/1997.

Last but not least, I wish to thank my dear friend Monika,
my parents, my friends, the Phachschaft Physik,
the Judo section of the DJK W\"{u}rzburg,
the Weltenwanderer W\"{u}rzburg e.V., and everyone else 
who made my private life more pleasant in the last years.


\ 
\end{document}